\documentclass[11pt,letterpaper]{article}
\usepackage{setspace}
\usepackage{natbib}
 \bibpunct[, ]{(}{)}{,}{a}{}{,}%
\usepackage{mathtools}              
 \usepackage{booktabs}
 \usepackage[shortlabels]{enumitem}
 \usepackage{multirow}
\usepackage[linesnumbered,ruled,noend,noline]{algorithm2e}
\usepackage{subfig}
 
\usepackage{amssymb,amsmath,amsthm}

\usepackage{csvsimple,booktabs,makecell}

\usepackage[dvipsnames]{xcolor}

\usepackage{hyperref} 

\hypersetup{colorlinks=true,breaklinks=true,
  citecolor = Blue}

\usepackage{siunitx, mhchem}
\usepackage{multirow}

\usepackage{appendix}
\usepackage{longtable} 
\usepackage{changepage}
\usepackage{tabularx}
\usepackage{setspace}
\newtheorem{definition}{Definition}

\newtheorem{theorem}{Theorem}

\newtheorem{proposition}{Proposition}
\newtheorem{corollary}{Corollary}

\usepackage[affil-it]{authblk}
\usepackage{longtable}
\DeclareMathOperator*{\argmin}{arg\,min}

\newtheorem{exe_thm}{Theorem}
\newtheorem{exe_cor}{Corollary}

\begin{document}
 
 \font\myfont=cmr12 at 14pt
\title{\vspace{0cm} Highly Connected Graph Partitioning: Exact Formulation and Solution Methods} 
\author[1]{Rahul Swamy\thanks{rahulswa@illinois.edu}}
\author[1]{Douglas M. King\thanks{dmking@illinois.edu}}
\author[2]{Sheldon H. Jacobson\thanks{shj@illinois.edu}}
\affil[1]{Department of Industrial and Enterprise Systems Engineering, University of Illinois Urbana-Champaign, 104 S Mathews Ave, Urbana, IL 61801.}
\affil[2]{Department of Computer Science, University of Illinois Urbana-Champaign, 201 N Goodwin Ave, Urbana, IL 61801.}
% \author{}
\date{}
\maketitle
\author{\vspace{-1cm}}
\begin{abstract} 
Graph partitioning (GP) and vertex connectivity have traditionally been two distinct fields of study.
This paper bridges this gap by introducing the \textit{highly connected graph partitioning} (HCGP) problem, which partitions a graph into compact, size balanced, and $Q$-(vertex) connected parts for any $Q\geq 1$. 
This problem is valuable in applications that seek cohesion and fault-tolerance within their parts, such as community detection in social networks and resiliency-focused partitioning of power networks.
Existing research in this fundamental interconnection primarily focuses on providing theoretical existence guarantees of highly connected partitions for a limited set of dense graphs, and do not include canonical GP considerations such as size balance and compactness.
This paper's key contribution is providing a general modeling and algorithmic approach for HCGP, inspired by recent work in the political districting problem, a special case of HCGP with $Q=1$.
This approach models $Q$-connectivity constraints as mixed integer programs for any $Q\geq 1$ and provides an efficient branch-and-cut method to solve HCGP.
When solution time is a priority over optimality, this paper provides a heuristic method specifically designed for HCGP with $Q=2$.
A computational analysis evaluates these methods using a test bed of instances from various real-world graphs.
In this analysis, the branch-and-cut method finds an optimal solution within one hour in $82.8\%$ of the instances solved.
For $Q=2$, small and sparse instances are challenging for the heuristic, whereas large and sparse instances are challenging for the exact method.
%is more effective in finding feasible solutions than the exact method in large instances.
%If giving priority to computational time over achieving optimality is the focus, our suggestion is to employ the exact method for small instances and opt for the heuristic in the case of larger instances.
Furthermore, this study quantifies the computational cost of ensuring higher connectivity using the branch-and-cut approach, compared to a baseline of ensuring $1$-connectivity.
Overall, the model and methods presented in this paper serve as effective tools to partition a graph into resilient and cohesive parts.
\end{abstract}

\maketitle
%%%%%%%%%%%%%%%%%%%%%%%%%%%%%%%%%%%%%%%%%%%%%%%%%%%%%%%%%%%%%%%%%%%%%%
 
%\section{Introduction.}\label{intro} %%1.
%\subsection{Duality and the Classical EOQ Problem.}\label{class-EOQ} %% 1.1.
%\subsection{Outline.}\label{outline1} %% 1.2.
%\subsubsection{Cyclic Schedules for the General Deterministic SMDP.}
%  \label{cyclic-schedules} %% 1.2.1
%\section{Problem Description.}\label{problemdescription} %% 2.

\section{Introduction} 

Graph partitioning (GP) is the fundamental problem of dividing a graph into smaller parts.
GP is important in several application domains such as community detection in social networks and decentralization of power and transportation networks.
In some applications, it is desirable that the subgraph induced by each part maintains a minimum level of (vertex) connectivity, $Q$.
Here, a graph's connectivity is measured by the minimum number of vertices whose removal disconnects the graph into $1$-connected (or \textit{simply connected}) components. 
For example, in community detection in social networks, each part (or community) is  cohesive if its connectivity is high \citep{orman2015overlapping}.
A cohesive community offers enhanced accessibility through multiple vertex-disjoint paths connecting different portions of the community.
Even if $Q-1$ community members are inactive or not responsive, the rest of the community remains simply connected.
In partitioning a power network, requiring minimum connectivity ensures that each part is fault-tolerant to vertex failures, i.e., it requires the failure of at least $Q$ vertices to disconnect a part \citep{abou2011floridian}.
Hence, high connectivity within graph partitioning is valuable in infusing cohesion and resilience into the parts.
In constraint-based modeling, requiring each part to be $Q$-connected for a pre-determined $Q\geq 1$ provides a lever in choosing the desired amount of cohesion and resilience to infuse within the parts.

%Hence, the problem of partitioning a graph into $Q$-connected parts, called a \textit{$Q$-proper partition}, is important for infusing resilience and cohesion into the parts.

%This paper studies an extension of GP that requires each part to be \textit{highly connected}.
%Here, a graph's \textit{connectivity} is measured by the minimum number of vertices whose removal disconnects the graph. 
%A highly connected graph has a pre-specified minimum level of connectivity ($Q\geq 1$) in each of the parts. 

%Highly connected graphs offer dual benefits.  
%First, they exhibit robustness against vertex failures, making them desirable in applications such as power systems where fault tolerance and network reliability are paramount.
%Second, they offer enhanced  accessibility through multiple vertex-disjoint paths connecting different parts of the graph, which is valuable in domains  such as social networks and transportation networks. 

%Why is it important to study graph partitioning with highly connected parts?

Despite its potential benefits, existing research on modeling and solving GP with $Q$-connectivity constraints is  limited.
%this variant of GP is understudied in existing existing research. 
The current body of research mainly focuses on providing theoretical conditions for the existence of a $Q$-\textit{proper partition} (i.e., a partition of a graph into $Q$-connected parts), primarily when the input graph is dense \citep{borozan2016partitioning,ferrara2013conditions}. 
%These studies generally aim to establish that any graph with a minimum degree above a certain threshold can be partitioned into highly connected parts. 
For example, \cite{borozan2016partitioning} showed that a graph with $n$ vertices can be partitioned into a bounded (but unspecified) number of $2$-connected parts if its minimum vertex degree is at least $\sqrt{n}$. 
While such findings shed light on the existence of a $2$-proper partition for this specific class of dense graphs, these findings do not extend to sparse graphs, i.e., graphs with a minimum vertex degree significantly smaller than $\sqrt{n}$, as is often the case in many real-world graphs such as in power and transportation networks. 
Even if these results are extended to sparse graphs, the resulting partition may not be suitable for practical scenarios.
For example, these works overlook fundamental GP considerations, such as pre-specifying the number of parts, incorporating size balance constraints or optimizing a compactness objective.
%, which are essential in many practical scenarios.   
Hence, a method to produce $Q$-proper partitions for practical applications must be able to handle any graph input  and accommodate fundamental GP considerations.  

%do not specify whether a certain number of parts is  

%Specifically, researchers \citep{borozan2016partitioning,ferrara2013conditions} have explored two key questions: Given a graph $G$ with $n$ vertices and minimum vertex degree $\delta$, (i) what is the minimum value of $\delta$ for which a $Q$-proper partition of the graph $G$ exists? (ii) can the number of parts in the partition be bounded above? 
%Here, a $Q$\textit{-proper partition} is a partition of $G$ where each part has a connectivity of at least $Q$.
%\cite{borozan2016partitioning} show that if $\delta \geq \sqrt{n}$, there exists a $2$-proper partition of $G$ into at most $n/\delta$ parts.
%Furthermore, if $\delta \geq \sqrt{c(Q-1)n}$, where $c = 2123/183$ and $Q\geq 2$, there exists a $Q$-proper partition of $G$ into at most $cn/\delta$ parts.
%These results provide general guarantees for highly dense graphs. 
 
An integer programming (IP) approach offers the flexibility to accommodate any specific graph instance while incorporating GP considerations. 
Existing work in IP approaches for GP has often studied the \textit{political districting problem} (PDP), which partitions a graph into compact, (weighted) size-balanced, and $1$-connected  parts \citep{swamy2023optimization}. 
While the IP  approaches for PDP in existing research have been introduced for political redistricting, some approaches are  application-agnostic and can be extended to GP with the aforementioned considerations.  
%PDP primarily focuses on political redistricting as an application, the models and methods presented in existing research in PDP are versatile and can be expanded to GP with the aforementioned considerations. 
%While PDP primarily focuses on political redistricting, the models and methods in existing work are application-agnostic and can be extended to GP with the aforementioned considerations. 
Particularly, a popular model for PDP \citep{hess1965nonpartisan,swamy2022} minimizes a widely-used compactness objective known as the \textit{k-median} or \textit{moment-of-inertia} objective, measured by the sum of weighted squared distances from a \textit{root} within each part; the root is a vertex that minimizes its sum of the weighted squared distances to the vertices within the same part.
Optimizing this objective ensures that the vertices within each part are close to the part's ``center of mass" \citep{elsner1997graph}. 
\cite{validi2020} noted that minimizing this objective encourages connectivity by bringing the vertices closer to their part's root.
While ensuring simple connectivity in PDP has historically posed computational challenges, 
recent advancements by \cite{oehrlein2017cutting} and \cite{validi2020} have introduced efficient IP formulations and branch-and-cut methods to alleviate these challenges. 
However, these formulations and methods cannot currently be used to find $Q$-proper partitions for $Q\geq 2$.

This paper formalizes the \textit{highly connected graph partitioning} (HCGP) problem, which partitions a graph into compact, balanced and $Q$-connected parts for any $Q\geq 1$.
The IP formulation for HCGP proposed in this study combines elements from IP formulations in two existing studies.
This work extends \cite{oehrlein2017cutting}'s IP formulation beyond simple connectivity to $Q$-connectivity for any $Q \geq 1$, thereby enabling fault-tolerant partitions.  
This formulation also extends \cite{buchanan2015integer}'s formulation for the fault-tolerant dominating set problem, extending their $Q$-connectivity constraint for a single subgraph to every subgraph within a graph partition.
Beyond the IP formulation, this paper offers efficient solution methods to solve HCGP.
To obtain an optimal solution for HCGP, we present a branch-and-cut method that solves the presented IP model, similar to a separator-based method in \cite{buchanan2015integer}.
A key challenge is that the number of $Q$-connectivity constraints in the IP model grows exponentially with the number of vertices.
%presented IP formulation for HCGP contains an 
% the number of constraints required to ensure $Q$-connectivity of the parts grows exponentially with the number of vertices. 
To tackle this computational challenge, the proposed branch-and-cut method adds violated cuts on-the-fly using an efficient integer separation algorithm.
To achieve this, we extend the procedure proposed by \cite{validi2020} for the $Q=1$ case. 
We show that the integer separation algorithm in the worst case 
%computational time for executing this separation algorithm is 
runs in polynomial-time in terms of the number of parts, vertices and edges, invariant of $Q$.
Furthermore, we introduce valid inequalities based on vertex degrees to strengthen the relaxation lower bounds.

This paper also presents a specialized heuristic, called the \textit{ear construction heuristic}, designed to solve HCGP with $Q=2$.
When prioritizing computational time over optimality, this heuristic  is valuable in tackling large instances where the exact method may be computationally intractable, as observed for the $Q=1$ case \citep{validi2020}.
%The goal of the heuristic is to provide a good feasible 
%This heuristic scales efficiently for large HCGP instances, while prioritizing computational time over optimality.
%, called the \textit{Bi-Connected Graph Partitioning} (BCGP) problem.
The ear construction heuristic aims to partition a graph into size balanced and $2$-connected parts while locally optimizing compactness. 
To accomplish this goal, we leverage the property that a $2$-connected graph contains an \textit{ear decomposition}, which comprises a cycle and a series of paths constructed sequentially.
Using this property, the heuristic first constructs $2$-connected subgraphs using multiple ear decompositions,  with each subgraph serving as a part in the partition.
Since any parts formed from the remaining graph may not be $2$-connected, this heuristic employs a repair stage that fixes parts that are not $2$-connected by locally reassigning vertices. 
Finally, two improvement stages improve the partition's size balance and compactness, respectively, using local search.
%The heuristic can handle graph instances of any size and density.

%The heuristic consists of four stages.
%The \textit{construction} stage builds an initial partition by finding multiple ear decompositions.
%The \textit{repair} stage performs local vertex reassignments to ensure that all the parts are $2$-connected.
%The \textit{balance} stage improves population balance until it is within the limits imposed by the balance constraint.
%Finally, the \textit{optimization} stage locally optimizes the compactness objective by reassigning vertices to neighboring parts one at a time.

To evaluate the computational performance of the proposed solution methods, we prepare a curated test bed of $42$ graph instances by altering graphs derived from various real-world domains. 
These domains encompass planar graphs such as transportation networks and census tract adjacency graphs, and non-planar graphs such as power and social networks.
The pre-processing stage alters these graphs to create diverse graphs that are at least $2$-connected. 
Ensuring this minimum level of input graph connectivity encourages the problem to have feasible partitions that are highly connected. 
Most of these processed graphs are sparse, with minimum degrees between two and four.
These minimum degrees are considerably lower than the $O(\sqrt{n})$ threshold required for existing research to guarantee the existence of a feasible $Q$-proper partition \citep{borozan2016partitioning,ferrara2013conditions}. 
The number of vertices in these processed graphs ranges from $20$ to $874$.
The diversity of this test bed in terms of domain, planarity and size enables us to make meaningful observations about the performance of the solution methods under different circumstances.
%%For context, none of the graphs are desnse e
 
The computational analysis in this paper evaluates the proposed solution methods' effectiveness in finding optimal or feasible solutions to HCGP.
For $Q=2$, this analysis delves into the impact of graph size and sparsity on the computational performance of the proposed methods.
%\textbf{Additionally, this analysis assesses the trade-offs between the computational advantage and the approximation gap in the compactness objective when using the heuristic for HCGP with $Q=2$.}
Furthermore, this analysis quantifies the \textit{costs} of enforcing higher connectivity (for $Q$ up to $4$) using the proposed branch-and-cut compared to the baseline case of $Q=1$ for the same graph.
These costs are measured in terms of the compromise in the compactness objective and the additional computational time required to solve the problem with higher connectivity. 
This analysis offers insights into the trade-offs in implementing the proposed IP formulation and solution methods for HCGP.

The rest of this paper is organized as follows.
Section \ref{sec_litrev} presents key literature interlinking graph partitioning, graph connectivity, and existing approaches to related problems.
Section \ref{sec_prelim_probdesc} describes graph preliminaries and formally defines HCGP.
Section \ref{sec_formulations} introduces an IP formulation for HCGP and provides a branch-and-cut approach to solve it.
%Focusing on $2$-connectivity, \ref{sec_repairheuristic} details the repair heuristic that generates feasible biconnected districts.
Section \ref{sec_ear_construction_heur} presents the ear construction heuristic for solving HCGP with $Q=2$.
Section \ref{sec_computationalresults} provides computational results from solving HCGP  for a variety of instances.
Finally, Section \ref{sec_conclusions} offers concluding remarks and potential future research directions. 
The appendix provides supplementary details on the proofs for theoretical results, pseudocodes for the heuristic, and tabulated data for the computational analysis.
 
%\newpage
\section{Literature Review} \label{sec_litrev}

%The highly-connected districting problem (HCDP) introduced in this paper connects several fundamental problems such as graph partitioning, network interdiction, clustering, among others.
%Each of this problems address different applications.
%In this section, we describe these existing problems and how they are connected to HCDP.
%We also discuss potential applications for HCDP.

%\citep{duchin2022explainer}

This section reviews key related literature to better understand models and methods at the intersection of graph partitioning and vertex connectivity.
Section \ref{subsec_litrev_GPHC} outlines literature on graph partitioning (GP) and the value of high connectivity in GP.
Section \ref{subsec_litrev_GPapproaches} details existing approaches to GP with high connectivity.
Section \ref{subsec_litre_exactformu} reviews IP approaches for high connectivity constraints and  the closely related political districting problem (PDP).
%Section \ref{subsec_litre_heuristic} provides a summary of heuristic methods for PDP.

%in the related problems, methods and applications.
%Section \ref{sec_litrev_relatedprobs} reviews existing work in HCGP and related problems.
%Section \ref{sec_litrev_methods} summarizes existing methods to solve the related problems.
%Section \ref{sec_litrev_applications} suggests potential applications for HCGP by drawing on existing applications of related problems.

%This paper connects two bodies of work related to graph partitioning and vertex connectivity.
%The motivations are two-fold.
%First, is the gap in the literature in addressing the problem of graph partitioning.

\subsection{Graph Partitioning and High Connectivity}\label{subsec_litrev_GPHC}
 
GP is an extensively studied  problem of dividing a graph into smaller parts \citep{bulucc2016recent,bichot2013graph}.
As an optimization problem, GP has a range of variants in terms of objectives and constraints.
A common objective is the compactness of the parts, which measures the shape of the parts.
While compactness can be measured in several ways, two popular formulations are to minimize the weighted number of edges that cross two parts \citep{hendrickson2000graph}, and to minimize the sum of weighted squared distances from each vertex to the ``center" of its part \citep{elsner1997graph}.  
In the latter case, the distances can be measured either by the Euclidean distance or the number of edges in a shortest path between two vertices \citep{diekmann2000shape,simon1991partitioning}.
Furthermore, it is common to ensure that the weighted sizes of the parts are relatively close, which is typically modeled as a constraint, called a \textit{size balance} constraint, and less frequently as an objective \citep{bulucc2016recent}.
These canonical GP considerations are common in a range of applications such as parallel computing \citep{hendrickson2000graph}, districting \citep{ricca2013political}, community detection in social networks \citep{bedi2016community}, and decentralization of power networks \citep{abou2011floridian}.

%\subsection{Connectivity}

Several application-specific studies have considered GP with high vertex connectivity within each part, particularly in community detection and enhancing resilience.
%GP applications in community detection in social networks, e.g., to maximize the internal connectivity within the parts 
In social networks, partitioning a graph into highly connected communities is valuable in applications such as mass communication or product recommendations \citep{bedi2016community}. 
The works by \cite{orman2015overlapping, orman2021multiple} and \cite{timofieiev2008social} have defined a $Q$-connected subgraph in a social network as a cohesive community of users, where the existence of $Q$ vertex-disjoint paths is desired.  
Maximizing the internal connectivity within each community can enhance communication and facilitate the flow of information in social networks.
In power networks, GP creates ``islands" or self-sufficient smaller networks  resilient to vertex failures. 
For example, \cite{abou2011floridian} introduced a clustering approach that maximizes the connectivities of the parts while partitioning a power network.
%\textbf{Add a sentence}
Furthermore, partitioning a transportation network is useful for distributed traffic management and computing.
In resilience-focused partitioning of a transportation network, existing work addresses the problem of dynamic re-partitioning after vertex disruptions \citep{ghavidelsyooki2017partitioning}.
%However, the problem of GP with resilient  
%However, there is no existing work that studies the static problem of partitioning a tra
These works  tailor  resilience considerations to their respective application domains, but do not provide a general solution framework for GP with high connectivity requirements.

\subsection{Approaches to GP With High Connectivity} \label{subsec_litrev_GPapproaches} 

%%We now review application-specific literature that link these concepts.

%Existing research finds a feasible $Q$-connected partition  of a graph into an arbitrary number of partitions, with no restrictions on the partition sizes. 

Existing research in GP with high connectivity requirements focuses on exploring theoretical conditions for the existence of a highly connected partition.
These works particularly focus on highly dense input graphs with a high minimum vertex degree $\delta$.
%Let us define a $Q$\textit{-pr oper partition} to a partition of $G$ where each part has a connectivity of at least $Q$; the number of parts and their sizes are arbitrary.
Researchers \citep{borozan2016partitioning,ferrara2013conditions} have explored two key questions: Given a graph $G$ with $n$ vertices and minimum vertex degree $\delta$, (i) what is the minimum value of $\delta$ for which a $Q$-proper partition of the graph $G$ exists, and (ii) can the number of parts in the partition be bounded from above?  
\cite{borozan2016partitioning} showed that if $\delta \geq \sqrt{n}$, there exists a $2$-proper partition of $G$ into at most $n/\delta$ parts.
Furthermore, if $\delta \geq \sqrt{c(Q-1)n}$, where $c = 2123/183$ and $Q\geq 2$, there exists a $Q$-proper partition of $G$ into at most $cn/\delta$ parts.
These results provide general existence guarantees for highly dense graphs. 
%Recent studies by \citep{ferrara2013conditions,borozan2016partitioning,chen20222} have investigated the feasibility of HCGP and provided upper bounds on the number of partitions.
% \cite{borozan2016partitioning} showed that for $Q=2$, there exists a $2$-connected partition  of $G$ into at most $(n-1)/\delta$ partitions if $\delta \geq \sqrt{n}$. 
%For a general $Q\geq2$, if the input graph is highly dense with $\delta \geq \sqrt{11.8(Q-1)n}$, then there exists a Q-connected partition of $G$ into at most $(11.8n)/\delta$ partitions.
%For example, for a graph with $500$ vertices and minimum degree $23$, this result guarantees the existence of a feasible $2$-partitioning into at most $22$ partitions.
However, these existence guarantees do not apply to partitioning problems on sparse graphs, nor can they enforce canonical constraints like guaranteeing a specific number of parts or limiting the size of each part.
Hence, an algorithmic approach to find a $Q$-proper partition in a general setting remains an open problem.

%We will now follow the progress of three such works.
%Given a graph $G$ with $n$ vertices, $m$ edges, and minimum degree $\delta$, \cite{ferrara2013conditions} showed that for $Q\geq 2$, if $\delta \geq 2Q\sqrt{n}$, there always exists a $Q$-connected partitioning of $G$.
%For $Q=2$, 
%\cite{chen20222} showed that if either $\delta \geq \sqrt{n}$, or ($\delta < \sqrt{n}$ and $\sigma \geq n/\delta + \delta - 1$), then there exists a $2$-connected partitioning of $G$ into at most $2(n-1)/\sigma$ partitions.
%For $Q = 2$, \cite{borozan2016partitioning} showed that if $\delta \geq \sqrt{n}$, then there exists a $2$-connected  partitioning of $G$ into at most $(n-1)/\delta$ partitions.
%For a general $Q \geq 2$, if the input graph is highly dense with  $\delta \geq \sqrt{11.8(Q-1)n}$, then there exists a $Q$-connected partitioning of $G$ into at most $(11.8n)/\delta$ partitions.
%\cite{borozan2016partitioning} improved this result: if $\delta \geq \sqrt{n}$, then there exists a $2$-connected $K$-partitioning of $G$, where $K \geq (n-1)/\delta$, and if $\delta \geq \sqrt{c(Q-1)n}$, then there exists a $Q\geq2$-connected $K$-partitioning of $G$, where $K \geq (cn)/\delta$.
%Here, $c = 2123/180 \approx 11.8$.
%However, these existence conditions do not extend to general graphs and when seeking a fixed number of partitions $K$.

A related algorithmic approach is the Highly Connected Subgraph clustering algorithm.
This method  partitions a graph while ensuring that each part's \textit{edge connectivity} (i.e., the minimum number of edges whose removal disconnects the graph) is greater than half the number of vertices in that part \citep{hartuv2000clustering}. 
This algorithm has been successful in applications involving biological networks, including clustering cDNA fingerprints and grouping protein sequences \citep{bliznets2017parameterized}.
Like the previously mentioned approaches for dense graphs, this algorithm does not enforce key GP constraints (e.g., requiring a particular number of parts). 
% such as having a fixed number of parts, size balance, compactness, among others.
Moreover, ensuring high edge connectivity may not translate to ensuring high vertex connectivity.
For example, consider a graph constructed as follows: consider two complete graphs, introduce an additional vertex, and add an edge from this vertex to every vertex in both graphs. 
In this composite graph, the edge connectivity is high, i.e., nearly equal to the number of vertices in either of the complete subgraphs.
However, the vertex connectivity is one, where the additional vertex acts as a cut vertex.
%For example, a simply connected graph could have a high edge connectivity: construct a graph composed of two fully connected subgraphs, and a vertex in the middle 
Hence, an alternative algorithmic approach is needed to ensure high vertex connectivity in graph partitioning.

%Connected and Balanced Graph Partitioning

%Another related problem is the Fault tolerant connected dominating sets (minimum $k$-connected $d$-dominating set problem).
%\citep{buchanan2015integer}

%Dense Graph Partitioning on sparse and dense graphs \cite{bazgan2021dense}.

%Concluding para

%\subsubsection{Graph Partitioning}

%Graph partitioning is the fundamental problem of dividing a graph into a number of sub-graphs.
%The relevant problem under study is partitioning of a graph into compact, balanced and $q$-connected partitions.

%\clearpage

\subsection{IP Models for High Connectivity and GP}\label{subsec_litre_exactformu}
 
%- High connectivity formulations - Fault-tolerant highly connected dominating set (Buchanan). Maybe critical node detection problem (salimi and buchanan)

%We now review exact formulations for the graph partitioning with a $k$-median objective.

%\cite{tyuryukanov2022new}
%
%\cite{vangerven2022parliament}
% 

%Among models that ensure connectivity, there are broadly three types of formulations -- those that assign vertices to a representative vertex (called a root), those that assign vertices to part labels, and those that decide whether two vertices are in the same part. 

Formulating high vertex connectivity constraints in an IP model has predominantly been studied with the goal of finding a highly connected dominating set in wireless networks.
In this problem, the goal is to construct a fault-tolerant \textit{backbone} subgraph, wherein the vertices in the backbone serve as intermediary transmission points.
Existing research in this area has proposed formulations for constructing a $2$-connected backbone \citep{li2012construction, dai2006constructing}, which have subsequently been extended to constructing a $Q$-connected backbone \citep{ahn2015optimization, buchanan2015integer} for any $Q\geq 2$. 
These models leverage vertex separator-based constraints to ensure $Q$-connectivity within the backbone subgraph.
%To adapt these  
However, the challenge faced by HCGP is that it requires all the part subgraphs to be $Q$-connected, rather than a single subgraph representing the backbone.  
Hence, it remains unclear how vertex separator-based constraints can  ensure $Q$-connectivity in a graph partitioning setting.

By examining the exact methods used for solving PDP, we can develop effective methods to address the challenge posed by HCGP. 
Notably, \cite{oehrlein2017cutting} and \cite{validi2020} have introduced state-of-the-art formulations and methods for ensuring $1$-connectivity in PDP, a longstanding challenge.
%In tandem with the flow model by \cite{shirabe2009districting}, these works present the state-of-the-art cutting planes methods for ensuring $1$-connectivity in PDP, which had been a longstanding challenge in solving large PDP instances.
\cite{oehrlein2017cutting} have formulated an IP based on vertex separators, and \cite{validi2020} have proposed an efficient branch-and-cut algorithm to solve this formulation by strengthening its valid cuts.
%Our paper extends both of these formulations for any $Q\geq 1$, thereby enabling potential applications that demand greater connectivity.
%\textbf{What are the implications of these methods?}
While these studies address the computational challenge of requiring $1$-connectivity in each part, 
the computational costs of extending these method to $Q$-connectivity for any $Q \geq 1$ (e.g., by leveraging the $Q$-connected backbone formulation from \cite{buchanan2015integer}) have not been studied.

%\subsection{Heuristic Approaches for PDP} \label{subsec_litre_heuristic}

%- Ear decomposition - used to sample a subgraph but not necessarily to partition - remains unclear how it can be used to partition a subgraph
If solution time is a priority over optimality, several heuristics have been proposed to solve PDP. 
A popular constructive heuristic is the multi-kernel growth method \citep{vickrey1961prevention}, which grows the parts by incrementally adding neighboring vertices. 
%\textbf{Why extend this to HCGP?}
To extend this approach to HCGP, it is necessary to ensure that each growth step maintains $Q$-connectivity within the parts. 
Furthermore, there are several improvement heuristics for PDP, including local search \citep{dobbs2023,swamyarizona}, simulated annealing \citep{d2002simulated}, tabu search \citep{bozkaya2003tabu}, iterative bisection \citep{ludden2022}, and multilevel algorithms \citep{swamy2022}. 
These heuristics modify an initial solution (e.g., one obtained with a constructive heuristic) to optimize an objective (e.g., compactness). 
To adapt these methods to HCGP, designing a heuristic that maintains $Q$-connectivity after every improvement step is essential.

\section{Problem Description} \label{sec_prelim_probdesc}

This section presents a mathematical description of the highly connected graph partitioning (HCGP) problem.
Section \ref{sec_prelim} defines mathematical terminology and notation related to graphs and connectivity.
Section \ref{sec_problem_description} applies these graph theoretic concepts to define HCGP formally. 
Section \ref{subsec_Hessmodel} provides a general integer programming (IP) model for HCGP.

\subsection{Graph Preliminaries}\label{sec_prelim}
%\textbf{cycle, tree, forest}

We introduce the graph terms and notation used in the rest of the paper.
An undirected simple graph $G=(V,E)$ comprises a vertex set $V$ and an edge set $E \subseteq \choose{V}{2}$.
Let the number of vertices and edges be $n=|V|$ and $m=|E|$, respectively.
%For each vertex $i \in V$, let $N(i) \coloneqq \{v \in V: (u,v)\in E\}$ denote the set of its neighbors.
For a subset of vertices $S \subseteq V$, let $N(S)\coloneqq \{v \in V\backslash S: \exists \ u \in S, (u,v)\in E \}$ denote  the set of its neighbors.
The \textit{degree} of a vertex $i \in V$ is its number of neighbors $|N(\{i\})|$.
The \textit{minimum} (\textit{average}) {degree} $\delta$ ($\overline{\delta}$) of a graph is the minimum (average) degree among all its vertices. 
Furthermore, for a subset of vertices $S \subseteq V$, let $G[S]$ denote the subgraph induced by $S$.
Inversely, for a subgraph $G'$ of $G$, let $V[G']$ and $E[G']$ denote the set of vertices and edges, respectively, in $G'$. 
Let $G-S \coloneqq G[V\backslash S]$ denote the subgraph induced by removing $S$ from $G$.
%Let $\kappa(G) \leq 0$ be the connectivity of $G$.
%Between every pair of vertices $i,j \in V$, an $i,j$-path is an ordered sequence of vertices $\{i,a_i,a_2,\dots,j\}$ for some $$

We now define concepts related to paths.
A \textit{path} $P$ is an ordered sequence of distinct vertices $(a_1,a_2,\dots,a_{|P|})$ such that $|P|\geq2$ and each vertex is a neighbor of its succeeding vertex, i.e., $a_{i+1} \in N(\{a_i\})$ for all $i = 1,2,\dots,|P|-1$.
In a path  $P$, the subset $P\backslash \{a_1,a_{|P|}\}$ constitutes its \textit{internal vertices}.
Given two vertices $s,t \in V$, an $s,t$-\textit{path} is a path originating from $s$ and terminating at $t$, i.e., $a_1 = s$, $a_{|P|} = t$.
Two $s,t$-paths $P$ and $P'$ are said to be \textit{internally disjoint} if they share no vertices besides $s$ and $t$, i.e., $P \cap P' = \{s,t\}$.
A \textit{cycle} $P$ is a path with at least three vertices and whose end-points are adjacent, i.e., $a_1 \in N(\{a_{|P|}\})$.
A \textit{tree} is a graph that does not contain a cycle.
\begin{figure}[!htbp]
\centering
\subfloat[$1$-connected \label{fig_example_1connected}]
{\includegraphics[width=0.2\textwidth]{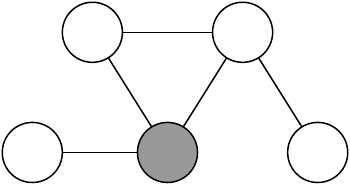}} 
%\vspace{.2cm}
\hspace{2cm}
%\hfloat
\subfloat[$2$-connected \label{fig_example_1connected}]
{\includegraphics[width=0.15\textwidth]{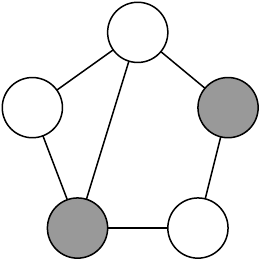}} 
%\vspace{.2cm}
\hspace{2cm}
\subfloat[$3$-connected \label{fig_example_1connected}]
{\includegraphics[width=0.15\textwidth]{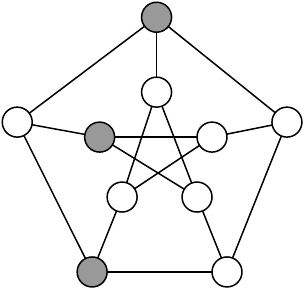}} 
%\vspace{.2cm}
%\hspace{0.1cm}
\caption{Graph examples illustrating vertex connectivity; shaded vertices constitute a minimum cutset.} \label{fig_examples_q_connectivity}
\end{figure}

We now formalize concepts related to graph connectivity.
A graph is said to be \textit{connected} if there is an $s,t$-path in $G$ between every pair of distinct vertices $s,(\neq)t \in V$; otherwise, the graph is \textit{disconnected}.
A \textit{component} of a graph is a maximal connected subgraph.
A \textit{forest} is a graph whose components are trees.
Given a graph $G$, a \textit{cutset} (or a \textit{vertex separator}) is a subset of vertices $C \subset V$ whose removal disconnects the remaining graph, i.e., $G-C$ is disconnected. 
If a cutset comprises a single vertex, that vertex is called a \textit{cut vertex}.
A \textit{minimum cutset} is a cutset of the smallest size.
%For a given $q\geq 1$, a $q$-cutset is a cutset of size $q$.
Let $\mathcal{C}(G)$ be the collection of all cutsets of $G$.
For a given positive integer $Q\geq 1$, a graph $G$ with $n$ vertices is $Q$-\textit{connected} if $n \geq Q+1$ and the size of every cutset of $G$ is at least $Q$, i.e., $|C| \geq Q$ for all $C \in \mathcal{C}(G)$; otherwise, it is called $Q$-\textit{disconnected}.
Figure \ref{fig_examples_q_connectivity} illustrates examples of $Q$-connected graphs for $Q\in[3]$; the notation $[n]$ refers to the set $\{1,2,\dots,n\}$.
%there is no cutset in $G$ of size less than $q$, i.e., $\argmin_{C \in \mathcal{C}(G)} |C| \geq q$.
Further, the \textit{connectivity} of graph $G$, given by $\kappa(G) \coloneqq \min_{C \in \mathcal{C}(G)} |C|$, is the size of its minimum cutset.
A complete graph $G$ is considered  $(n-1)$-connected, and $\kappa(G)=n-1$.
Furthermore, a disconnected graph $G$ is $0$-connected, and $\kappa(G)=0$.
%A $2$($3$)-connected graph is called a \textit{bi(tri)-connected} graph.

%For a given integer $Q \geq 1$, the \textit{highly connected graph partitioning} (HCGP) problem partitions a graph $G$ into subgraphs that are $Q$-connected.
%There are no limits on the number of partitions or their sizes.

\subsection{Highly Connected Graph Partitioning (HCGP)}\label{sec_problem_description}

The \textit{highly connected graph partitioning} (HCGP) problem divides a graph into a given number of parts, denoted by $K \geq 2$, that are balanced in size, highly connected, and optimally compact.
HCGP is formally defined as follows.
% finds a $Q$-connected $K$-partition of a graph that is balanced in weighted size and is optimally compact. 
Consider a graph $G = (V,E)$ with vertex set $V$ and edge set $E$. 
%HCGP divides $G$ into $K$ parts.
Each vertex $i \in V$ has an associated weight $p_i \geq 0$.
For a subset of vertices $S\subseteq V$, let $size(S)\coloneqq \sum_{i \in S} p_i$ denote the total weighted size  of $S$.
HCGP requires that the weighted size of each part should be within given limits.
Let $L$ and $U$ be the lower and upper limits on the weighted sizes, respectively.
Further, HCGP requires that the connectivity of each part meets a minimum threshold, denoted $Q \geq 1$ 
(e.g., when $Q = 1$, each part must be simply connected as in PDP).
% of each part denoted as $L \in [0,\sum_{i\in V} p_i/K]$ and $U \in [L,\sum_{i\in V} p_i]$, respectively.

A feasible partition to HCGP is now defined.
A \textit{$K$-partition} $\mathcal{P}$ of $G=(V,E)$ is a collection of vertex subsets $V^{(1)},V^{(2)},\dots,V^{(K)}$ that cover $G$, i.e., $\cup_{k=1}^K V^{(k)} = V$, 
and are pair-wise disjoint, i.e., for $k,k'\in[K]$, $V^{(k)} \cap V^{(k')} \neq \emptyset \Leftrightarrow k=k'$.
\begin{definition}
A $K$-partition $\mathcal{P}$ is said to be:
\begin{itemize}
\item \textnormal{L,U-balanced} if the weighted size of each part is within the range $[L,U]$, i.e., $ size(V^{(k)})  \in [L,U]$ for all $k=1,2,\dots,K$.
\item $Q$-\textnormal{proper} if the subgraph $G[V^{(k)}]$  is $Q$-connected for all $k=1,2,\dots,K$. 
\end{itemize}
\end{definition}

HCGP finds an $L,U$-balanced $Q$-proper $K$-partition of $G$ that  minimizes a compactness objective defined as follows.
%Given a $K$-partition, its compactness measures the shape of the parts defined as follows.
%k\textbf{-median objective, give application examples}
For every pair of vertices $i$ and $j$, let $w_{ij} \geq 0$ denote the distance between them.
Given a $K$-partition $\mathcal{P} = \{V^{(k)}\}_{k \in [K]}$,
 for each part $k \in [K]$, a \textit{root} $r_k \in V^{(k)}$ is a vertex that minimizes the sum of distances to the rest of the part, i.e., $r_k \coloneqq \argmin_{i \in V^{(k)}} \sum_{j \in V^{(k)}} w_{ij}$. 
The \textit{compactness} of a partition is calculated as the sum of distances from the roots to the rest of their respective parts, given as
\begin{align}
\phi_{comp}  (\mathcal{P})\coloneqq \sum_{k=1}^K \sum_{j \in V^{(k)}} w_{r_kj}. \label{eq_compactness_obj}
\end{align} 
%\textbf{population times distance}
This objective is commonly referred to as the $k$-median or $p$-median objective.
As prescribed in PDP models, such as in \cite{hess1965nonpartisan}, we set the pair-wise vertex distance $w_{ij} \coloneqq p_j d^2_{ij}/size(V)$, where $d_{ij}$ is the number of edges in a shortest path between $i$ and $j$ in $G$; the denominator $size(V)$ is a normalization term.
If the geometric embedding of the vertices is known, it is also common for $d_{ij}$ to measure the Euclidean distance between the coordinates of $i$ and $j$.
However, we consider the shortest path-based distance for an embedding-agnostic graph instance.
%Further, as a special case when $Q=2$, HCGP will be referred to as the \textit{Bi-Connected graph partitioning} (BCGP) problem.

\begin{figure}[!htbp]
\centering
\subfloat[Graph $G$. \label{fig_hcgp_example1}]
{\includegraphics[width=0.27\textwidth]{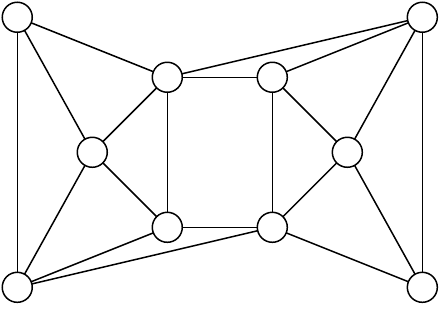}} 
%\vspace{.2cm}
\hspace{.3cm}
%\hfloat
\subfloat[Optimal solution for $Q=1$. \label{fig_hcgp_example2}]
{\includegraphics[width=0.27\textwidth]{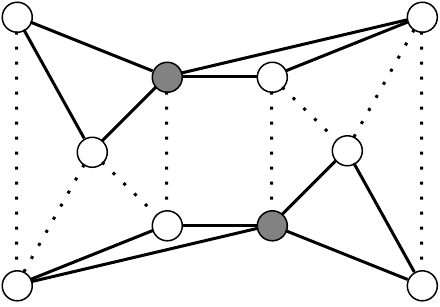}}  
\hspace{.3cm}
%\hfloat
\subfloat[Optimal solution for $Q=3$. \label{fig_hcgp_example3}]
{\includegraphics[width=0.27\textwidth]{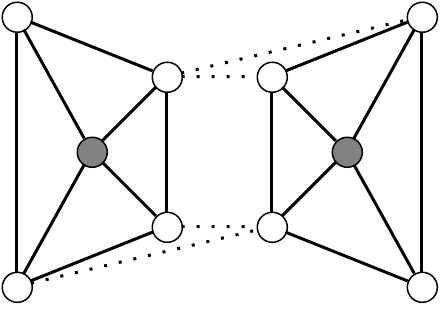}}   
\caption{$1$-connected versus $3$-connected parts: An example illustrating optimal $5,5$-balanced $Q$-proper $2$-partitions of a graph $G$ for $Q \in \{1,3\}$; assume each vertex's weight to be one.
%solutions to HCGP given a graph $G$ with ten vertices, $K=2$, $Q \in \{1,3\}$, and $L=U=5$
Solid lines are internal edges in each part and dotted lines are inter-part edges.
Each part's root vertex is shaded.
} \label{fig_hcgp_example}
\end{figure}

Figure \ref{fig_hcgp_example} illustrates optimal solutions to HCGP with $Q=1$ and with $Q=3$ for a sample graph instance depicted in Figure \ref{fig_hcgp_example1}.
Compared to the $1$-proper partition in Figure \ref{fig_hcgp_example2}, the $3$-proper partition in Figure \ref{fig_hcgp_example3} exhibits more robust parts with multiple vertex-disjoint paths within each part.  
Hence, ensuring $3$-connectivity within the parts is essential to guarantee a robust partition.
Further, note that both partitions have the same compactness value of $0.8$. 
Hence, for this specific graph, ensuring $3$-connectivity does not inflict any cost in terms of compactness compared to ensuring $1$-connectivity.

\subsection{Modeling HCGP as an Integer Program}\label{subsec_Hessmodel}
We now present an IP formulation for HCGP.
HCGP can be modeled as an extension of the connectivity-free PDP model given by \cite{hess1965nonpartisan}, referred to as the \textit{Hess model} by \cite{validi2020}.
This model captures the balance constraint and the compactness objective in HCGP, but not the $Q$-connectivity constraints.
The primary decision variables in this IP select $K$ roots and assign each vertex to a root, denoted as 
\begin{align*} \nonumber
x_{ii} &= \left\{\begin{array}{rl}
1, &\mbox{\textrm{if vertex $i \in V$ is chosen as a  root}},\\
0, &\mbox{\textrm{otherwise},}
      \end{array}\right. \\
x_{ij} &= \left\{\begin{array}{rl}
1, &\mbox{\textrm{if vertex $j\ (\neq i) \in V$ is assigned to root $i \in V$}},\\
0, &\mbox{\textrm{otherwise}.}
      \end{array}\right. 
\end{align*} 
The Hess model is given as
\begin{subequations}
\begin{align}
\textrm{(PDP) Minimize } &\sum_{i,j \in V} w_{ij} x_{ij} \label{eq_obj}\\
\textrm{such that }
 L x_{ii} &\leq \sum_{j\in V} p_j x_{ij}  \leq U x_{ii} &\forall i \in V, \label{const_popbal}\\
\sum_{i \in V} x_{ii} & = K, & \label{const_K_assns}\\
\sum_{i \in V} x_{ij} & = 1 &\forall j \in V, \label{const_unique_assn}\\
x_{ij} &\leq x_{ii} &\forall i, j \in V, \label{const_only_if_root}\\
%\textbf{x} & \textrm{ is $Q$-connected},   & \label{const_qconnected_Hess}\\
\textbf{x} & \in \{0,1\}^{|V|\times|V|}. & \label{const_defining_x_hess}
\end{align}
\end{subequations}
The objective in (\ref{eq_obj}) minimizes compactness.
Constraints (\ref{const_popbal}) ensure that the weighted size of every part is within the given range $[L,U]$.
Constraint (\ref{const_K_assns}) requires exactly $K$ vertices to be chosen as roots, and constraints (\ref{const_unique_assn}) ensure that every vertex is assigned to exactly one root.
Constraints (\ref{const_only_if_root}) ensure that a vertex $j$ can be assigned to a vertex $i$ only if $i$ is chosen as a root.  
%Constraint (\ref{const_qconnected_Hess}) ensures that the connectivities of the part subgraphs is at least $Q$.
%Constraints (\ref{flow_constr_1}) ensure that for every $i,j \in V$, if $j$ is assigned to $i$ (i.e., $x_{i,j} = 1$), then vertex $j$ absorbs one vertex of flow that originated from $i$ for $i \neq j$; for $i = j$, vertex $i$ sends out a quantity of flow equal to the number of vertices in that part  (i.e., $\sum_{v \in V | i = j} x_{iv}$).
%Constraints (\ref{flow_constr_2}) ensure that vertex $j$ receives a positive amount of flow (from its neighbors) originating from $i$ only if $j$ is assigned to part root $i$.
%Constraints (\ref{flow_constr_1}) and (\ref{flow_constr_2}) together ensure that  the parts are contiguous.
Constraint (\ref{const_defining_x_hess}) defines the binary nature of the variables.
We denote $\mathcal{F}_{HESS} \coloneqq \{\textbf{x} \in \{0,1\}^{|V|\times|V|}: \textbf{x} \ \textrm{ satisfies constraints } (\ref{const_popbal})-(\ref{const_only_if_root})\}$ as  the polytope of feasible solutions to the Hess model.

The Hess model finds an optimally compact $L,U$-balanced $K$-partition of a graph.
HCGP additionally requires the subgraph induced by each part to be $Q$-connected (for $Q\geq 1$), which is the focus of this paper.
%We now formalize what it means for a solution to the Hess model to be $Q$-connected.
Given a solution $\hat{\textbf{x}} \in \mathcal{F}_{HESS}$, let $R(\hat{\textbf{x}})\coloneqq \{i \in V: \hat{x}_{ii}=1\}$ denote the set of all roots.
For each root $r \in R(\hat{\textbf{x}})$, let $V_r \coloneqq \{i \in V: \hat{x}_{ri}=1\}$ denote the set of vertices assigned to $r$.
Then, $\hat{\textbf{x}}$ is $Q$-\textit{proper} if the subgraph $G[V_r]$  is $Q$-connected for every root $r \in R(\hat{\textbf{x}})$.
HCGP is generally expressed as
\begin{subequations}
\begin{align}
\textrm{(HCGP) Minimize } &\sum_{i,j \in V} w_{ij} x_{ij} \label{eq_obj_HCGP}\\
\textrm{such that }
\textbf{x} &\in \mathcal{F}_{HESS}, \\ 
\textbf{x} & \textrm{ is $Q$-proper}. \label{const_qconnected_Hess}
\end{align}
\end{subequations}
%This formulation has $O(n^2)$ variables and $O(n^2)$ constraints, and has strong linear programming (LP) lower bounds that facilitate rapid convergence to an optimal solution \citep{salazar2011bi}.   
In existing research \citep{shirabe2009districting,oehrlein2017cutting,validi2020},
formulating constraints (\ref{const_qconnected_Hess}) as a (mixed) integer program  has been limited to the $Q=1$ case.
This paper provides an IP formulation for constraints (\ref{const_qconnected_Hess}) for any $Q\geq 1$.

\section{Exact Method} \label{sec_formulations}

This section presents an IP formulation for HCGP and a branch-and-cut method to solve it efficiently.
To enable the IP formulation, Section \ref{subsec_characterization} presents a characterization of a $Q$-proper partition (for $Q\geq1$) based on vertex separators.
Using this characterization, Section \ref{subsec_formulation_qcut} presents a quadratic-constrained IP formulation for the $Q$-connectivity constraints (\ref{const_qconnected_Hess}).
Section \ref{subsec_degree_valid_constraints} presents  valid inequalities based on the property that the degree of a $Q$-connected subgraph should be at least $Q$.
Finally, Section \ref{subsec_cuttingplanes_overall} presents an integer separation algorithm  solved within a branch-and-cut algorithm for HCGP.
Appendix \ref{appendix_proofs} contains proofs for the theoretical results presented  in this section.

\subsection{Characterization of a $Q$-Proper Partition}\label{subsec_characterization}
 
%\cite{oehrlein2017cutting} provided an MILP formulation for $1$-connectivity, which we extend to $Q$-connectivity.
%Next, we provide a cut-based formulation for ensuring that each district is $Q$-connected.
%But first, some definitions.

\begin{figure*}[!h]
\centering
\includegraphics[width=0.3\textwidth]{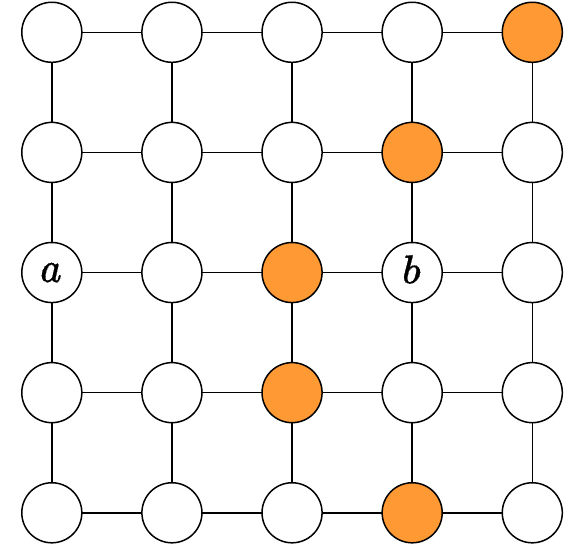}
\caption{A minimal $a,b$-separator represented by shaded vertices in a $5\times 5$ grid graph. 
} \label{fig_multilevel_schematic}
\end{figure*}

%We identify conditions under which a given partition is $Q$-connected.
In a $Q$-proper partition, by definition, the size of every vertex separator within each part must be at least $Q$. 
%A $Q$-proper partition, by definition, is one where every part remains connected after the removal of any less than $Q$ vertices from it.
This requirement can be expressed by considering pairs of vertices in a graph and verifying this condition for each vertex separator that disconnects the given pair. 
%To address this, let us introduce the concept of separators.
Given two vertices $a$ and $b$ in a graph $G$, an $a,b$-\textit{separator} in $G$  is defined as follows.
\begin{definition}\label{defn_abseparator}
\textit{$a,b$-separator in $G(V,E)$ \citep{oehrlein2017cutting}}. For $a,b \in V$, a subset of vertices $C \subseteq V \backslash \{a,b\}$ is an $a,b$-separator in $G$ if there is no $a,b$-path in $G-C$.
\end{definition}
Let $\mathcal{C}_{a,b,G}$ denote the collection of all $a,b$-separators in $G$.
Note that $\mathcal{C}_{a,b,G} = \emptyset$ if $a=b$ or $a\in N(\{b\})$.
In a complete graph, and only in a complete graph, 
$\mathcal{C}_{a,b,G} = \emptyset$ for every distinct $a,b \in V$.
Furthermore, a separator is \textit{minimal} if it cannot be reduced in size while it remains a separator.
Figure \ref{fig_multilevel_schematic} illustrates a minimal $a,b$-separator for two vertices $a,b$ in a $5\times 5$ grid graph.

\cite{oehrlein2017cutting} gave Definition \ref{defn_abseparator} to formulate $1$-connectivity requirements, which we now extend to $Q$-connectivity. 
The goal is to characterize each part's $Q$-connectivity with separators defined on the whole graph.
Given the set of all separators for all pairs of vertices in graph $G$, a $Q$-proper partition can be characterized as follows. 
A $K$-partition is $Q$-proper if and only if each part $k \in [K]$ has at least $Q+1$ vertices, and for any pair of vertices $a$ and $b$ within part $k$, every $a,b$-separator in $G$ intersects with at least $Q$ vertices within part $k$. 
This claim is shown in the following theorem.
%Let $\mathcal{C}$ be the set of all $a,b$-separators.
%Consider a connected solution $x$.
%We want to generate a separating cut. 
%The following inequalities for $1$-connectivity were proposed by \cite{oehrlein2017cutting}.
%For every pair of non-adjacent vertices $a,b \in V$ and $a,b$-separator $C \in \mathcal{C}$, 
%\begin{subequations}
%\begin{align}
%\textbf{x}  & \in \mathcal{P}_{HESS}, \label{const_hess_constraints_CUT} \\
%x_{a,b} &\leq \sum_{c \in \mathcal{C}} x_{a,c} \quad \forall \ (a,b,C). \label{eq_1CUTs} 
%\end{align}
%\end{subequations}

%Let $\mathcal{P}_{CUT} \coloneqq \{\textbf{x} \in \mathcal{P}_{HESS}: \textbf{x} \textrm	{ satisfies constraints (\ref{eq_1CUTs})}\}$ be the polyhedron represented by the solutions that satisfy constraints (\ref{eq_1CUTs}).
%\cite{validi2020} showed that $\mathcal{P}_{CUT} \subseteq \mathcal{P}_{conn}^{(1)}$.

%\subsection{Cut Formulation ($q$-$CUT$) for $q\geq 1$}

%\subsection{Separation Algorithm}

%For each district with root $a$, let $V_a\coloneqq \{i \in V | x^*_{ai} = 1\}$ be the set of vertices assigned to $a$.
%Let $A$ be the set of articulation points in the component of $G[V_a]$ containing $a$.
%Consider an articulation point $v \in A$ and a component $G_0$ of $G[G_a-\{v\}]$ not containing $a$.
%Let $b \in G_0$ be an arbitrarily chosen vertex, and let $\mathcal{C}$ be an $a,b$-separator.

\begin{theorem}
 \label{thm_characterization} (Characterization of a $Q$-proper partition)
Given a graph $G$, positive integers $K \geq 2$ and $Q \geq 1$, a $K$-partition $\mathcal{P} = \{V^{(k)}\}_{k \in [K]}$ is $Q$-proper if and only if for every $k \in [K]$, 
\begin{enumerate}[(i)]
\item $|V^{(k)}| \geq Q+1$, and 
\item for every distinct $a,b \in V^{(k)}$, $|C\cap V^{(k)}| \geq Q$ for every $a,b$-separator $C \in \mathcal{C}_{a,b,G}.$
\end{enumerate}
\end{theorem}

Theorem \ref{thm_characterization} is conceptually similar to \cite{buchanan2015integer}'s result that characterizes a $Q$-connected $d$-dominating set for $d\geq Q$; here, a $d$-dominating set is a subset of vertices $S$ such that every vertex in $G-S$ has $d$ neighbors in $S$.
The similarity is in applying \cite{buchanan2015integer}'s result to each part's vertices (i.e., $S=V^{(k)}$ for $k \in [K]$). 
However, unlike \cite{buchanan2015integer}'s requirement for a subset to be $d$-dominating (for $d\geq Q$), Theorem \ref{thm_characterization} does not impose the dominating constraint on each part's vertices.
Furthermore, \cite{buchanan2015integer}'s result is stated for any minimum cutset $C$, whereas Theorem \ref{thm_characterization} only considers a subset of cutsets, $\mathcal{C}_{a,b,G}$, for vertex pairs $a,b$ within each part. 
Hence, for characterizing a $Q$-proper partition, Theorem \ref{thm_characterization} differs from \cite{buchanan2015integer}'s consideration of subsets by relaxing the dominating constraint, but restricts the number of cutsets examined.

\subsection{$Q$-CUT Formulation}\label{subsec_formulation_qcut}

We express the $Q$-connectivity constraints (\ref{const_qconnected_Hess}) as an integer program using the characterization of a $Q$-proper partition in Theorem \ref{thm_characterization}.
%We now formulate $Q$-connectivity constraints (\ref{const_qconnected_Hess}) as an integer program.
%First, the following corollary establishes quadratic constraints for HCGP.
%\begin{corollary} \label{corollary_qcut_quadratic}
%Given a graph $G$, positive integers $K$, $Q$ and a solution $\hat{\textbf{x}} \in \mathcal{P}_{HESS}$, $\hat{\textbf{x}}$ belongs to $\mathcal{P}_{Q}$ if and only if for all $a,b,r \in V$ and $C \in \mathcal{C}_{a,b,G}$,
%$$\sum_{c \in C} \hat{x}_{rc} \geq Q \hat{x}_{ra}\ \hat{x}_{rb}.$$
%\end{corollary}
%\textbf{Proof.}
%($\Rightarrow$) We are given a solution $\hat{\textbf{x}}$ such that $\sum_{c \in C} \hat{x}_{rc} \geq Q \hat{x}_{ra}\ \hat{x}_{rb}.$ for all $a,b,r \in R$, $C \in \mathcal{C}_{a,b,G}$.  
%Assume the contrary, that there exists a district with root $r \in R(\hat{\textbf{x}})$ such that $\kappa(G[V_r]) = q$ for some $q\in [Q-1]$. 
%$\blacksquare$ 
Our formulation is a hybrid of the $1$-connectivity constraints for graph partitioning from \cite{oehrlein2017cutting} and the $Q$-connectivity constraints for the $Q$-connected $d$-dominating set problem from \cite{buchanan2015integer}.
\cite{oehrlein2017cutting}'s $CUT$ formulation is given as
%The following constraints are defined for each pair of distinct vertices $a,b \in V$ and $a,b$-separator $C \in \mathcal{C}_{a,b}$.
\begin{subequations}
\begin{align}
\textbf{x}  & \in \mathcal{F}_{HESS}, \label{const_hess_constraints_CUT} \\
\sum_{c \in C} x_{r,c} &\geq x_{r,a}  \quad \forall \ r \in V,\ a\in V\backslash\{r\},\ C\in \mathcal{C}_{r,a,G}.  \label{eq_1connected_CUT}
\end{align}
\end{subequations}
This formulation defines a constraint for every vertex separator between every vertex and its root.
Constraint  (\ref{const_hess_constraints_CUT}) defines a feasible solution to the Hess model.
Constraints (\ref{eq_1connected_CUT}) ensure that for every pair of vertices $r,a\in V$ and for every $r,a$-separator $C$ in $G$, if $r$ is a root and $a$ is assigned to it, then at least one of the vertices in $C$ must also be assigned to $r$.
This constraint ensures that the subgraph induced by each part is connected (i.e., the partition is $1$-proper).
%We show that $\mathcal{P}_{2CUT} \subseteq \mathcal{P}_{conn}^{(2)}$.
  
Furthermore, \cite{buchanan2015integer}'s formulation creates a $Q$-connected subgraph as follows.
Binary variables select a subset of vertices: for $i \in V$, $z_i=1$ if $i$ is selected; $z_i=0$, otherwise. 
Constraints (10) from \cite{buchanan2015integer} are given as \begin{align}
\sum_{c \in C} z_{c} &\geq Q z_{a} \ z_{b}   \quad \forall \ a \in V,\ b\in V\backslash\{a\}, \ C\in \mathcal{C}_{a,b,G}. \label{eq_buchanan}
\end{align} 
For every pair of selected vertices $a$ and $b$, this constraint ensures that at least $Q$ vertices in every $a,b$-separator are also selected.
Collectively, these constraints select vertices that induce a $Q$-connected subgraph.

Blending constraints (\ref{eq_1connected_CUT}) and (\ref{eq_buchanan}), and exploiting the separator-based characterization of a $Q$-proper partition in Theorem \ref{thm_characterization}, we create the $Q$-$CUT$ formulation for HCGP given as
%First, we define new quadratic variables as follows.
%\begin{align*} \nonumber
%y^r_{ab} = x_{ra}\ x_{rb} &= \left\{\begin{array}{rl}
%1, &\mbox{\textrm{if vertices $a,b \in V$ are both assigned to root $r \in V$}},\\
%0, &\mbox{\textrm{otherwise}.}
%      \end{array}\right.
%\end{align*}
%Then, the $Q$-$CUT$ formulation is given by,
\begin{subequations}
\begin{align}
\textbf{x}  & \in \mathcal{F}_{HESS}, \label{const_hess_constraints_qCUT}  \\
%y^r_{ab} &\leq x_{ra}   \quad \forall \ a \in V,\ b\in V\backslash\{a\},\ r\in V, \label{eq_Y_x_1}\\
%y^r_{ab} &\leq x_{rb}   \quad \forall \ a \in V,\ b\in V\backslash\{a\},\ r\in V, \label{eq_Y_x_2}\\
%y^r_{ab} &\geq x_{ra} + x_{rb} - 1 \quad \forall \ a \in V,\ b\in V\backslash\{a\},\ r\in V, \label{eq_Y_x_3}\\
\sum_{c \in C} x_{rc} &\geq Q x_{ra}\ x_{rb}  \quad \forall \ a \in V,\ b\in V\backslash\{a\},\ r\in V,\ C\in \mathcal{C}_{a,b,G}, \label{eq_qconnected} \\
\sum_{i \in V} x_{ri} &\geq (Q+1) x_{rr}  \quad \forall \ r\in V. \label{eq_atleastQ} 
%y^r_{ab} &\geq 0 \quad \forall \ a \in V,\ b\in V\backslash\{a\},\ r\in V. \label{eq_Y_nonneg}
\end{align}
\end{subequations}
This formulation defines a quadratic constraint for every separator between every pair of vertices assigned to the same root.
Constraint (\ref{const_hess_constraints_qCUT}) defines a feasible solution to the Hess model.
%For every pair of 
%Constraints (\ref{eq_Y_x_1})-(\ref{eq_Y_x_3}) ensure the quadratic relationship $y^r_{ab} = x_{ra}\ x_{rb}$.
Constraints (\ref{eq_qconnected}) ensure that for every trio of vertices $a,b,r \in V$ such that $a\neq b$ and for every $a,b$-separator $C$ in $G$, if $r$ is a root and both $a$ and $b$ are assigned to it, then at least $Q$ of the vertices in $C$ must also be assigned to $r$.
%We denote $\mathcal{P}_{Q-\textrm{CUT}} \coloneqq \{\textbf{x} \in \mathcal{P}_{HESS: \textbf{x} \ \textrm{ satisfies constraints } (\ref{eq_qconnected})\}$ as  the polytope of feasible solutions to the $Q$-CUT model.
Constraints (\ref{eq_atleastQ}) ensure that at least $Q+1$ vertices are assigned to each part.
The following corollary shows that constraints (\ref{eq_qconnected})-(\ref{eq_atleastQ}) ensure $Q$-connectivity in each part. 

\begin{corollary}\label{corollary_Q_proper}
Given a graph $G$ and $K \geq 2$, consider a solution to the Hess model $\hat{\textbf{x}} \in \mathcal{F}_{HESS}$.
For $Q\geq 1$, $\hat{\textbf{x}}$ is $Q$-proper if and only if $\hat{\textbf{x}}$ satisfies constraints (\ref{eq_qconnected})-(\ref{eq_atleastQ}).
\end{corollary}

Constraints (\ref{eq_qconnected}) are quadratic constraints that can be linearized using McCormick envelopes \citep{mccormick1976computability} by introducing a new variable $y^r_{ab} = x_{ra}\ x_{rb}$ 
and additional constraints given by
\begin{subequations}
\begin{align}
y^r_{ab} &\leq x_{ra}  \quad \forall \ a \in V,\ b\in V\backslash\{a\},\ r\in V, \label{eq_McCormick1}  \\
y^r_{ab} &\leq x_{rb}  \quad \forall \ a \in V,\ b\in V\backslash\{a\},\ r\in V, \label{eq_McCormick2} \\
y^r_{ab} &\geq x_{ra} + x_{rb} - 1  \quad \forall \ a \in V,\ b\in V\backslash\{a\},\ r\in V.\label{eq_McCormick3}
\end{align}
\end{subequations}
Similar to the $CUT$ formulation for $Q=1$, the $Q$-$CUT$ formulation for $Q\geq 1$ has an exponential number of constraints due to the number of possible separators.
Instead of enumerating all separators, Section \ref{subsec_cuttingplanes_overall} provides a branch-and-cut method that solves the Hess model, and adds violated constraints on-the-fly for solutions that are not $Q$-proper.

\subsection{Minimum Degree Valid Inequalities}
\label{subsec_degree_valid_constraints}
We now introduce valid inequalities to the formulation using the following property.
% that every vertex in a $Q$-connected graph has at least $Q$ neighbors, stated as the following proposition.
\begin{proposition}
For $Q\geq 1$, the minimum vertex degree of a $Q$-connected graph is at least $Q$.
\end{proposition}
The correctness of this property can be seen by assuming the contrary that the degree of some vertex is less than $Q$.
Then, the neighborhood set of this vertex is a cutset of size less than $Q$, which contradicts the fact that the given graph is $Q$-connected.
Using this property, the degree-constraints ensure that every vertex has at least $Q$ neighbors in the same part, given as
\begin{align} 
\sum_{j \in N(\{i\})} x_{rj} &\geq Q x_{ri}  \quad \forall \ i \in V, \ r\in V. \label{eq_degree_valid_constraints}  
\end{align} 

There are $n^2$ constraints in (\ref{eq_degree_valid_constraints}), comparable in size to the Hess model with $n^2+2n+1$ constraints. 
Furthermore, note that constraints  (\ref{eq_degree_valid_constraints}) ensure that each part contains $Q+1$ vertices (e.g., the root and at least $Q$ of its neighbors), and therefore implies (\ref{eq_atleastQ}).
Hence, the $Q$-$CUT$ formulation can be expressed by the Hess model (\ref{eq_obj})-(\ref{const_defining_x_hess}), and constraints (\ref{eq_qconnected}) and (\ref{eq_degree_valid_constraints}).

\subsection{Separation Algorithm}\label{subsec_cuttingplanes_overall}

We propose a branch-and-cut approach to solve HCGP, which addresses the challenges posed by the exponential number and quadratic nature of the separator constraints (\ref{eq_qconnected}). 
This approach starts by relaxing the separator constraints and solving the Hess model (\ref{eq_obj})-(\ref{const_defining_x_hess}), and the minimum degree constraints (\ref{eq_degree_valid_constraints}). 
By solving the relaxed problem, this approach avoids adding the exponential number of constraints (\ref{eq_qconnected}). 
When the branch-and-cut approach encounters an integer feasible solution that is not $Q$-proper, it solves a separation problem that separates integer feasible solutions.
This approach identifies a set of violated separator constraints and adds them on-the-fly.
To address the quadratic nature of separator constraints (\ref{eq_qconnected}), this approach adds linearized versions of these violated constraints; the correctness of the linearization is shown in  Theorem \ref{thm_separationalgocorrectness}.
This dynamic and linearized addition of constraints effectively handles the computational intractability of the model to find optimal or near-optimal solutions to HCGP.
%
%This approach involves relaxing the $Q$-connectivity constraints and solving the Hess model (\ref{eq_obj})-(\ref{const_defining_x_hess}), augmented by the valid inequalities (\ref{eq_degree_valid_constraints}).
%Within the branch-and-cut, when an integer feasible solution that is not $Q$-proper is encountered, an integer separation problem is solved to identify a set of violated constraints (\ref{eq_qconnected}). 
%A linearized version of these violated constraints is dynamically added to the model and the branch-and-cut proceeds.
%Employing this on-the-fly addition of constraints avoids the need to include the exponentially large number of constraints a priori.
%%Commercial solvers such as Gurobi or CPLEX offer functionality called \textit{lazy constraints}, which enables the inclusion of these dynamic on-the-fly constraints.

This section presents an integer separation algorithm for HCGP, extending the approach by \cite{validi2020} for the $Q=1$ case to accommodate any $Q\geq 1$. 
This approach also extends  \cite{buchanan2015integer}'s separator-based cutting planes method for the fault-tolerant dominating set problem.
Given an integer feasible solution $\hat{\textbf{x}}$ to the relaxed model, if $\hat{\textbf{x}}$ is not $Q$-proper,
this algorithm identifies a subset of separator constraints (\ref{eq_qconnected}) that $\hat{\textbf{x}}$ violates.
Each such violated constraint is identified by selecting the corresponding vertices $a$, $b$ and $r$ (where $a \neq b$), and an $a,b$-separator $C$ in $G$.
To find a minimal $a,b$-separator in $G$, we use the procedure outlined in \cite{fischetti2017thinning}, initially proposed by \cite{validi2020} for the $Q=1$ case.
We demonstrate how this procedure extends to any $Q\geq 1$.

%The proposed integer separation algorithm adds 
%The integer separation algorithm operates based on the following idea.  
The computational efficiency of the proposed integer separation algorithm stems from examining the structure of a $Q$-disconnected part. 
When a part with a root $r \in R(\hat{\textbf{x}})$ is connected but not $Q$-connected, it contains a minimum cutset $D$ such that $|D|< Q$. 
Removing $D$ from that part disconnects it into more than one component.
A naive implementation of the algorithm proceeds as follows: from every pair of such components, select arbitrary vertices $a$ and $b$.
Then, identify a minimal $a,b$-separator in $G$, and add the corresponding violated constraint.   
However, this approach can potentially iterate through $O(n^2)$ pairs of components in the worst case. 

Alternatively, assuming that the root is not a part of the cutset (i.e., $r \notin D$), an alternative approach is to set $a=r$. 
Then, from every component that does not contain $r$, an arbitrary vertex $b$ is selected. 
A minimal $r,b$-separator in $G$ is identified, and the corresponding violated constraint is added. 
This alternative approach would add $O(n)$ violated constraints in the worst case.
Hence, to enhance computational efficiency, the separation algorithm first checks whether $r \notin D$, in which case that part is called \textit{root-resilient}.
If the part is root-resilient, it adds the $O(n)$ constraints;
otherwise, it adds the pair-wise set of $O(n^2)$ constraints.

If the part with root $r$ is not connected, the same procedure is applied to the component of the part that contains $r$ rather than the entire part (i.e., if this component is not $Q$-connected, then the integer separation algorithm finds a minimum cutset $D$ of the component, classifies the part as root-resilient if $r \notin D$, and adds the corresponding set of violated constraints for that component); to promote $Q$-connectivity for the entire part, the algorithm also adds violated constraints between the component containing $r$ and the other components.
This approach diverges from \cite{buchanan2015integer}'s approach by employing \cite{fischetti2017thinning}'s separator construction and by incorporating root-resilience.

\begin{algorithm}[!ht]
\SetKwInOut{Input}{Input}
\SetKwInOut{Output}{Output}
\SetKwProg{myproc}{Procedure}{}{}
\For{$r \in R(\hat{\textbf{x}})$}{
%	let $V_r \coloneqq \{i \in V | \hat{x}_{ri} = 1\}$\\
	\For{\textrm{every component $G_0$ of $G[V_r]$ not containing $r$}}{\tcc{the part is disconnected}
		let $b$ be an arbitrary vertex of $G_0$ \\
		$C \leftarrow $ Minimal separator($G$, $V[G_0]$, $r$, $b$)\\	
		add constraint $Q\hat{x}_{rb} \leq \sum_{c \in C} \hat{x}_{rc}$  to the model \\
	}
	let $G_r$ be the component of $G[V_r]$ that contains $r$\\
	let $D$ be a minimum cutset in $G_r$\\
	\If{$|D| < Q$}{ 
%		let $A$ be the collection of all $(q-1)$-sized cut sets in $G_a$\\
%		\If{$r \notin D$}{
%	%		\For{cut set $D \in A$}
%		}
		\If{$r \notin D$}{  \tcc{the part is $Q$-disconnected and root-resilient}
			\For{\textrm{every component $G_0$ of $G_r-D$ that does not contain $r$}}{
					let $b$ be an arbitrary vertex of $G_0$ \\
					$C \leftarrow$ Minimal separator($G$, $V[G_0]$, $r$, $b$)\\
					add constraint $Q\hat{x}_{rb} \leq \sum_{c \in C} \hat{x}_{rc}$  to the model \\
				}
		}
		\Else{\tcc{the part is $Q$-disconnected and not root-resilient}
			\For{every pair of components $G_1$ and $G_2$ of $G_r-D$}{
				let $a$ and $b$ be arbitrary vertices in $G_1$ and $G_2$, respectively \\
				$C \leftarrow$ Minimal separator$(G, V[G_2], a, b)$\\
				add constraint $Q(\hat{x}_{ra}+\hat{x}_{rb}-1) \leq \sum_{c \in C} \hat{x}_{rc}$ to the model
			}
		
		}
	} 
} \vspace{4mm} 
\myproc{Minimal separator ($G$, $S$, $a$, $b$) \citep{fischetti2017thinning}}{
\tcc{here, $S$ is a subset of vertices of $G$ containing $b$, and $a \notin S$} 
define the neighbors of $S$ to be $\mathcal{N} = N(S)$ \\
delete all edges in $E[S \cup N(S)]$ from $G$\\
find the set $\mathcal{R}$ of vertices that can be reached from $a$ in $G$\\
\Return{$\mathcal{N} \cap \mathcal{R}$}
}
%\Input{Adjacency graph $G = (V, E)$, distance function $d$, number of districts $K \geq 1$, number of levels $L \geq 1$}
%\Output{Approximate solution $\textbf{x}^{(0)}$} 
%\Return{$\textbf{x}^{(0)}$}
\caption{Integer $Q$-Connectivity Separation (graph $G$, integer feasible solution to Hess model $\cup$ constraints (\ref{eq_degree_valid_constraints}) $\hat{\textbf{x}}$, integer $Q \geq 1$)} \label{algo_intseparation_biconnected}
\end{algorithm}

Algorithm \ref{algo_intseparation_biconnected} presents a pseudocode for the separation algorithm for a given integer solution to the relaxed model $\hat{\textbf{x}}$.
Recall that $R(\hat{\textbf{x}})$ denotes the set of roots and $V_r$ denotes the set of vertices assigned to a root $r \in R(\hat{\textbf{x}})$.
The algorithm iterates (in line 1) over each part with a root $r$.
First, the algorithm checks whether the part is disconnected. 
If it is, lines 2-5 add violated constraints for each $r,b$-separator, where $b$ is an arbitrary vertex selected from each component of the part subgraph  that does not contain $r$. 
Next, lines 6 and 7 find a minimum cutset $D$ in the component $G_r$ of the part containing $r$.
Note that $G_r=G[V_r]$ if the part is connected.
In the special case that $G_r$ is a complete graph, $D$ returns the vertices in $G_r$ without $r$, rather than a traditionally-defined cutset of $G_r$.
Line 8 verifies whether the size of $D$ is less than $Q$. If it is, indicating that the part is not $Q$-connected, lines 9-18 add violated constraints.
If the part is root-resilient, lines 10-13 add a constraint between $r$ and an arbitrary vertex $b$ in every component of $G_r-D$. 
Otherwise, lines 15-18 add a constraint between every pair of vertices $a$ and $b$ from different components of $G_r-D$.

\begin{figure}[!htbp]
\centering
\subfloat[The part is disconnected; the shaded regions compose the part with root $r$ and the hatched region is $C$, a minimal $r,b$-separator in $G$. \label{fig_abstract_rep_1}]
{\includegraphics[width=0.3\textwidth]{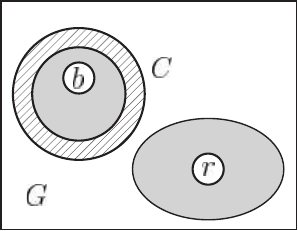}} 
%\vspace{.2cm}
\hspace{.5cm}
%\hfloat
\subfloat[The part is root-resilient; the shaded region (light and dark)  is its component $G_r$ containing $r$, the dark region is a minimum cutset $D$, and the hatched region is $C$, a minimal $r,b$-separator in $G$.\label{fig_abstract_rep_2}]
{\includegraphics[width=0.3\textwidth]{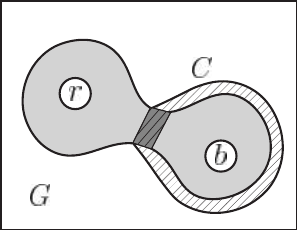}} 
%\vspace{.2cm}
\hspace{.5cm}
%\hfloat
\subfloat[The part is not root-resilient; the shaded region (light and dark)  is  its component $G_r$ containing $r$, the dark region is a minimum cutset $D$, and the hatched region is $C$, a minimal $a,b$-separator in $G$.\label{fig_abstract_rep_3}]
{\includegraphics[width=0.3\textwidth]{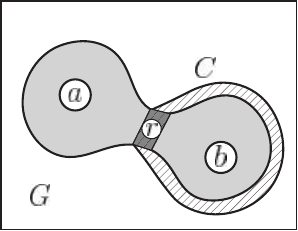}} 
%\vspace{.2cm}
%\hspace{0.1cm}
\caption{Abstract representations of a $Q$-disconnected part with root $r$ in graph $G$ considering three cases corresponding to the three sets of constraints added in Algorithm \ref{algo_intseparation_biconnected}. 
%The lightly shaded regions in (a) represent the part subgraph $G[V_r]$ and the lightly shaded regions in (b) and (c) represent the component $G_r$ of $G[V_r]$ containing $r$. 
%The darker regions in (b) and (c) represent a minimum cutset $D$  in $G_r$. The hatched region is $C$, a minimal $r,b$-separator in (a) and (b), and a minimal $a,b$-separator in (c).
} \label{fig_abstract_rep}
\end{figure}

Figure \ref{fig_abstract_rep} depicts an abstract representation of the three cases  when Algorithm \ref{algo_intseparation_biconnected} adds violating constraints. 
The correctness of Algorithm \ref{algo_intseparation_biconnected} is shown in Theorem \ref{thm_separationalgocorrectness}.
Let $\mathcal{F'}_{HESS}$ denote the polytope of feasible solutions to the relaxed model given by the Hess model (\ref{eq_obj})-(\ref{const_defining_x_hess}) and constraints (\ref{eq_degree_valid_constraints}).
%The following Theorem shows that the algorithm adds a violated $Q$-connectivity constraint if and only if the given solution to the relaxed model is not $Q$-proper.
%Figure \ref{fig_abstract_rep}  can be used as a visual reference to follow the proof of Theorem \ref{thm_separationalgocorrectness}.
Note that in Theorem \ref{thm_separationalgocorrectness}, a violated separator constraint is a subset of separator constraints (\ref{eq_qconnected}) and (\ref{eq_McCormick1})-(\ref{eq_McCormick3}) violated by a solution to the relaxed model.

\begin{theorem}\label{thm_separationalgocorrectness}
Given a graph $G=(V,E)$, an integer $Q \geq 1$, and an integer feasible solution $\hat{\textbf{x}} \in \mathcal{F'}_{HESS}$, Algorithm  \ref{algo_intseparation_biconnected} adds violated separator constraints if and only if $\hat{\textbf{x}}$ is not $Q$-proper.
\end{theorem}

To analyze the run-time of Algorithm \ref{algo_intseparation_biconnected}, we consider two cases. 
First, if all the parts  are root-resilient to the minimum cutsets found in line 8, adding violated constraints is less time-intensive as it only executes the first 13 lines of Algorithm \ref{algo_intseparation_biconnected}.
Second, if some parts are not root-resilient, lines 15-18 are executed, which increases the solution time.
The following theorem provides the worst case run-time for both cases. 

\begin{theorem} \label{thm_separationalgo_time}
Given a graph $G=(V,E)$ with $|V|=n$ and $|E|=m$, integers $Q \geq 1$ and $K\geq 2$, and a solution $\hat{\textbf{x}} \in \mathcal{F'}_{HESS}$, Algorithm \ref{algo_intseparation_biconnected} takes $O(Kmn^{5/3})$ time if all the parts of $\hat{\textbf{x}}$ are root-resilient to the cutsets in line 8 or $K\geq \sqrt[3]{n}$; otherwise, it takes $O(mn^2)$ time. 
\end{theorem}
%\textit{Proof.}
%First, finding a minimal separator from \cite{fischetti2017thinning} takes $O(m)$ time \citep{validi2020}. 
%Since every vertex $b$ (in line 3) is visited exactly once among all roots $r$ and all components $G_0$, lines 1-5 take $O(mn)$ time.
%Next, in lines 6 and 7, finding a minimum cutset in $G_r$  takes $O(mn^{5/3})$ time using Algorithm 11 in \cite{esfahanian2013connectivity}.
%Lines 6 and 7 take $O(Kmn^{5/3})$ time when iterating over all the $K$ roots.
%Next, in lines 9-13, every vertex $b$ is visited exactly once since each $b$ is present in exactly one component of $G_r-D$ among all roots $r$.
%Hence, lines 9-13 take $O(mn)$ time using the same reasoning for lines 1-5.
%If all the parts are root-resilient, only lines 1-13 are executed, and hence the overall run-time is $O(Kmn^{5/3})$.
%
%If some part is not root-resilient, lines 15-18 are also executed.
%Here, each pair of vertices $a$ and $b$ are visited exactly once among all roots $r$ and all pairs of components $G_1$ and $G_2$, resulting in $O(n^2)$ iterations of the for-loop in line 15, where each iteration takes $O(m)$ to generate a minimal separator and add the cut.
%Lines 15-18 take $O(mn^2)$ time.
%Hence, if some parts are not root-resilient, executing lines 1-18 takes $O(Kmn^{5/3}+mn^2)$ time, which is $O(Kmn^{5/3})$ time if $K \geq \sqrt[3]{n}$; otherwise, it takes $O(mn^2)$ time.
%$\square$

%The argument is that each vertex $b$ in line 12 appears in a component created by at most $O(n^q)$ $q$-cutsets.

While the worst case run time of the separation algorithm is independent of $Q$, the if condition in line 8 may return True for more solutions for larger $Q$ values.
As a result, lines 9-18 would generate and add more violated cuts, leading to a longer overall run time.
Furthermore, in the special case where $G$ is planar (i.e., $m = O(n)$), Algorithm \ref{algo_intseparation_biconnected} takes $O(Kn^{8/3})$ time if all the parts in $\hat{\textbf{x}}$  are root-resilient or $K\geq \sqrt[3]{n}$; otherwise, it takes $O(n^3)$ time.

%The number of computational steps in practice is typically much less than the steps accounted for in calculating the theoretical worst-case run-time.

Even though the theoretical worst case run-time of a single run of the separation algorithm is a large polynomial of $m$ and $n$, empirical results suggest that the algorithm performs well in practice. 
Specifically, on average over all the instances solved in Section \ref{sec_computationalresults}, the total run-time of all the runs of the separation algorithm within a single execution of the branch-and-cut only accounts for approximately $14.9\%$ of the branch-and-cut time.
Hence, the separation algorithm efficiently enforces $Q$-connectivity for practical instances of HCGP.
%In the next section, we provide a computational analysis of solving theHCDP using instances from political redistricting.
 
%11.97, 21.28, 24.86, 17.97

%It takes $O(2^qn^3)$ time to find all minimum size vertex separators, where $q$ is the connectivity of the graph and $n$ is the number of vertices in the graph  \citep{kanevsky1993finding}.

%\clearpage
\section{An Ear-Construction Heuristic for HCGP with $Q=2$}
\label{sec_ear_construction_heur}
 
For large instances of HCGP, a heuristic approach can be more effective if the goal is to find a feasible solution in a shorter time than solving the branch-and-cut method.
This section presents a heuristic approach to solve HCGP with $Q=2$.
Motivated by heuristic approaches to the political districting problem (PDP), this heuristic first constructs and then improves a solution to HCGP.
%Recall that this problem finds a compact, $2$-proper and size-balanced partition of a graph.
% $2$-proper $L,U$-balanced $K$-partition of a graph $G$. 
The inputs to the heuristic are graph $G$, the number of parts $K$, and the lower and upper bounds on the part sizes ($L$ and $U$, respectively).
This heuristic has four stages.
The first three stages generate a feasible solution, i.e., a $2$-proper $L,U$-balanced $K$-partition of $G$, and the fourth stage improves the compactness objective using local search.
%The heuristic  generates a feasible solution to BCGP through a four-stage process.
%Each stage progressively achieves specific feasibility requirements of BCGP in the following order:
% creating $K$ simply connected parts, improving the connectivity of each part to two, improving the size balance to be within the specified bounds, and finally locally optimizing the compactness objective.
The stages are as follows.
\begin{enumerate}[(i)]
\item The \textit{construct} stage generates an initial partition of $G$ into $K$ connected parts. 
First, it iteratively creates $2$-connected parts using a procedure called \textit{ear construction}, and the rest of the graph comprises a set of connected parts.
If the number of parts exceeds $K$, it iteratively merges neighboring parts until there are $K$ parts; if the number of parts is smaller than $K$, it restarts the construct stage.  
If some parts are $2$-disconnected, the algorithm proceeds to the \textit{repair} stage; otherwise, the algorithm proceeds to the third stage. 
\item The \textit{repair} stage fixes the $2$-connectivity of each $2$-disconnected part by reassigning certain vertices between neighboring parts. 
These vertices are chosen with the goal of reducing the number of cut vertices within a part.
After the repair stage, if all the parts are $2$-connected, the heuristic proceeds to the size balance improvement stage; otherwise, the algorithm restarts from the first stage.
\item The \textit{size balance improvement} stage uses local search (using vertex reassignments) to improve a size balance objective until all the part sizes are within $[L,U]$. 
If local search terminates before achieving this size balance (i.e., there are no reassignments that improve size balance while also retaining each part’s  $2$-connectivity), the algorithm restarts from the first stage.
\item The \textit{compactness improvement} stage uses local search to optimize the compactness objective, producing a locally optimal $2$-proper and size-balanced partition.
\end{enumerate} 

%In these four stages, the heuristic progressively generates a feasible solution to BCGP that satisfies $2$-connectivity and size balance, and has a locally optimal compactness.

%The resulting partition is feasible to BCGP and is locally optimal with respect to compactness.

Each stage incorporates several random steps.
For example, the construct stage initially constructs a set of random $2$-connected subgraphs of $G$. 
If a stage does not produce a partition that satisfies its designed criteria, the heuristic restarts from the first stage.  
Furthermore, ordering the first three stages ensures connectivity, $2$-connectivity, and size balance in that order.
Preliminary computational investigations revealed that this ordering produced a feasible solution faster than other permutations, such as ensuring size balance before $2$-connectivity.

%\begin{algorithm}[H]
%\SetKwInOut{Input}{Input}
%\SetKwInOut{Output}{Output}
%\SetKwProg{myproc}{Procedure}{}{}   
%$\mathcal{P} \leftarrow$ Construct an Initial Partition ($G$, $K$, $L$) \\
%%$\mathcal{P}_{merged} \leftarrow$ Merge districts in $\mathcal{P}_{ED}$ to get $K$ districts \\
%\If{$\mathcal{P}$ is not $2$-proper}{
%	$\mathcal{P} \leftarrow$ Repair $2$-Connectivity ($G$, $\mathcal{P}$) \\
%	\If{$\mathcal{P}$ is not $2$-proper}{
%		Go to line 1\\	
%	}
%}
%\If{$\mathcal{P}$ is not $L,U$-balanced}{
%	$\mathcal{P} \leftarrow$ Local Search ($G$, $\mathcal{P}$, $\phi_{bal}$)  \\
%	\If{$\mathcal{P}$ is not $L,U$-balanced}{
%		Go to line 1\\
%	}
%}
%$\mathcal{P} \leftarrow$ Local Search ($G$, $\mathcal{P}$, $\phi_{comp}$) \\
%\Return{$\mathcal{P}$}
%\caption{Ear-Construction Heuristic ($G$, $K$, $L$, $U$)} \label{algo_heuristic_overview}
%\end{algorithm} 

The rest of this section is organized as follows.
Section \ref{subsec_earconstruction} describes the ear construction procedure to construct a $2$-connected subgraph from a given graph.
Sections \ref{subsec_constructinitial} and \ref{subsec_repair} describe the construct and repair stages, respectively.
Section \ref{subsec_improve} defines a size balance objective  and outlines the local search-based improvement stages.
The pseudocodes for the overall heuristic and the individual stages are given in Appendix \ref{appendix_pseudocode}.

\subsection{Ear Construction} \label{subsec_earconstruction}

\begin{figure*}[!h]
\centering
\includegraphics[width=1\textwidth]{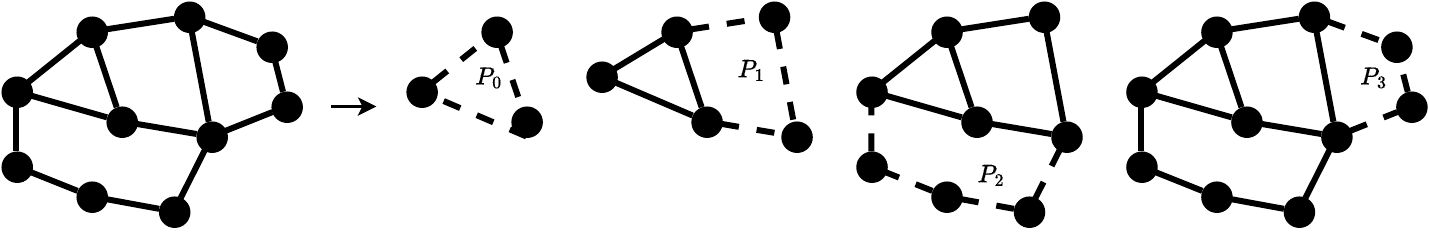}
\caption{An open ear decomposition of a $2$-connected graph with $10$ vertices into four ears. 
} \label{fig_ear_decomposition}
\end{figure*}

We now present a procedure to grow a $2$-connected subgraph from a given graph if one exists.
Such a $2$-connected subgraph composes a $2$-connected part in the initial partition assembled in the construct  stage described in the next section.
%Every subgraph built using this procedure constitutes a guaranteed $2$-connected part in the partition generated in the construct  stage described in the following section.
To construct a $2$-connected subgraph, we utilize a graph structure known as an \textit{open ear decomposition}, defined as follows.
%An \textit{ear} $P$ is a sequence of vertices 
%$(a_1,a_2,\dots,a_{|P|})$ that either constitute a path, or $(a_1,a_2,\dots,a_{|P|-1})$ is a path and $a_1 = a_{|P|}$.
%path with distinct internal vertices, although the endpoints may coincide.
An \textit{ear}  is an ordered sequence of vertices that have the properties of a path with the only exception that the end-points may coincide.
An \textit{open ear} is an ordered sequence of distinct vertices identical to a path.
%an ear with distinct endpoints, identical to a path.
An \textit{open ear decomposition}, illustrated in Figure \ref{fig_ear_decomposition}, is constructively defined as follows.
%a partition of the edge set of a graph into a sequence of ears, where the first ear is a cycle, and the two endpoints of each subsequent open ear belong to earlier ears.
%A constructive definition of an open ear decomposition is provided below.
\begin{definition}\label{defn_open_ear_decomp}
Given a graph $H$, an \textit{open ear decomposition} of $H$ is a sequence of ears $P_0,P_1,\dots,P_s$ (for some integer $s \geq 0$) in $H$ such that
\begin{enumerate}
\item $P_0$ is a cycle in $H$, and
\item if $s\geq 1$, for $i = 1,2,\dots,s$, $P_i$ is an open ear in $H$ with end-points in $H[P_0 \cup P_1 \cup \dots \cup P_{i-1}]$.
\end{enumerate}
\end{definition}

%Figure \ref{fig_ear_decomposition} depicts an example of an open ear decomposition. 
The constructive procedure in Definition \ref{defn_open_ear_decomp} can be used to create a $2$-connected part subgraph in the initial partition of the graph.
To achieve this, we use the classical result from \cite{whitney1931non}, which states that a graph with at least two edges is $2$-connected if and only if it has an open ear decomposition. 
In our procedure, rather than decomposing a given graph, we employ the constructive procedure in Definition \ref{defn_open_ear_decomp} to generate a sequence of ears to constitute the part subgraph.
From \cite{whitney1931non}'s result, the vertices traversed by these ears induce a $2$-connected subgraph.

To create an initial part, beyond $2$-connectivity, it is also desirable to ensure that the part size is at least $L$.
In our procedure, ears are added one at a time.
%The algorithm parameters are the termination criterion (given by the number of ears added) and how each ear is chosen.
To ensure the minimum size requirement, this iterative addition of ears terminates either when the traversed set of vertices is large enough, i.e., its size is at least $L$, enough to form a balanced part, or is maximal, i.e., no additional ears can be added.
Furthermore, each ear is constructed to increase the part size incrementally to avoid excessively increasing the part size within a single iteration. 
This goal is achieved by finding a vertex-weighted shortest path (weighted by the vertex sizes) between two randomly chosen vertices that already appearing in the part.

\subsection{Construct an Initial Partition Using Ear Construction} \label{subsec_constructinitial}
The construct  stage creates an initial $K$-partition of $G$.
This procedure is analogous to the multi-kernel growth method to find a feasible $1$-proper partition for PDP \citep{vickrey1961prevention}.
This procedure first constructs a series of $2$-connected parts by iteratively employing the ear construction procedure in Section \ref{subsec_earconstruction}, removing that $2$-connected subgraph, and proceeding with the remaining graph.
This iterative procedure continues until either the remaining graph is empty or is a forest, i.e., there are no more $2$-connected subgraphs.
%The procedure assembles connected parts using these $2$-connected subgraphs and any existing trees in the forest.
Let $K'$ denote the total number of $2$-connected subgraphs and trees in the forest. 
This stage terminates if $K'=K$, indicating that we already have $K$ parts; in the ideal case, all $K$ parts are $2$-connected.
If $K' \neq K$, there are two possible cases:
\begin{itemize}
\item $K'<K$, indicating that there are fewer parts than desired, in which case the construct  stage is restarted, or
\item $K'>K$, indicating an excess of parts, in which case the procedure iteratively merges a part with a neighboring part until the desired number of $K$ parts are obtained. 
Each iteration merges the pair of parts with the smallest total size, thereby prioritizing smaller parts to be merged first.
%in the non-decreasing order of part sizes, i.e., merging parts with small sizes are prioritized first.
\end{itemize} 

This stage returns a partition with $K$ connected parts, some of which may be $2$-disconnected.

%
%Algorithm \ref{algo_find_2con_districts} outlines a pseudocode for this procedure.
%Throughout this procedure, let $G'$ (initialized with $G$) denote the remaining graph.
%Lines 3-6 iteratively find a set of $2$-connected subgraphs using Algorithm \ref{algo_2_connected_subgraph}.
%After these iterations, if $G'$ is a forest, lines 7-9 designate each tree in the forest as an individual part.
%If $K'<K$, line 11 directs the algorithm to restart.
%If $K' > K$, line 13 proceeds to merge parts iteratively until $K'=K$. 
%Finally, Algorithm \ref{algo_find_2con_districts} returns a partition with $K$ connected parts, some of which may be $2$-disconnected.
%
%
%\begin{algorithm}[!ht]
%\SetKwInOut{Input}{Input}
%\SetKwInOut{Output}{Output}
%\SetKwProg{myproc}{Procedure}{}{} 
%$G'(V',E') \leftarrow G(V,E)$: Remaining graph \\
%$K' \leftarrow 0$: Initiate the label for the current part\\
%\While{$G'$ contains a cycle}{ 
%	$K' \leftarrow K'+1$ \\ 
%	$V^{(K')} \leftarrow$ Ear Construction($G'$, $L$) \\	 
%	$G' \leftarrow G' - V^{(K')}$ \\
%}
%\For{tree $T$ of $G'$}{
%	$K' \leftarrow K'+1$	\\
%	$V^{(K')} \leftarrow T$ \\
%}
%\If{$K' < K$}{
%	Go to Line 1 \\
%}
%\While{$K' > K$}
%{
%	Select two neighboring parts and merge; proceed in the non-increasing order of pair-wise total part sizes \\
%} 
%\Return{$\{V^{(k)}\}_{k=1}^{K}$}  
%\caption{Construct an Initial Partition ($G$, $K$, $L$)} \label{algo_find_2con_districts}
%\end{algorithm} 

\subsection{Repair $2$-Connectivity by Reassigning Small Components} \label{subsec_repair} 

\begin{figure*}[!h]
\centering
\includegraphics[width=.5\textwidth]{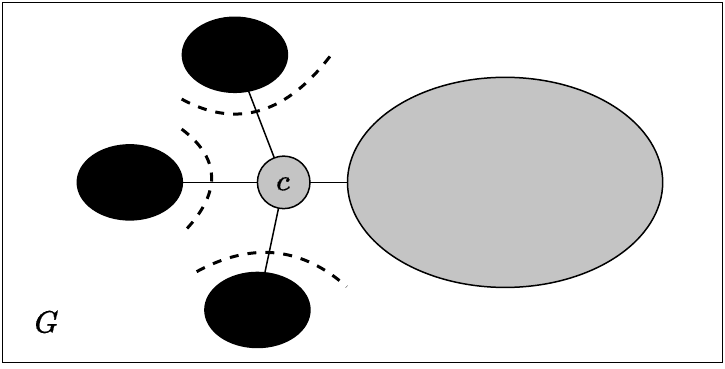}
\caption{Repairing $2$-disconnectivity: An abstract representation of a $2$-disconnected part (shaded region) with a cut vertex $c$.
The hypothetical removal of $c$ from the part disconnects the part into one large component (lightly shaded region) and several smaller components (dark region) which are reassigned to neighboring parts.
} \label{fig_repair_stage}
\end{figure*}

If the partition found in the first stage is not $2$-proper, the repair stage attempts to modify the partition to achieve a $2$-proper partition.
%This modification is achieved by identifying a cut vertex and reassigning the components of the part upon hypothetically
Our preliminary empirical investigations revealed a pattern in a typical $2$-disconnected part found in the first stage: removing a cut vertex $c$ of the part would typically result in one large component and one or more smaller components. 
The repair stage reassigns these smaller components to neighboring parts while leaving $c$ in its original part, at the end of which $c$ is no longer a cut vertex of that part.    
Figure \ref{fig_repair_stage} depicts an abstract representation of this reassignment for a given cut vertex $c$. 
After repeating this procedure for all the cut vertices of a part, that part becomes $2$-connected.    
%Motivated by this observation, our proposed repair procedure is as follows.
%First, identify the cut vertices within a randomly chosen $2$-disconnected part.
%For each randomly chosen cut vertex, reassign a small component of the part (upon hypothetically removing the cut vertex) to a randomly chosen neighboring part. 
%Continue this procedure until all the parts are $2$-connected.
%Reassess the connectivities of the parts, and continue this procedure.
 
There are two potential issues with this reassignment procedure.
First is the possibility that a small component $S$ may not have a neighboring part, in which case the vertices in $S$ cannot be reassigned.
Note that this scenario arises only when the given graph $G$ is $1$-connected, i.e., the cut vertex $c$ of a $2$-disconnected part is also a cut vertex of $G$.
Second, in some cases, the reassignment procedure may introduce cut vertices to the part that receives the vertices. 
%A caveat to this reassignment procedure is that in some cases, it may introduce cut vertices to the part that receives the vertices. 
Hence, cycling may occur, where the same subset of vertices is repeatedly reassigned.
Here, cycling is heuristically assessed by checking whether the previous $100$ iterations are exclusively composed of repeated reassignments. 
If either of these issues is detected (i.e., a smaller component $S$ does not have a neighboring part or if the cycling condition is true), the procedure terminates and restarts the construct stage.
Otherwise, this procedure continues until all the parts become $2$-connected.

\subsection{Improvement Stages: Local Search to Optimize Size Balance and Compactness} \label{subsec_improve}
 
Once the heuristic obtains a $2$-proper $K$-partition, it executes two improvement stages.  
%We now describe the local search-based improvement stages.
First, starting from a $2$-proper partition, this procedure optimizes a size balance objective until all the part sizes are within $[L,U]$.
Second, this procedure locally optimizes compactness by minimizing the objective $\phi_{comp}$ given in Equation (\ref{eq_compactness_obj}).
%Given a $2$-proper partition, the heuristic employs local search to first optimize population balance, followed by optimizing compactness. 
For a given partition $\mathcal{P}=\{V^{(k)}\}_{k \in [K]}$, the size balance objective is 
\begin{align}
\phi_{bal} (\mathcal{P}) \coloneqq \max \{\max_{k \in [K]}\{0, size(V^{(k)})-U\},  \max_{k \in [K]}\{0, L-size(V^{(k)})\}\}. \label{popbal_objective}
\end{align} 
Since $L,U \geq 0$, observe that $\phi_{bal}$ is  zero if and only if all the part sizes are within $[L,U]$; otherwise, $\phi_{bal}$ measures the extent of size imbalance.
Therefore, to create an $L,U$-balanced partition, the local search procedure minimizes $\phi_{bal}$ until it reaches zero, or when no more feasible reassignments improve $\phi_{bal}$. In the latter case, the local search has not balanced the parts, in which case the heuristic will restart from the construct stage.

%Furthermore, the compactness minimization objective, $\phi_{comp}$, is given in Equation (\ref{eq_compactness_obj}).

The local search procedure uses the popular \textit{Flip} method for PDP \citep{ricca2008local}.
This procedure makes small adjustments to a partition by iteratively reassigning vertices, one at a time, from their current part to a neighboring part. 
Each potential reassignment is \textit{feasible} if the current and receiving parts maintain their $2$-connectivity after the reassignment.
Here, $2$-connectivity is assessed using Algorithm 11 in \cite{esfahanian2013connectivity}.
A feasible reassignment is accepted if it improves the given objective.
At the end of the two improvement stages, the heuristic returns an $L,U$-balanced $2$-proper $K$-partition of $G$ with locally optimal compactness. 

\section{Computational Results} \label{sec_computationalresults}
 
This section presents a computational investigation of the branch-and-cut and heuristic methods for HCGP on various problem instances.
We consider $42$ graph instances, three values for the number of parts $K \in\{2,3,4\}$, two size balance settings, and four values for the minimum connectivity requirement $Q\in [4]$.
First, for $Q=2$, we analyze the effectiveness of both methods in achieving optimal or feasible solutions to HCGP within a limited time. 
This analysis offers insights into the strengths and weaknesses of each algorithm by assessing their performance with different classes of instances, considering factors such as size and density.
Next, using the branch-and-cut method, we empirically measure the cost of ensuring high connectivity for $Q \in \{2,3,4\}$.
To achieve this goal, we compare the optimal compactness values and the computational time required to optimally solve instances with $Q \in \{2,3,4\}$ against instances with $Q=1$.
Taken together, these two analyses provide an empirical understanding of the effectiveness of the methods in achieving optimal or near-optimal solutions to HCGP.
%This cost is measured in terms of the computational time and the optimal compactness objective.

All computations in this study were performed on a MacBook with an 8-core M1 chip, 16GB of RAM, 3.2GHz CPU, and up to 12MB of cache per core. The algorithms were implemented in Python 3.7.
The IP instances were solved sequentially using Gurobi 10.0, which used all eight cores of the M1 chip.
%The computational times report the clock time (in seconds).
The separation algorithm was implemented within Gurobi's callback functionality.
When comparing the computational performance of two algorithms across a set of instances, we take the geometric mean of the ratios of the computational times across the instances, as prescribed by \cite{koch2022progress}.
  
The rest of this section is organized as follows.
Section \ref{subsec_results_datasets} details the datasets and the process used to create instances for HCGP.  
To determine an efficient implementation strategy for the branch-and-cut method used in this section, Section \ref{sec_preliminaryanalysis} presents a preliminary investigation on the number and type of violated constraints added in the separation algorithm.
% to deduce the number and type of violated constraints added by the separation algorithm.
Section \ref{subsec_results_BCGP} investigates the performance of the exact and heuristic methods for HCGP with $Q=2$.
Section \ref{subsec_results_higherconn} expands the investigations of the branch-and-cut method  for $Q$ values of three and four.
Drawing on the results from the previous section, Section \ref{subsec_costofhigherconn} empirically quantifies the cost of ensuring high connectivity (when using the branch-and-cut method) compared to a baseline of ensuring $1$-connectivity.
Section \ref{subsec_results_specialgraphs} illustrates visual examples of highly connected partitions of two graphs.

%%To simulate real-world applications of the highly connected graph partitioning problem (HCGP), we solve 
%apply the the cutting planes method described in Section \ref{sec_cuttingplanes} is applied to real-world instances. 
%To ensure diversity in the instances, three application domains of HCGP are considered. 
%The first domain involves political redistricting where data from the Census Bureau is utilized to partition a large geographical region into electoral districts. 
%The second domain involves decentralized transportation monitoring where data from five benchmark graphs described in \cite{ghavidelsyooki2017partitioning} is used to partition a large transportation network into smaller networks. 
%The third domain focuses on cloud computing where data from Cisco Secure Workload is used to partition a large (non-planar) network of machines into clusters, each of which is resilient to failures. 
%The first two are geographic applications with instances that have a relatively low vertex degree, but exhibit planarity.
%The third application establishes relationships between machines that are non-planar, and have a higher vertex connectivity.
%%These instances exhibit varying structural graph properties such as minimum vertex degree and planarity. 
%The performance of the cutting planes method is investigated on these classes of instances.

\subsection{Datasets and Instance Preparation}\label{subsec_results_datasets}

This section creates a test bed of instances for HCGP derived from real-world graphs in several domains.
First, we describe the $42$ graphs compiled from various data sources.
Next, to encourage each graph instance to contain feasible highly connected partitions, a pre-precessing stage increases the connectivities of these graph instances.
% using vertex deletion or edge addition operations. 
We now detail the graphs selected from the various data sources and the pre-processing stage.

\subsubsection{Raw Datasets}  
We use real-world graphs spanning several domains: transportation networks, census tract adjacency graphs,  power networks, social networks, interaction networks, animal social networks, and other miscellaneous graphs.
%These graphs exhibit differing levels of graph densities, quantified by the average vertex degrees.
These graphs are obtained from a variety of publicly available data repositories.
We select graph instances with fewer than $1,000$ vertices for computational tractability. 
We now describe each data source by domain.

%For transportation networks, we use the data published in \cite{transpo_networks_data}, where each network corresponds to a region's road network.
%The nodes represent locations of road intersections and the edges represent road links.
%This collection of networks has been popularly used by several studies for the traffic assignment problem \citep{bar2002origin}, which predicts route choices of travelers given their origin and destination. 
%The full dataset contains 22 regions with number of nodes ranging from $24$ to $33,837$, from which we focus on the nine regions with less than $1,000$ vertices each.
A transportation network models road intersections as vertices and road links as edges. 
\textit{Transportation Networks} \citep{transpo_networks_data} is a repository of graphs containing urban road networks from cities worldwide, such as Anaheim and Barcelona.
%This collection has been popularly used in research in the traffic assignment problem \citep{bar2002origin}, which predicts route choices of travelers given their origin and destination. 
We use nine graphs from this repository for our test bed. 

A census tract adjacency graph models census tracts in the U.S. as vertices, and edges represent whether two census tracts share a boundary of non-zero length. 
Each graph corresponds to the census tract graph within a U.S. state.
These graphs are used in various applications, most notably in political redistricting.
Beyond the adjacency information, \cite{census_2020} also provided the population in each census tract.  
Among the $50$ states, we consider $12$ states that each (a) have fewer than $1,000$ census tracts, and (b) were apportioned at least two congressional districts based on the 2020 census. 

A power network represents an electricity transmission system comprising buses and generators as vertices, and edges representing transmission lines.
We use the \textit{IEEE 300-Bus} network with $247$ vertices from  \cite{ICSEG} and a power network with $494$ vertices from \textit{Network Repository} \citep{networkdatarepository}, a repository with graphs from more than $30$ domains.  

Social networks are modeled using graphs where vertices represent anonymized social networking platform users, and edges represent pairs of connected users.
The \textit{Stanford Network Analysis Project} (SNAP) dataset \citep{snapnets} contains Facebook networks created by \cite{leskovec2012learning}, and we use six graphs from this dataset that have fewer than $1,000$ vertices.

Further, we use interaction and animal social networks from the \textit{Network Repository} \citep{networkdatarepository}.
Interactive networks model people as vertices and edges represent physical or virtual interactions. 
We select five such graphs, including a patient proximity network in a hospital ward and an email communication network.
In animal social networks, vertices represent animals and edges represent whether two animals have been in physical proximity. 
We select three graphs representing an ant colony, wild birds, and dolphins.
  
Finally, we use five miscellaneous graphs from the \textit{SuiteSparse Matrix Collection} \citep{sparsematrix_data}, formerly known as the \textit{University of Florida Sparse Matrix Collection}. 
%This collection is a well-known set of sparse matrix benchmarks collected from a several applications, and it is used by the numerical linear algebra commy for the development and performance evaluation of sparse matrix algorithms. 
Three graphs are Mycielskian graphs with the special properties of being triangle-free and having a high chromatic number.
Mycielskian graphs have been used as test instances for problems requiring high connectivity \citep{buchanan2015integer}.
The other two graphs are highly connected graphs with connectivities of $19$ and $81$.

\subsubsection{Processing Instances}
The pre-processing stage aims to increase the connectivities of the input graphs derived from these data sources.
%We now describe the pre-processing stage that prepares a test bed of instances.
%This goal aids in the goals of the computational study, which is to prepare
We conjecture that highly connected instances are more likely to contain highly connected partitions.
Hence, to prepare a test bed of instances for HCGP likely to contain feasible highly connected partitions, we modify each graph instance using vertex deletion or edge addition operations.
Note that in the graph creation, at most one edge is allowed to exist between any two vertices,  even if the problem domain typically permits the presence of multiple edges, e.g., multiple boundary segments for a pair of census tracts. 
%Moreover, the raw data from some of these data sources are sometimes not suitable as problem instances for HCGP.
%For example, seven of the $42$ graphs are disconnected.
%Even though it is possible to partition a disconnected graph into  highly connected partitions, we observe that solving the problem directly using these instances yields an infeasibility certificate even for $Q=1$.
%Hence, the pre-processing stage alters the raw graphs to create higher connected versions.

We introduce two different pre-processing methods based on the graph's planarity.
The instances from transportation networks and  census tract adjacency graphs are planar, while those from the rest of the domains are non-planar.
For planar instances, the goal is to produce a $2$-connected graph instance while retaining its planarity.
To achieve this goal, we select the largest $2$-connected component of each graph, and delete the rest of the vertices from the graph.
%For the rest of the graphs, we increase the connectivity to be at least $4$, with the exception of power graphs, whose connectivity is increased to at least $2$. 
%First, if the graph is disconnected, we arbitrarily select a pair of vertices, one from each disconnected component, and introduce an edge between them; this is repeated until the graph becomes connected.
%Next, we consider the cut-sets of size three (or one for power graphs) or less. For each cut-set, we add edges between arbitrary pairs of vertices across the components created by hypothetically removing the cut-set from the graph.
%By following this process, we obtain a graph with a connectivity of at least four (and two for power graphs), which is added to the test bed.
Furthermore, for the non-planar graph instances, the goal is to achieve $4$-connected versions, except for power networks where we achieve $2$-connected versions due to their sparsity; increasing the connectivity beyond two would significantly alter the structure of a power network.
To achieve this goal, we alter a given graph into a $Q$-connected graph for $Q \in \{2,4\}$ using the following iterative procedure.
In each iteration, we identify the cutsets of sizes less than $Q$, and the components of the graph that result from the removal of each cutset. 
Then, we arbitrarily select two vertices (namely the first two vertices in the arrangement of vertex labels) across two different components, and add an edge between them.
After adding edges across all pairs of components created by each cutset, and iterating this procedure until there are no cutsets of size less than $Q$, this graph becomes $Q$-connected.
%We apply this procedure to produce $2$-connected versions of power networks, and $4$-connected versions of the social, interaction, and animal social networks.
Note that this procedure does not alter the miscellaneous graphs since their connectivities are above five.

\begin{table}[!ht]
\centering
\csvreader[ 
before reading=\footnotesize\sisetup{round-mode=places,round-precision=2,round-integer-to-decimal},
tabular=cccccc,
table head= \toprule  
Domain &	 {\begin{tabular}[c]{@{}c@{}} Number \\ of graphs ($G$) \end{tabular}} & Size ($n$) & {\begin{tabular}[c]{@{}c@{}} Minimum \\ degree ($\delta$) \end{tabular}} &  {\begin{tabular}[c]{@{}c@{}} Average \\ degree ($\overline{\delta}$) \end{tabular}}  & Connectivity ($\kappa(G)$)  \\
\midrule,
  late after last line=\\  \midrule,
%    late after line=\\\midrule,
]{tables_tex/overview_of_instances.csv}{}%
{\csvcoli & \csvcolii & \csvcoliii & \csvcoliv & \csvcolv & \csvcolvi}
\caption{An overview of $42$ processed graph instances in each domain with the ranges in the number of vertices, minimum and average vertex degrees, and the graph connectivities. \label{table_dataset_overview}}
\end{table}

Table \ref{table_dataset_overview} summarizes the processed graph instances.
It reports the graph instances' domain-wise size, degree and connectivity ranges.
Table \ref{table_dataset_full} in Appendix \ref{appendix_overview_table} details the number of vertices removed or the edges added in the pre-processing stage.
For the planar instances, $1\%$ to $45\%$ of vertices were removed (median: $9\%$), and for the non-planar instances, $0\%$ to $38\%$ of edges were added (median: $1\%$).
In the rest of this section, the transportation, census tract adjacency and power networks are referred to as \textit{sparse} graphs due to their relatively low minimum degrees (i.e., less than four), and the rest are \textit{dense} graphs.
Further, we designate graphs with more than $500$ vertices as \textit{large}, while the rest are \textit{small}.
Among the $42$ graphs, $23$ are sparse and $11$ are large.
%Power networks are the sparsest, while animal social networks exhibit the highest densities.

Each graph instance includes vertex weights necessary to incorporate the size-balance constraints.
For census tract adjacency graphs, we treat the population residing in each census tract as its vertex weight.
For the rest of the graphs, we set the vertex weights to be one.
Furthermore, the analysis in this section considers two settings for size balance: one without the size balance constraints, and one requiring that the part sizes do not deviate more than $10\%$ from the average part size, denoted by $\overline{p}\coloneqq size(V)/K$.
This section represents these two settings by a size balance deviation threshold: $\tau = \infty$ without size balance constraints, and  $\tau = 0.1$ when requiring $L=0.9\ \overline{p}$ and $U=1.1\ \overline{p}$.

%To ensure tractability of solving HCGP, we choose appropriate values for the number of districts $K$, population deviation $\tau$, and $Q$. 
%We observe that $K$ values greater than five become intractable. 
%As is customary in political redistricting algorithms, we set $\tau$ to be $1\%$ for the political redistricting instances and $10\%$ for the remaining instances.
%In planar instances, $Q$ is limited to $2$, while in social networks, we explore up to $Q=4$. 
%For the instances in the rest of the domains, we explore $Q$ up to the connectivity of the input graph, which can be as high as $80$.

%Using the largest urban network, namely the Chicago instance, we create $100$ smaller instances each with $300$-$500$ vertices using the ear-construction procedure described in Section \ref{}.
%These instances are labeled $Chicago\_subgraphs$. 

\subsection{Preliminary Analysis of Separation Algorithm}
\label{sec_preliminaryanalysis}
To solve an instance of HCGP, the branch-and-cut method adds on-the-fly constraints generated by the separation algorithm outlined in Algorithm \ref{algo_intseparation_biconnected}.
%Each execution of Algorithm \ref{algo_intseparation_biconnected} adds a set of violated constraints.
In this section, we provide a preliminary analysis of the number and choice of violated constraints added in each run of the separation algorithm.
The goal is to determine the settings that result in the fastest computational time.
First, we investigate whether adding all the violated constraints generated by Algorithm \ref{algo_intseparation_biconnected} is superior to adding just one violated constraint.
%This preliminary analysis aims to empirically determine the preferable approach. 
Next, recall that Algorithm \ref{algo_intseparation_biconnected} assesses whether each part is root-resilient; if it is,  the algorithm adds $O(n)$ violated constraints in lines 10-13 for root-resilient parts, while it adds $O(n^2)$ violated constraints in lines 15-18 for parts that are not root-resilient.
If the root-resilience consideration was not implemented, the default method would add $O(n^2)$ constraints for every integer feasible solution.
%The incorporation of root-resilience consideration offers an advantage in terms of theoretical worst case computational time.    
We conduct comparative experiments by running the algorithm with and without root-resilience considerations to verify its computational benefit.

 \begin{table}[!ht]
\centering
\csvreader[ 
before reading=\footnotesize\sisetup{round-mode=places,round-precision=2,round-integer-to-decimal},
tabular=cccccc,
table head = \toprule  
{\begin{tabular}[c]{@{}c@{}}  \\ Graph ($G$) \end{tabular}}  & 
{\begin{tabular}[c]{@{}c@{}}  \\ $n$ \end{tabular}}  &
\multicolumn{2}{c}{Time taken with root-resilence (s)} & 
\multicolumn{2}{c}{Time taken without root-resilence (s)} \\ \cmidrule(lr){3-4} \cmidrule(lr){5-6} 
%\\
 & & all & one & all & one \\
\midrule,
  late after last line=\\  \midrule,
]{tables_tex/table_prelim_analysis.csv}{}%
{\csvcoli & \csvcolii & \csvcoliii & \csvcoliv & \csvcolv & \csvcolvi}
\caption{Computational times for solving HCGP with seven graph instances, $K=2$, $Q=2$ and $\tau = \infty$; the total time spent in callbacks is in parenthesis. 
Each run of the separation algorithm is executed $with$ versus $without$ root-resilience considerations, and with adding $all$ versus $one$ violated constraint. *No feasible solution was found within the time limit. \label{table_prelim_separation}}
\end{table} 

For this analysis, we select seven graph instances representing the median number of vertices in each of the seven domains. 
We solve HCGP with $K=2$, $Q=2$, and without the balance constraints.  
Table \ref{table_prelim_separation} presents the computational times for the four algorithm settings: \textit{with} versus \textit{without} root-resilience, and including \textit{all} versus \textit{one} violated constraint in each execution of Algorithm \ref{algo_intseparation_biconnected}.
Each run of the branch-and-cut method was given a time limit of one hour.
 
Among these four settings, we observe that incorporating root-resilience and adding all the violated constraints produces the fastest computational time in all seven instances. 
%in the majority of the cases, incorporating root-resilience is more efficient that not incorporating it, and adding all the violated constraints is more efficient than adding them one at a time.
%In particular, the combination of settings with incorporating root-resilience and adding all the violated constraints produces the fastest computational time in all seven instances. 
Therefore, this setting is used for the remainder of this section. 
Furthermore, we make three observations.
First, when solving without root-resilience, adding all constraints is more efficient than adding one constraint in the minority ($3$ out of $7$) of instances;
the most significant difference is with the mycielskian9 instance.
In contrast, when solving with root-resilience, in all seven instances, adding all constraints is more efficient than adding one constraint.
This observation suggests that adding all the $O(n^2)$ constraints is computationally expensive without root-resilience, as observed in the callback time.
Hence, we recommend adding one constraint at a time when solving without root-resilience.
Second, as expected, incorporating root-resilience yields a faster computational time in most (i.e., $12$ out of $14$) instances.
Third, as expected, in most (i.e., $11$ out of $14$) instances, the algorithm spends less time within the callbacks when one constraint is added than when all constraints are added. 
However, despite reducing the time spent in callbacks, the branch-and-cut method's overall computational time increases.
A possible explanation is that adding all constraints improves the best lower bound, which could enable the algorithm to prune more sub-problems in the branch-and-cut tree, contributing to a faster overall run-time.

\subsection{Results for  HCGP with $Q=2$}\label{subsec_results_BCGP}

In this section, we investigate the computational performance of the branch-and-cut and heuristic methods for  HCGP with $Q=2$.
We solved this problem for the $42$ graph instances, three values of the number of parts $K \in \{2,3,4\}$, and two balance settings, $\tau = \infty$ and $\tau = 0.1$.
Overall, these combinations of graphs, number of parts and balance settings produce $252$ instances. 
Both methods were given a time limit of one hour.

%\begin{table}[!ht]
%\centering
%\csvreader[ 
%before reading=\footnotesize\sisetup{round-mode=places,round-precision=2,round-integer-to-decimal},
%tabular=ccccccc,
%table head= \toprule  
%Domain &	 {\begin{tabular}[c]{@{}c@{}} Number \\ of graphs ($G$) \end{tabular}} & Size ($n$) & {\begin{tabular}[c]{@{}c@{}} Minimum \\ degree ($\delta$) \end{tabular}} &  {\begin{tabular}[c]{@{}c@{}} Average \\ degree ($\overline{\delta}$) \end{tabular}}  & Connectivity ($\kappa(G)$) & \\
%\midrule,
%  late after last line=\\  \midrule,
%%    late after line=\\\midrule,
%]{results/table_solutions_summary.csv}{}%
%{\csvcoli & \csvcolii & \csvcoliii & \csvcoliv & \csvcolv & \csvcolvi & \csvcolvii}
%\caption{ . \label{table_solutions_overview}}
%\end{table}
   
\begin{table}[!ht]
\small
\centering
\begin{tabular}{@{}ccccccc@{}}
\toprule
{\begin{tabular}[c]{@{}c@{}}  Solution \\ outcome \end{tabular}} 
 & \multicolumn{2}{c}{All instances (252)} & \multicolumn{2}{c}{Sparse instances (138)} & \multicolumn{2}{c}{Large instances (66)} \\ \cmidrule{2-3} \cmidrule{4-5} \cmidrule{6-7}
             & Exact       & Heuristic       & Exact         & Heuristic        & Exact        & Heuristic       \\ \midrule
Optimal      & 204         & 9               & 102           & 0                & 42           & 0               \\
Feasible     & 13         & 219             & 2           & 118              & 2           & 65              \\
Infeasible   & 9           & 0 & 9             & 0             & 0            & 0 \\
Inconclusive & 26          & 24              & 25            & 20               & 22           & 1               \\ \bottomrule
\end{tabular}
\caption{Distribution of instances based on solution outcomes for the two methods: optimal, feasible (and suboptimal), infeasible, or inconclusive, categorized by instance type.
% sparse instances have a minimum vertex degree less than four, and large instances have more than $500$ vertices.
 \label{table_solutions_overview}}
\end{table}
 
Table \ref{table_solutions_overview} summarizes the performance of both methods by listing the number of instances with outcomes in each of four categories: optimal, feasible (but suboptimal), infeasible, and inconclusive.
An instance is \textit{inconclusive} if a method terminates within the time limit with neither a feasible solution nor an infeasibility certificate.
Note that for an infeasible instance, the heuristic cannot produce an infeasibility certificate by design and would instead time-out without a feasible solution.  
Tables \ref{table_crosstabulated_Q2} and \ref{table_results_Q=2} in Appendix \ref{appendix_results_HCGP} present more details on these results.
Overall, as expected, the heuristic performed better in finding a feasible solution (i.e., categorized as either optimal or feasible in Table \ref{table_solutions_overview}) within the time limit, while the exact method performed much better in finding an optimal solution.

The methods' ability to find a feasible solution within the time limit depends on the instance's sparsity and size.
Of the $26$ instances categorized as inconclusive by the exact method, $25$ are sparse, and $22$ are large (and sparse).
%Hence, large and sparse instances are challenging for the exact method.
Note that $23$ of these $26$ instances were solved by the heuristic.
Further, among the $24$ instances categorized as inconclusive by the heuristic, $20$ are sparse and $23$ are small.
%Hence, small and sparse instances are challenging for the heuristic. 
Taken as a whole, these results show that even though both methods were able to solve many of the sparse instances, some sparse instances were computationally challenging (i.e., produced inconclusive results);
while large and sparse instances are challenging for the exact method ($22$ of its $26$ inconclusive outcomes), small and sparse instances are challenging for the heuristic ($19$ of its $24$ inconclusive outcomes).

%Therefore, large instances pose a greater challenge for the exact method than the heuristic, whereas sparse instances are challenging for both methods.

%The exact method successfully terminated with an optimal (feasible) solution in $204$ ($217$) instances. 
%In nine instances, the exact method produced an infeasibility certificate; all of these are sparse graph instances.
%In $26$ instances, called the \textit{inconclusive} instances, the exact method reached the time limit without finding either a feasible solution or an infeasibility certificate; in $13$ of these instances, it terminated without a non-zero lower bound on the compactness objective.
%Furthermore, 
%%Among the $243$ instances where the exact method did not provide an infeasibility certificate, 
%the heuristic was able (unable) to find a feasible solution within the one-hour time limit in $228$ ($24$) instances; it found an optimal solution in nine instances.

%Table \ref{table_Q2_heuristic_OPT} in the Appendix provides results from solving all the instances for $Q=2$, i.e., for $41$ graphs and three choices of $K=2,3,4.$

Applying the heuristic presents a trade-off between the cost in the compactness objective and the advantage in computational time.
% particularly for large graphs.
%We classify an instance as large if the graph has more than $500$ vertices.
To quantify the overall computational time advantage of the heuristic over the exact method for each instance, we compute the ratio of the exact method's time to the heuristic's time. 
%For instances where the exact method reached the one-hour time limit, we treat its computational time to be one hour; hence, this ratio serves as a lower bound on the actual ratio.
Then, we calculate the geometric mean of the ratios across the instances.
Further, we compute the \textit{approximation gap} for each instance to measure the cost in the compactness objective. 
Let $OPT$ and $HEUR$ be the compactness of the exact and heuristic solutions, respectively.
The approximation gap, quantified by $(HEUR-OPT)/OPT$, measures the relative compromise of the heuristic's compactness objective compared to the optimal compactness. 
If the exact method terminated without an optimal solution, the best known lower bound is used in place of $OPT$, in which case this gap serves as an upper bound to the heuristic's approximation gap.
%We then calculate the arithmetic mean of the approximation gaps across all (and the large)  instances.  
%The trade-off offered by the heuristic can be quantified.
We focus on the $205$ (and $44$ large) instances for which both methods found a feasible solution.
%There $205$ such instances, among which $44$ are large.
On average, in these and in the large instances, the heuristic is $3.80$ and $13.86$ times faster than the exact method, respectively, while exhibiting an average approximation gap of $30.07\%$ and $38.46\%$, respectively.
Hence, the computational time advantage of the heuristic over the exact method is more evident in solving large instances than all instances, even though a relatively high approximation gap accompanies this advantage.

While the heuristic provides a computational advantage on average, the exact method outperforms it  in small instances.
The exact method is faster than the heuristic in $80$ out of all $252$ instances, all of which are small.
Among the $66$ large instances, the heuristic is faster in $64$ instances.
%Furthermore, in the $26$ instances deemed inconclusive by the exact method, $22$ are large.
%Hence, the heuristic outperforms the exact method in large and sparse graphs.
This observation aligns with the earlier finding that the heuristic finds more feasible solutions among the larger instances compared to the exact method.
Hence, if computational time is prioritized over achieving optimality, we recommend using the exact method for small instances and the heuristic for large instances.

%%Among the $252$ instances, $66$ were large. 
%%Tallying the instances, the heuristic is faster than the exact method in $165$ out of all $252$ instances, and in $64$ out of the $66$ large instances.
%Conversely, the exact method is faster than the heuristic in $80$ out of all $252$ instances, and in none of the large instances.
%%Among the $228$ instances for which the heuristic found a feasible solution, and specifically the $65$ large instances,  the heuristic exhibits an average approximation gap of $25.21\%$ and $16.59\%$, respectively.  
%%Hence, the heuristic's advantage over the exact method is more evident in solving large instances than all instances.
%%If optimality is preferred at the cost of computational time, the exact method is the effective choice.

\subsection{Results for HCGP with $Q \in \{3,4\}$}\label{subsec_results_higherconn}

In this section, we investigate the performance of the branch-and-cut method for HCGP with $Q \in \{3,4\}$  applied to the $19$ dense graphs; sparse graph instance results are not presented here since they yielded infeasibility certificates for $Q \in \{3,4\}$.
Similar to the settings used in the previous section, we solved for three different values of $K \in \{2,3,4\}$, and two balance settings, $\tau = \infty$ and $\tau = 0.1$.
In total, we solved $228$ instances, with each run of the method given a time limit of one hour.

Table \ref{table_dataset_full_higherconnected} in Appendix \ref{appendix_results_HCGP} presents the results from solving these $228$ instances. 
The method terminated with an optimal (feasible) solution in $171$ ($192$) instances. 
In $17$ instances, the method produced an infeasibility certificate. 
In $19$ instances, the method was inconclusive within the allotted time limit.
Among the $17$ infeasible instances, two ($15$) instances are for $Q=3$ ($4$).
None of these graph instances produced an infeasibility certificate for $Q=2$.
This trend demonstrates that the feasible region of HCGP becomes smaller when the required connectivity is increased from two through four. 

%One of the miscellaneous graphs with 

%
%\subsection{Performance of Separation Algorithm} 
% 
%- Time in callback, number of solutions traversed, number of root-resilient solutions
%What does the time spent in callback depend on?

%- time in call back, separation algo, number of root resilience solutions, and how that depends on $Q$

\subsection{Costs of Higher Connectivity}\label{subsec_costofhigherconn}

Ensuring highly connected partitions potentially incurs two costs: (i) a compromise in the optimal compactness objective due to the reduction in the solution space of feasible partitions, and (ii) an increase in computational time required to solve the associated integer program.
%Beyond $Q=1$, requiring the parts be higher connected reduces the feasible region, and therefore potentially reduces the achievable optimal compactness of the partition.
In this analysis, we examine the optimal compactness value and the computational time for $Q\in\{2,3,4\}$ and contrast them with the baseline case of $Q=1$.
This comparison provides insight into the costs of ensuring each additional level of connectivity (beyond one).
 
To quantify the baseline for these comparisons, Table \ref{table_results_Q1} in Appendix \ref{appendix_results_HCGP} presents the results for $Q=1$ across the $252$ instances solved in Section \ref{subsec_results_BCGP} with a time limit of one hour.
The branch-and-cut method terminated with an optimal (feasible) solution in $231$ ($236$) instances. 
In $16$ instances, the method was inconclusive; all of these instances are sparse graphs.  
None of these instances yielded an infeasibility certificate.
\begin{figure}[!ht]
\centering
\subfloat[$\tau=\infty$. \label{fig_n_instances_versus_Q_inf}] {\includegraphics[width=0.45\textwidth]{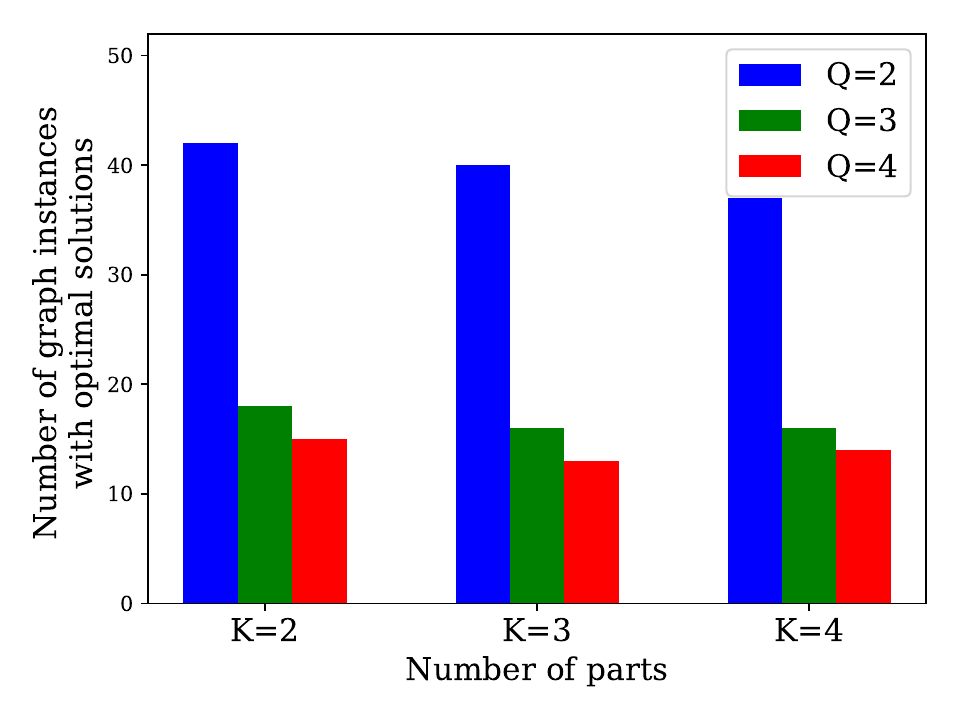}}
\hspace{.05cm}
\subfloat[$\tau=0.1$. \label{fig_n_instances_versus_Q_01}] {\includegraphics[width=0.45\textwidth]{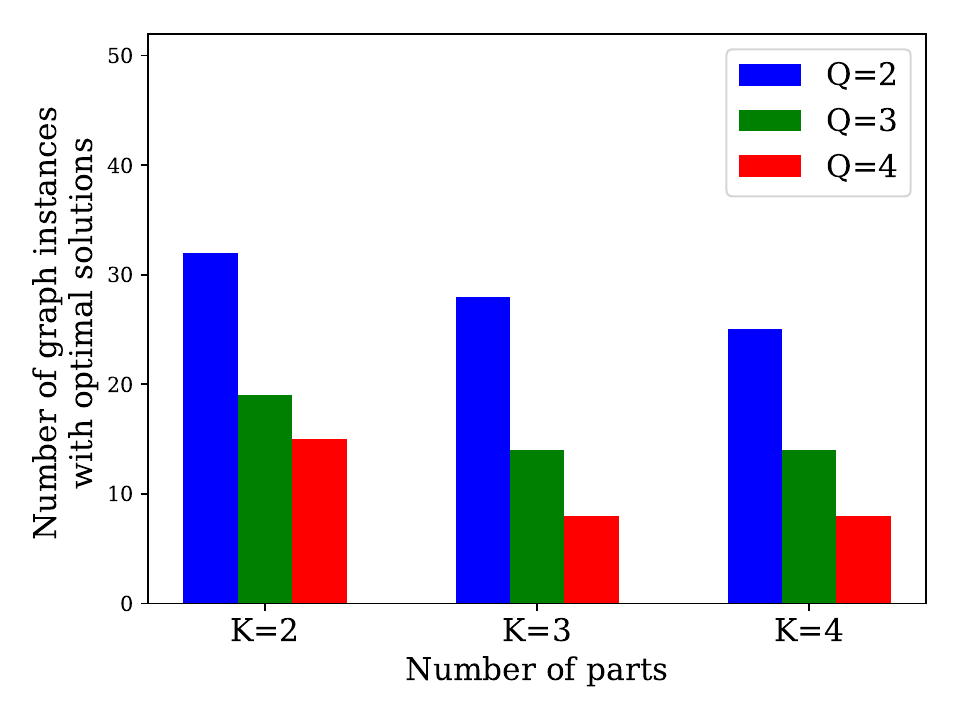}}
\caption{
Number of instances in which the branch-and-cut method found optimal solutions for $Q=1$ and subsequently found optimal solutions for each $Q \in \{2, 3, 4\}$; categorized for each $K \in \{2, 3, 4\}$ and balance $\tau \in \{0.1, \infty\}$.
}
\label{fig_n_instances_versus_Q}
\end{figure}

\begin{figure}[!ht]
\centering
\subfloat[$\tau=\infty$. \label{fig_compactness_versus_Q_inf}] {\includegraphics[width=0.45\textwidth]{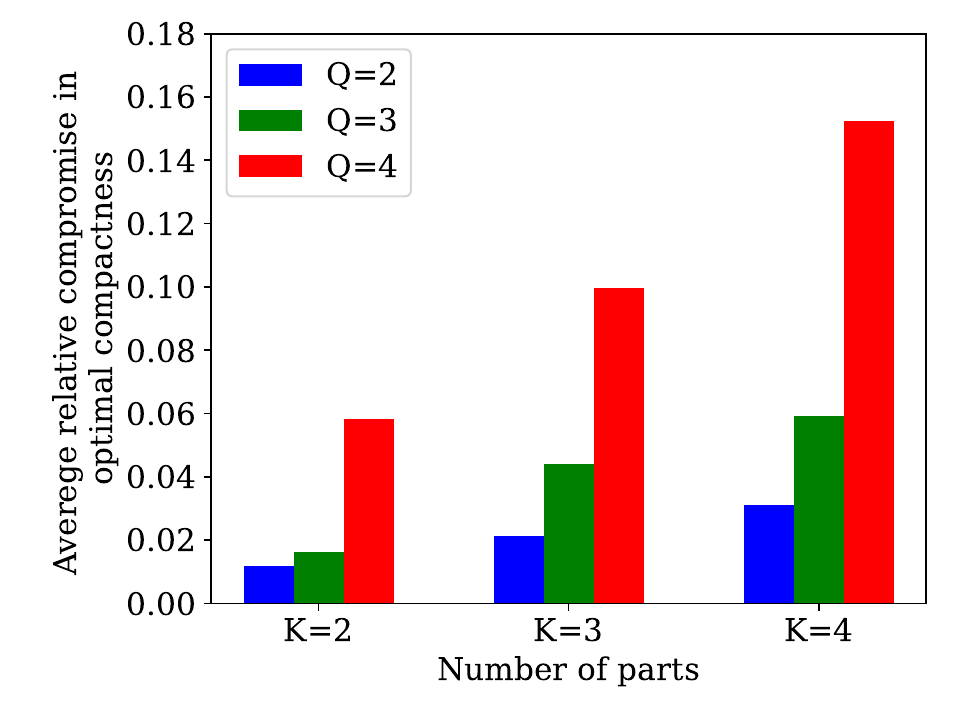}}
\hspace{.05cm}
\subfloat[$\tau=0.1$. \label{fig_compactness_versus_Q_01}] {\includegraphics[width=0.45\textwidth]{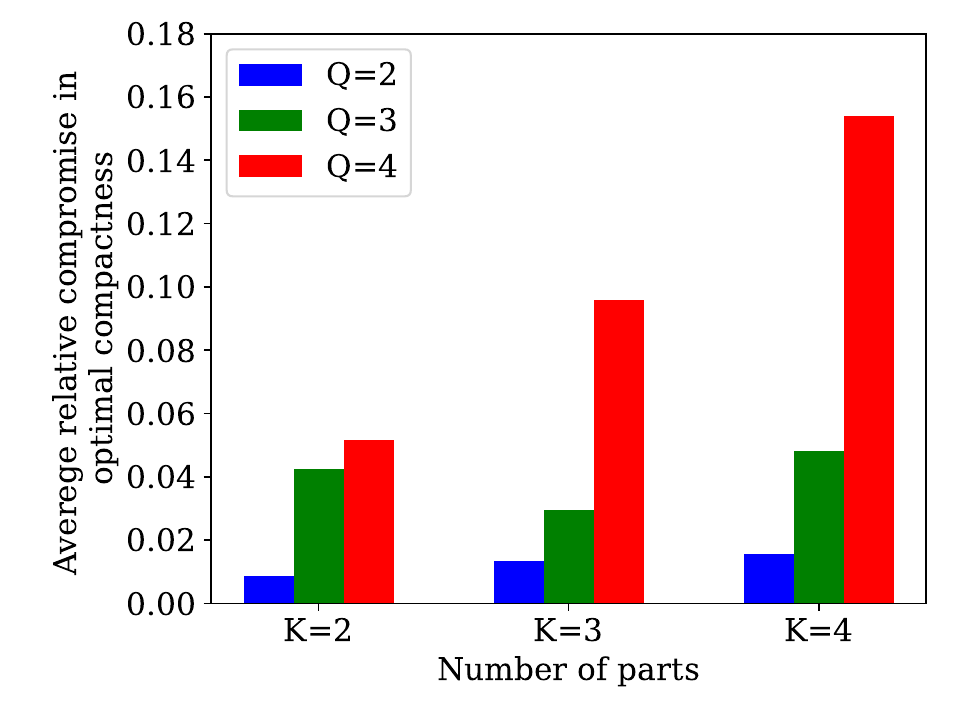}}
\caption{Average relative compromise in optimal compactness for $Q \in \{2,3,4\}$ compared to the baseline of $Q=1$; categorized for each $K \in \{2, 3, 4\}$ and balance $\tau \in \{0.1, \infty\}$.}
\label{fig_compactness_versus_Q}
\end{figure}

\begin{figure}[!ht]
\centering
\subfloat[$\tau=\infty$. \label{fig_timeratio_versus_Q_inf}] {\includegraphics[width=0.45\textwidth]{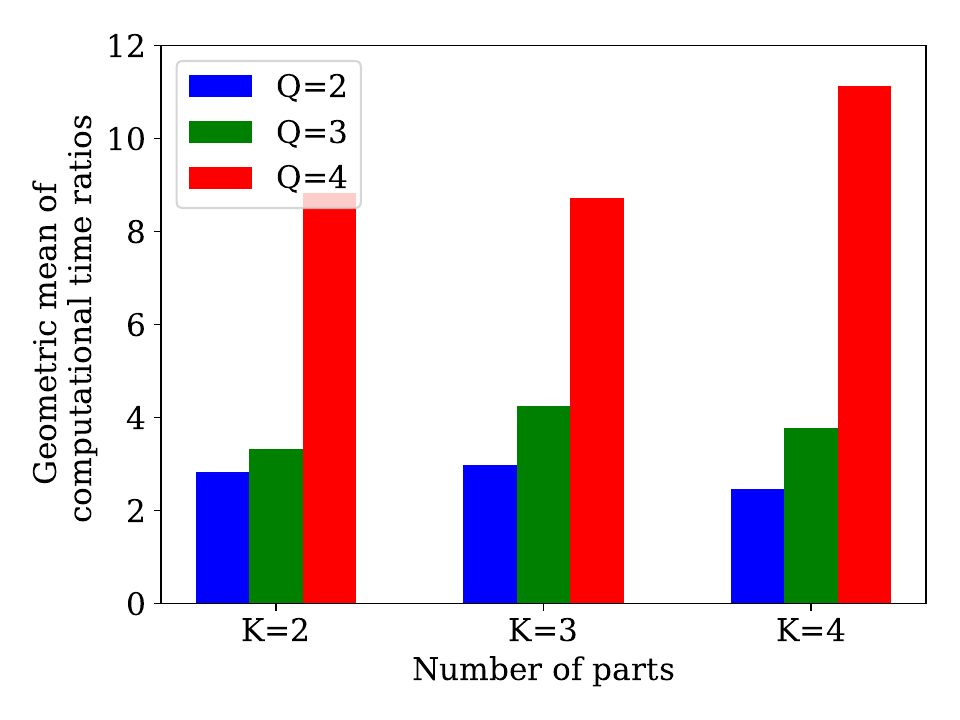}}
\hspace{.05cm}
\subfloat[$\tau=0.1$. \label{fig_timeratio_versus_Q_01}] {\includegraphics[width=0.45\textwidth]{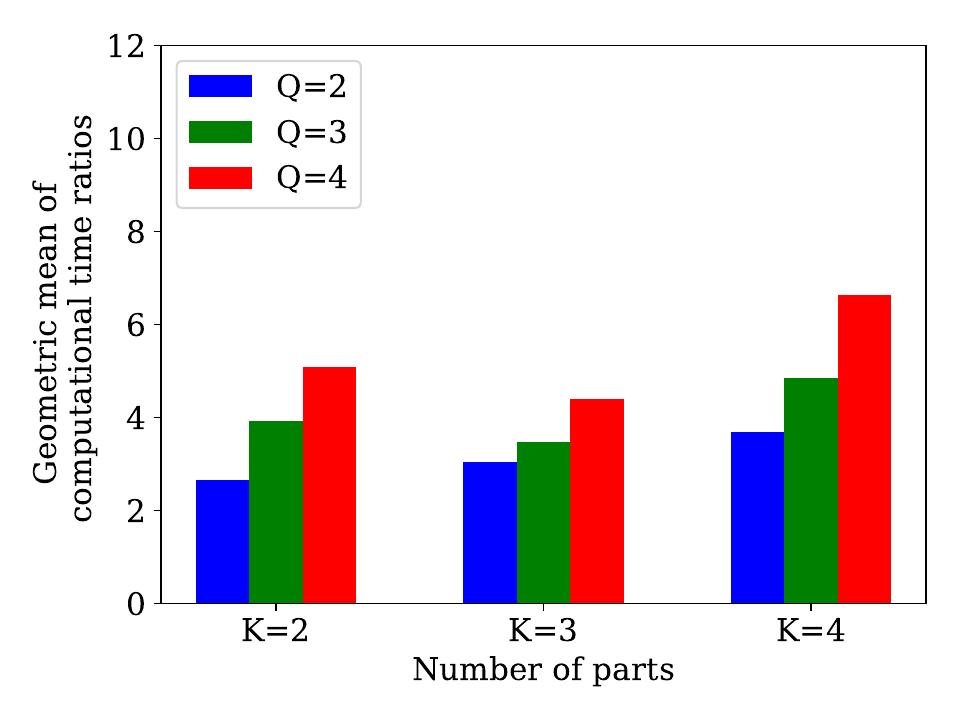}}
\caption{Geometric mean of computational time ratios for $Q \in \{2,3,4\}$ compared to the baseline of $Q=1$; categorized for each $K \in \{2, 3, 4\}$ and balance $\tau \in \{0.1, \infty\}$.}
\label{fig_timeratio_versus_Q}
\end{figure}

In our analysis, we compare the optimal compactness values and the computational times for each value of $Q \in \{2,3,4\}$ with the same for $Q=1$. 
%To achieve a fair comparison, we only include optimal solutions, and not sub-optimal solutions.
% instances for which the branch-and-cut method found an optimal solution for each setting of $K$, $\tau$ and $Q$.
For a given $Q \in \{2,3,4\}$, let $\mathcal{I}_{1,Q}$ be the subset of instances for which the branch-and-cut method found an optimal solution for both $Q=1$ and for the given $Q$.
For each $Q \in \{2,3,4\}$, Figure \ref{fig_n_instances_versus_Q} depicts the number of instances in $\mathcal{I}_{1,Q}$, categorized by $K$ and $\tau$.
Furthermore, for each instance, let  $OPT_Q$ denote the optimal compactness for HCGP with $Q\geq 1$.
For every  $Q \in \{2,3,4\}$, we measure the \textit{relative compromise} in optimal compactness from the baseline $Q=1$ case as $(OPT_Q-OPT_1)/OPT_1$.
Figure \ref{fig_compactness_versus_Q} depicts the average relative compromise in optimal compactness (among the instances in $\mathcal{I}_{1,Q}$ for $Q \in \{2,3,4\}$), categorized by $K$ and $\tau$. 
Additionally, to quantify the effect of $Q$ on the computational time, we calculate the ratio between the computational times for each $Q \in \{2,3,4\}$ and for $Q=1$. 
Figure \ref{fig_timeratio_versus_Q} presents the geometric mean of the ratios (among the instances in $\mathcal{I}_{1,Q}$ for $Q \in \{2,3,4\}$), categorized by $K$ and $\tau$.
%requiring $4$-connectivity raises the computational time of the exact method by $11.2$ times compared to requiring simple connectivity. 

These empirical results suggest that increasing the required connectivity level $Q$ leads to more infeasible instances, while also eroding the optimal compactness objective and increasing computational time among the feasible instances.  
%As expected, increasing $Q$ reduces the number of instances with known optimal solutions as depicted in Figure \ref{fig_n_instances_versus_Q}.
%The average relative compromise in Figure \ref{fig_compactness_versus_Q} shows a monotonic increase with $Q$ for each setting of $K$ and $\tau$. 
%Moreover, the computational time ratio in Figure \ref{fig_timeratio_versus_Q} demonstrates a monotonic increase with $Q$.
For example, for $Q=4$, $K=4$ and $\tau = \infty$, on average among the $14$ instances in $\mathcal{I}_{1,4}$, an optimal $4$-proper partition exhibits a relative compromise in compactness objective of $15.24\%$, and requires approximately $11.2$ times the computational effort compared to finding an optimal $1$-proper partition.

\subsection{Visual Examples}\label{subsec_results_specialgraphs} 
\begin{figure}[!ht]
\centering
\subfloat[$Q=1$. \label{figure_iowa_Q1}] {\includegraphics[width=0.3\textwidth]{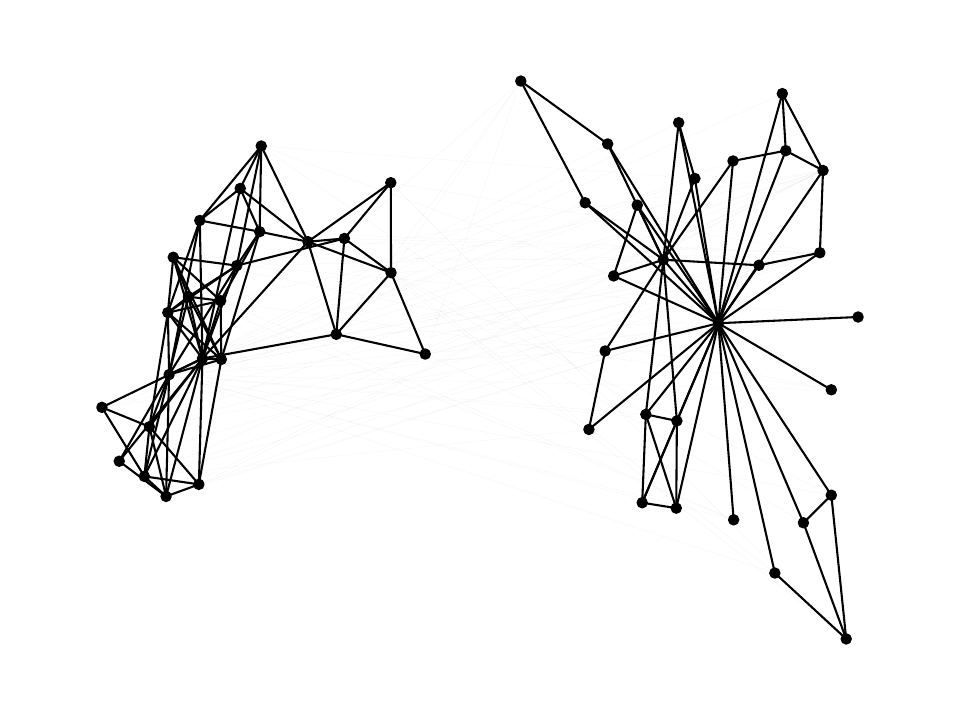}}
\hspace{.2cm}
\subfloat[$Q=2$. \label{figure_iowa_Q1}] {\includegraphics[width=0.3\textwidth]{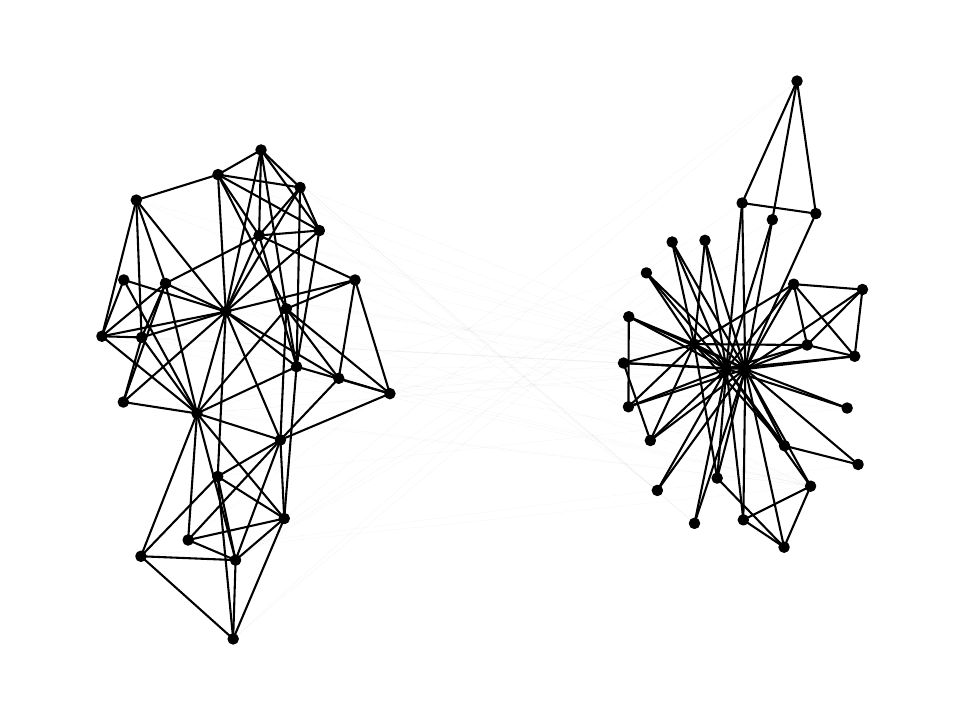}}
\hspace{.2cm}
\subfloat[$Q=3$. \label{figure_iowa_Q1}] {\includegraphics[width=0.3\textwidth]{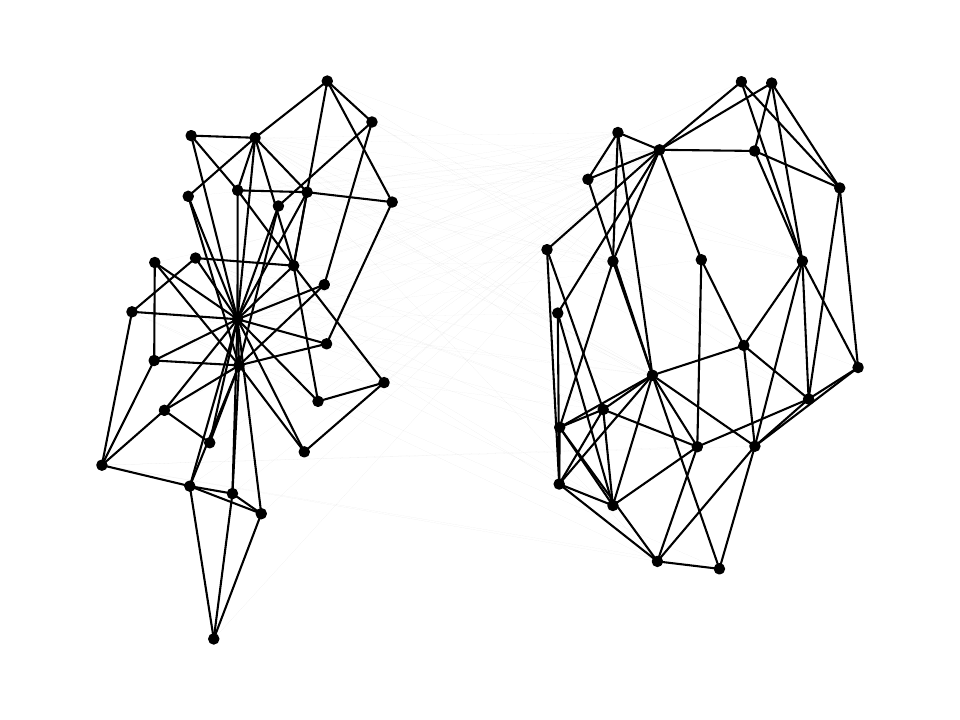}} 
\caption{Optimally compact and balanced (with $\tau=0.1$) $2$-partitions of a small social network with $52$ vertices and $201$ edges into $Q$-connected subgraphs for $Q\in [3]$; each part has a Kamada Kawai graph embedding.} \label{fig_socnet_exame_3980}
\end{figure}

Finally, we present sample visualizations of optimal solutions from two graph instances.
Figure \ref{fig_socnet_exame_3980} illustrates optimally compact and balanced $Q$-proper $2$-partitions of a social network comprising $52$ vertices for $Q\in [3]$.
For $Q=1$, notice the presence of a cut vertex in one of the parts. 
In the context of community detection in social networks, a partition with higher connectivity represents greater cohesion and robustness with multiple connections across the vertices within each part, as observed in the $Q=3$ case.
  
\begin{figure}[!ht]
\centering
\subfloat[$Q=1$. \label{figure_iowa_Q1}] {\includegraphics[width=0.4\textwidth]{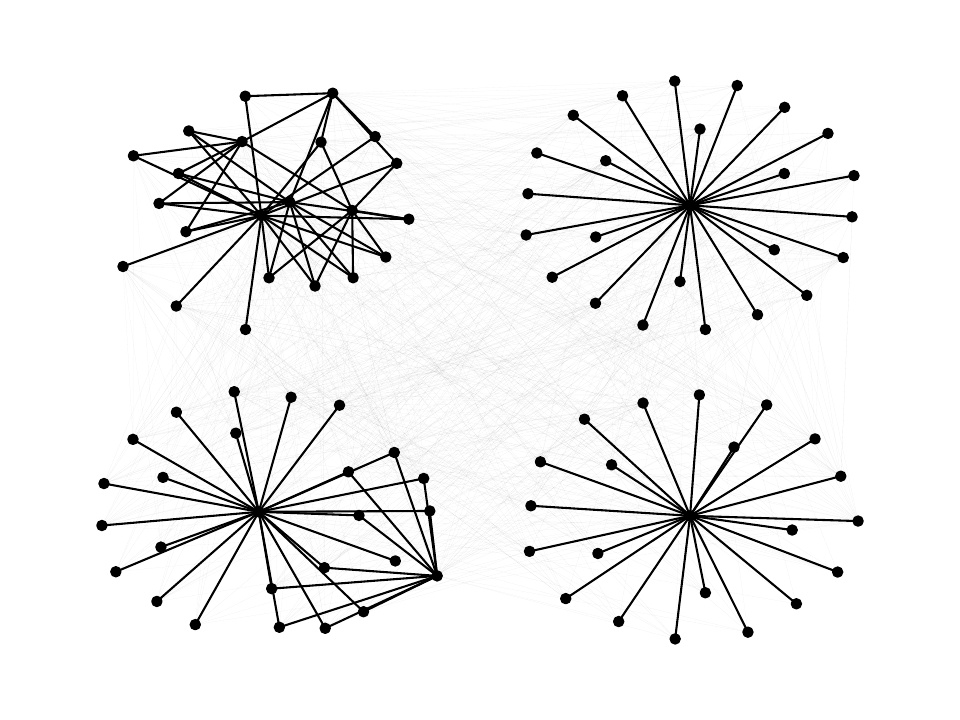}}
\hspace{.2cm}
%\subfloat[$Q=2$. \label{figure_iowa_Q1}] {\includegraphics[width=0.4\textwidth]{figures_HCGP/fig_misc_mycielskian7_K4_Q2_eps0.pdf}} \\
%\subfloat[$Q=3$. \label{figure_iowa_Q1}] {\includegraphics[width=0.4\textwidth]{figures_HCGP/fig_misc_mycielskian7_K4_Q3_eps0.pdf}} 
%\hspace{.2cm}
\subfloat[$Q=4$. \label{figure_iowa_Q1}] {\includegraphics[width=0.4\textwidth]{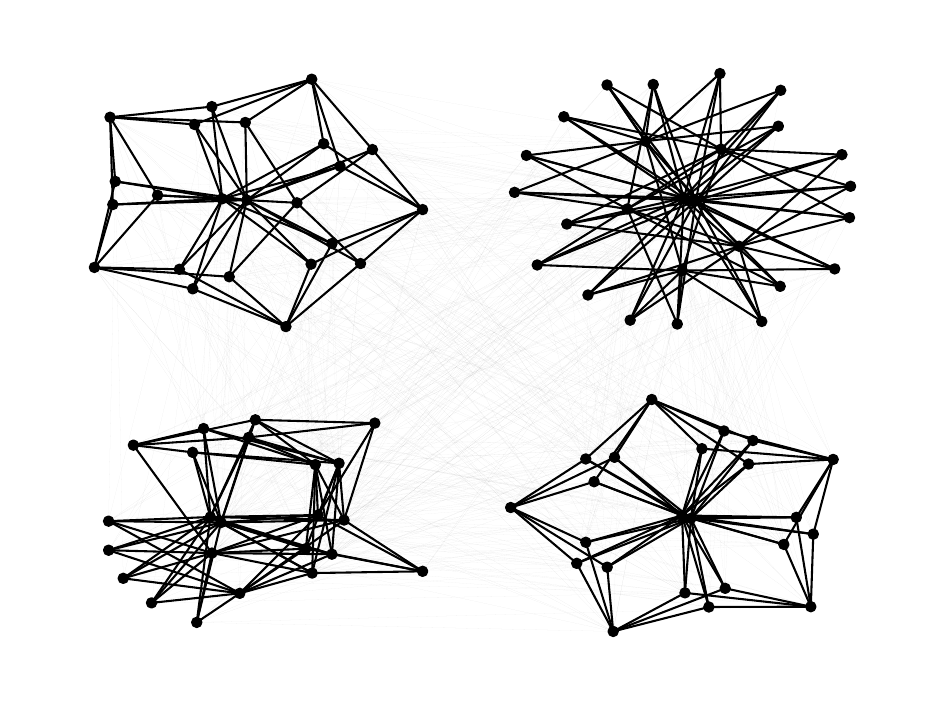}} 
\caption{Optimally compact and balanced (with $\tau=0.1$) $Q$-proper $4$-partitions of a Mycielskian graph with $95$ vertices and $755$ edges for $Q\in \{1,4\}$; each part has a Kamada Kawai graph embedding.} \label{fig_mycielskian}
\end{figure}

%\begin{figure}[!ht]
%\centering
%\subfloat[$Q=1$. \label{figure_iowa_Q1}] {\includegraphics[width=0.3\textwidth]{results/socnetwork/solutions/698_K2_Q1.png}}
%\hspace{.2cm}
%\subfloat[$Q=2$. \label{figure_iowa_Q2}] {\includegraphics[width=0.3\textwidth]{results/socnetwork/solutions/698_K2_Q2.png}}  
%\hspace{.2cm}
%\subfloat[$Q=3$. \label{figure_iowa_Q3}] {\includegraphics[width=0.3\textwidth]{results/socnetwork/solutions/698_K2_Q3.png}}
%\caption{Optimal partitions of a graph with $61$ nodes and $303$ edges into $K=2$ districts each being $Q=1,2,3$-connected.} \label{fig_example_compact}
%\end{figure}

Next, Figure \ref{fig_mycielskian} illustrates optimally compact and balanced $Q$-proper $4$-partitions of a Mycielskian graph with $95$ vertices for $Q\in \{1,4\}$.
This graph has the special property that it is triangle-free and has a chromatic number of seven.
A graph with a high chromatic number has a complex structure where the vertices are interconnected so that they cannot be colored with a small number of colors without violating the rule of adjacent vertices having different colors.
Observe that the $1$-proper partition includes two parts with star-like structures, wherein removing the central vertex from the part would disconnect that part into singleton components.
Hence, a vertex disruption to the central vertex would maximally degrade that part.
In contrast, the $4$-proper partition exhibits a higher level of connectivity across its parts, thereby being more resilient to vertex failures.

\section{Conclusions}\label{sec_conclusions}

In summary, this study combines two fundamental graph concepts of graph partitioning and vertex connectivity.
This paper presents an IP-based formulation for HCGP and solution methods to solve it.
Through computational analysis, we evaluate the performance of these methods across a diverse set of instances.
The results  showcase the effectiveness of the proposed methods in generating optimal or feasible solutions for HCGP.
%\textbf{Anything specific to say about the solutions?}
The analysis presented in this work serves as a starting point for studying the interconnection between graph partitioning and vertex connectivity.

%\begin{figure}[!ht]
%\centering
%\subfloat[$Q=1$. \label{figure_iowa_Q1}] {\includegraphics[width=0.3\textwidth]{results/specialgraphs/solutions/mycielskian7_K3_Q1.png}}
%\hspace{.2cm}
%\subfloat[$Q=3$. \label{figure_iowa_Q1}] {\includegraphics[width=0.3\textwidth]{results/specialgraphs/solutions/mycielskian7_K3_Q3.png}}
%\hspace{.2cm}
%\subfloat[$Q=5$. \label{figure_iowa_Q1}] {\includegraphics[width=0.3\textwidth]{results/specialgraphs/solutions/mycielskian7_K3_Q5.png}} 
%\caption{Optimally compact and balanced $3$-partitions of a Mycielskian graph with $95$ nodes and $755$ edges into $Q$-connected subgraphs for $Q=1,3,6$.} \label{fig_socnet_exame_3980}
%\end{figure}

This work lays the foundation for several potential directions for future research. 
First, investigating alternative formulations for the $Q$-connectivity constraints, such as using multi-commodity flows, could provide alternative ways to model and solve HCGP.
Second, investigating new techniques to improve the efficiency of the solution methods could enable practitioners to solve large  instances of HCGP.
For example, variable fixing techniques such as in \cite{validi2020} could improve the branch-and-cut method's efficiency.
For the heuristic setting, a constructive procedure to generate a $Q$-connected partition for $Q\geq3$ could be explored.
Third, studying the implications of adopting HCGP for practical applications would be valuable, while incorporating application-specific objectives and constraints.
For example, partitioning a power network  typically incorporates constraints that regulate the amount of power transmitted within each part \citep{sun2016network}. 
Hence, tailoring HCGP for application-specific needs would contribute to a positive impact of HCGP in practice.
%In summary, this work opens up multiple avenues for future research in designing efficient modeling and algorithmic approaches for HCGP. 

%\clearpage  
%
%\section{Conclusions}
% 
%
%Contributions:
%Limitations: Only for small instances - small $K$, number of nodes less than $1000$, etc.
%The heuristic is fast in hard instances, but the compactness objective is pretty poor. 
%
%Future research directions: 
%- higher connectivity could improve the number of inter-part edges, this relationship could be further studied.
%
%- Improving the branch-and-cut algorithm: incorporating techniques such as variable-fixing to strengthen the lower-bound. to do so, you need to take advantage of the structure of higher connected graphs.
%- To improve the performance of the heuristic, need to improve how $2$-connectivity is incorporated in the heuristic.
%Also, means to construct $\geq Q$ connected partitions should be explored.

%\newpage
\bibliographystyle{informs2014} % outcomment this and next line in Case 1
\bibliography{bibliography_HCGP.bib} % if more than one, comma separated

\newpage
\begin{appendices}

\section{Proofs for Theoretical Results}\label{appendix_proofs}
This appendix provides proofs for the theoretical results from Section \ref{sec_formulations}.
\subsection{Proof for Theorem \ref{thm_characterization}}

\begin{exe_thm}
 \label{thm_appendix_characterization} (Characterization of a $Q$-proper partition)
Given a graph $G$, positive integers $K \geq 2$ and $Q \geq 1$, a $K$-partition $\mathcal{P} = \{V^{(k)}\}_{k \in [K]}$ is $Q$-proper if and only if for every $k \in [K]$, 
\begin{enumerate}[(i)]
\item $|V^{(k)}| \geq Q+1$, and 
\item for every distinct $a,b \in V^{(k)}$, $|C\cap V^{(k)}| \geq Q$ for every $a,b$-separator $C \in \mathcal{C}_{a,b,G}.$
\end{enumerate} 
\end{exe_thm}
\textit{Proof:} 
($\Rightarrow$) 
Consider any part $k \in [K]$. 
To show the first condition, we note that since $\mathcal{P}$ is $Q$-proper, $G[V^{(k)}]$ is $Q$-connected and therefore $|V^{(k)}| \geq Q+1$ by definition. 
To show the second condition, we first consider the case where part $k$ forms a complete graph; then $\mathcal{C}_{a,b,G}=\emptyset$ for every $a,b\in V^{(k)}$, and so the second condition holds by default since there are no separators to consider. 
If part $k$ is not a complete graph, then we first claim that for every $k \in [K]$, every distinct $a,b \in V^{(k)}$ and $C \in \mathcal{C}_{a,b,G}$, $C \cap V^{(k)}$ is an $a,b$-separator in $G[V^{(k)}]$.
We proceed using proof by contradiction; assume the contrary that for some $k \in [K]$, $a,b \in V^{(k)}$ and $C \in \mathcal{C}_{a,b,G}$,  $C \cap V^{(k)}$ is not an $a,b$-separator in $G[V^{(k)}]$.
%First, consider the case that (i) is true.
Then, there exists an $a,b$-path $P$ in $G[V^{(k)}]-C$.
Since $V^{(k)} \subseteq V$, path $P$ is also in $G - C$.
This is a contradiction since $C$ is an $a,b$-separator in $G$.
Hence, $C \cap V^{(k)}$ is an $a,b$-separator in $G[V^{(k)}]$.
Therefore, $|C \cap V^{(k)}| \geq Q$ since $G[V^{(k)}]$ is $Q$-connected.

($\Leftarrow$) 
We are given that for every $k \in [K]$, (i) $|V^{(k)}| \geq Q+1$, and (ii) for every distinct $a,b \in V^{(k)}$ and $C \in \mathcal{C}_{a,b,G}$, $|C\cap V^{(k)}| \geq Q$.
We need to show that every part is $Q$-connected.
If a part $k\in [K]$ forms a complete graph, $\kappa(G[V^{(k)}]) \geq Q$ since it contains at least $Q+1$ vertices, and is therefore $Q$-connected.
If a part $k\in [K]$ is not a complete graph, we proceed using proof by contradiction; assume the contrary that $G[V^{(k)}]$ is not $Q$-connected, i.e., there exists a cutset $S \subset V^{(k)}$ with $|S| = Q' < Q$ that disconnects $G[V^{(k)}]$ into two parts, say $A$ and $B$.
Select arbitrary vertices $a \in A$ and $b \in B$.
By construction, $S$ is an $a,b$-separator in $G[V^{(k)}]$, but $S$ may not be an $a,b$-separator in $G$.
Define $W = V\backslash V^{(k)}$, and $C = S \cup W$.
Any $a,b$-path $P$ in $G$ must pass through $C$
(i.e., since either $P$ lies entirely in $V^{(k)}$ and must include a vertex from $S \subset C$, or $P$ leaves $V^{(k)}$ and must include a vertex from $W \subset C$), so $C \in \mathcal{C}_{a,b,G}$.
%Clearly, $C$ is an $a,b$-separator in $G$.
By construction, $|C \cap V^{(k)}| = |S| < Q$, which is a contradiction. 
$\square$

\subsection{Proof for Corollary \ref{corollary_Q_proper}}

\begin{exe_cor}\label{corollary_appendix_Q_proper}
Given a graph $G$ and $K \geq 2$, consider a solution to the Hess model $\hat{\textbf{x}} \in \mathcal{F}_{HESS}$.
For $Q\geq 1$, $\hat{\textbf{x}}$ is $Q$-proper if and only if $\hat{\textbf{x}}$ satisfies constraints (\ref{eq_qconnected})-(\ref{eq_atleastQ}).
\end{exe_cor}
\textit{Proof:}
This proof uses Theorem \ref{thm_characterization}.
Given $\hat{\textbf{x}}$, we first construct the corresponding $K$-partition $\mathcal{P} = \{V^{(k)}\}_{k\in[K]}$ as follows.
Recall from Section \ref{subsec_Hessmodel} that $R(\hat{\textbf{x}})$ denotes the set of roots in $\hat{\textbf{x}}$, and  $V_r$ denotes the set of vertices assigned to root $r \in R(\hat{\textbf{x}})$.
Since $\hat{\textbf{x}}$ satisfies constraints (\ref{const_K_assns}), there are exactly $K$ roots.
Each root $r\in R(\hat{\textbf{x}})$ is labeled with a part $k \in [K]$ and  set $V^{(k)} = V_r$.
From constraints (\ref{const_unique_assn}) and (\ref{const_only_if_root}), we can see that $\cup_{k\in[K]} V^{(k)} = V$ and the parts are mutually disjoint.
Hence, $\mathcal{P} = \{V^{(k)}\}_{k\in[K]}$ is the $K$-partition of $G$ that corresponds to the solution $\hat{\textbf{x}}$.
Note that $\mathcal{P}$ can be equivalently written as $\{V_r\}_{r \in \mathcal{R}(\hat{\textbf{x}})}$.
We now prove both directions of the theorem.

%The rest of this proof focuses on each part $k \in [K]$ with a corresponding root $r$.
%The goal is to show that part $k$ is $Q$-connected if and only if $\hat{\textbf{x}}$ satisfies constraints (\ref{eq_qconnected})-(\ref{eq_atleastQ}) for the corresponding $r$ (i.e., $\sum_{c \in C} \hat{x}_{rc} \geq Q \hat{x}_{ra} \hat{x}_{rb} \ \forall \ a \in V,\ b\in V\backslash\{a\}, \ C\in \mathcal{C}_{a,b,G}$ and $\sum_{i \in V} \hat{x}_{ri} \geq (Q+1) \hat{x}_{rr}$).

($\Rightarrow$) 
We are given that $\hat{\textbf{x}}$ is $Q$-proper.
Hence, its representation $\mathcal{P}$ is $Q$-proper and  Theorem \ref{thm_characterization} implies that for every $r \in \mathcal{R}(\hat{\textbf{x}})$, (i) $|V_r| \geq Q+1$, and (ii) for every distinct $a,b \in V_r$, $|C\cap V_r| \geq Q$ for every $a,b$-separator $C \in \mathcal{C}_{a,b,G}$.
To show that $\hat{\textbf{x}}$ satisfies constraints (\ref{eq_qconnected}), consider each $r \in V$, $a \in V,\ b\in V\backslash\{a\}, \ C\in \mathcal{C}_{a,b,G}$, and the corresponding constraint is $\sum_{c \in C} \hat{x}_{rc} \geq Q \hat{x}_{ra} \hat{x}_{rb}$.
Note that this constraint does not exist for cases where $\mathcal{C}_{a,b,G} = \emptyset$ (i.e., $a$ and $b$ are neighbors in $G$).
There are two cases: (i) $r \in \mathcal{R}(\hat{\textbf{x}})$ and $a,b \in V_r$, in which case
constraint (\ref{eq_qconnected}) can be written as $|C\cap V_r| \geq Q$, which is satisfied by Theorem \ref{thm_characterization}, or (ii) either $r \notin \mathcal{R}(\hat{\textbf{x}})$ or $\{a,b\} \not\subset V_r$, in which case the right hand side of the constraint (i.e., $Q \hat{x}_{ra} \hat{x}_{rb}$) is zero and the constraint is always satisfied. 
To show that $\hat{\textbf{x}}$ satisfies constraint (\ref{eq_atleastQ}) for every $r \in V$, consider two cases: (i) if $r \in \mathcal{R}(\hat{\textbf{x}})$, the constraint is satisfied since $\hat{x}_{rr}=1$ and $\sum_{i \in V} \hat{x}_{ri} = |V_r| \geq Q+1$, or (ii) $r \notin \mathcal{R}(\hat{\textbf{x}})$, and so the right hand side of the constraint (i.e., $(Q+1)\hat{x}_{rr}$) is zero and the constraint is always satisfied.

%Note that this constraint is only for vertices $a$ and $b$ for which $\mathcal{C}_{a,b,G} \neq \emptyset$.

($\Leftarrow$) 
We are given that $\hat{\textbf{x}}$ satisfies constraints (\ref{eq_qconnected})-(\ref{eq_atleastQ}).
To show that $\hat{\textbf{x}}$ is $Q$-proper (or equivalently $\mathcal{P}$ is $Q$-proper), from Theorem \ref{thm_characterization}, it is sufficient to show that for every $r \in \mathcal{R}(\hat{\textbf{x}})$, (i) $|V_r| \geq Q+1$, and (ii) for every distinct $a,b \in V_r$ and $C \in \mathcal{C}_{a,b,G}$, $|C\cap V_r| \geq Q$.
%It is straight-forward to see that constraints (\ref{eq_qconnected}) and (\ref{eq_atleastQ}) imply these two conditions.
 Constraints (\ref{eq_qconnected}) imply that for every $r \in \mathcal{R}(\hat{\textbf{x}})$, for every distinct $a,b \in V_r$ and $C \in \mathcal{C}_{a,b,G}$, $|C\cap V_r| = \sum_{c \in C} \hat{x}_{rc} \geq Q \hat{x}_{ra} \hat{x}_{rb} =  Q$.
Constraints (\ref{eq_atleastQ}) imply that for every $r \in \mathcal{R}(\hat{\textbf{x}})$, $|V_r| = \sum_{i \in V} \hat{x}_{ri} \geq (Q+1)\hat{x}_{rr} = Q+1$. $\square$

\subsection{Proof for Theorem \ref{thm_separationalgocorrectness}}

\begin{exe_thm}\label{thm_appendix_separationalgocorrectness}
Given a graph $G=(V,E)$, an integer $Q \geq 1$, and a solution $\hat{\textbf{x}} \in \mathcal{F'}_{HESS}$, Algorithm  \ref{algo_intseparation_biconnected} adds violated separator constraints if and only if $\hat{\textbf{x}}$ is not $Q$-proper.
\end{exe_thm}
\textit{Proof.}
First, consider the case when $\hat{\textbf{x}}$ is $Q$-proper.
Equivalently, for every root $r \in R(\hat{\textbf{x}})$, the subgraph $G[V_r]$ is $Q$-connected $\Rightarrow$ the size of every cutset of $G[V_r]$ is at least $Q$ and $|V_r|\geq Q+1$.
Since $Q \geq 1$ implies that $G[V_r]$ is connected, the single component of $G[V_r]$ must contain $r$, and so lines 2-5 are skipped. 
In line 6, this single component is termed $G_r$, which is exactly $G[V_r]$. 
There are two cases for the set $D$ returned in line 7: either (i) $G[V_r]$ is a complete graph, in which case the algorithm returns $D = V_r \backslash \{r\}$, and hence $|D|= |V_r|-1\geq (Q+1) -1=Q$, or (ii) $G[V_r]$ is not a complete graph, and the algorithm returns $D$ as a minimum cutset for $G[V_r]$, and hence $|D| \geq Q$ since $G[V_r]$ is $Q$-connected. In either case, we have $|D| \geq Q$, and lines 9-18 are skipped. 
Therefore, Algorithm \ref{algo_intseparation_biconnected} does not add any violated $Q$-connectivity constraints.

\begin{figure}[!htbp]
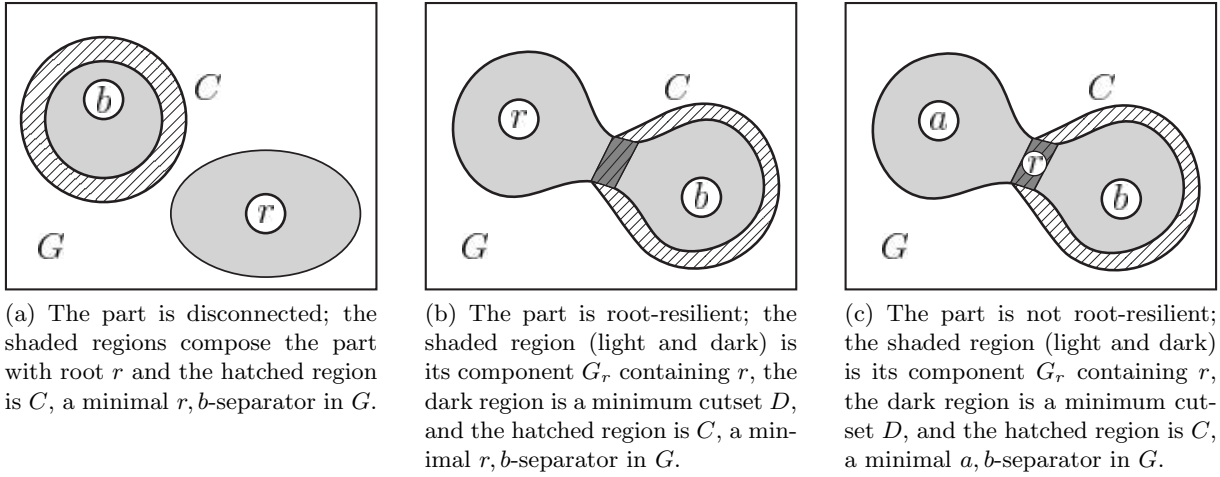

\centering
\subfloat[The part is disconnected; the shaded regions compose the part with root $r$ and the hatched region is $C$, a minimal $r,b$-separator in $G$. \label{fig_abstract_rep_1}]
{\includegraphics[width=0.3\textwidth]{figures_HCGP/Asset21.pdf}} 
%\vspace{.2cm}
\hspace{.5cm}
%\hfloat
\subfloat[The part is root-resilient; the shaded region (light and dark)  is its component $G_r$ containing $r$, the dark region is a minimum cutset $D$, and the hatched region is $C$, a minimal $r,b$-separator in $G$.\label{fig_abstract_rep_2}]
{\includegraphics[width=0.3\textwidth]{figures_HCGP/Asset28.pdf}} 
%\vspace{.2cm}
\hspace{.5cm}
%\hfloat
\subfloat[The part is not root-resilient; the shaded region (light and dark)  is  its component $G_r$ containing $r$, the dark region is a minimum cutset $D$, and the hatched region is $C$, a minimal $a,b$-separator in $G$.\label{fig_abstract_rep_3}]
{\includegraphics[width=0.3\textwidth]{figures_HCGP/Asset29.pdf}} 
%\vspace{.2cm}
%\hspace{0.1cm}
\caption{Abstract representations of a $Q$-disconnected part with root $r$ in graph $G$ considering three cases corresponding to the three sets of constraints added in Algorithm \ref{algo_intseparation_biconnected}. 
%The lightly shaded regions in (a) represent the part subgraph $G[V_r]$ and the lightly shaded regions in (b) and (c) represent the component $G_r$ of $G[V_r]$ containing $r$. 
%The darker regions in (b) and (c) represent a minimum cutset $D$  in $G_r$. The hatched region is $C$, a minimal $r,b$-separator in (a) and (b), and a minimal $a,b$-separator in (c).
} \label{fig_abstract_rep_appendix}
\end{figure}

Next, consider the case when $\hat{\textbf{x}}$ is not $Q$-proper.
For the rest of this proof, consider an arbitrary root $r \in R(\hat{\textbf{x}})$ such that $G[V_r]$ is not $Q$-connected.
Note that $G[V_r]$ is not a complete graph  since $|V_r|\geq Q+1$ from constraints (\ref{eq_degree_valid_constraints}); a complete graph would have connectivity $|V_r| - 1\geq Q$ and hence be $Q$-connected. 
First, we show that $\hat{\textbf{x}}$ violates at least one of the three sets of constraints added in lines 5, 13, and 18.
If $G[V_r]$ is disconnected, we show that $\hat{\textbf{x}}$ violates the constraints added in line 5; else, it violates either the constraints added in line 13 or the ones in line 18.
Then, we show that these sets of constraints are subsets of constraints (\ref{eq_qconnected}) and (\ref{eq_McCormick1})-(\ref{eq_McCormick3}).
%First, we show that $\hat{\textbf{x}}$ violates 
Figure \ref{fig_abstract_rep_appendix} depicts an abstract representation of the three cases  when Algorithm \ref{algo_intseparation_biconnected} adds violating constraints. 

The first set of constraints is added (in line 5) if $G[V_r]$ is disconnected, i.e., there exists a component $G_0$ of $G[V_r]$ that does not contain $r$.
In this case, line 3 first selects a vertex $b$ in $G_0$.
Line 4 finds a minimal $r,b$-separator $C$.
We now show that none of the vertices in $C$ are assigned to root $r$, i.e., $C \cap V_r = \emptyset$.
Consider the contrary, that there exists some vertex $u \in C \cap V_r$.
Line 23 implies that $u \in N(V[G_0])$, and hence does not belong to $G_0$, but is a neighbor of the vertices in $G_0$.
The fact that $u \in V_r$ is a contradiction since $G_0$ is a component of $G[V_r]$.
Hence, $C \cap V_r = \emptyset$, and therefore $\hat{x}_{rc}=0$ for all $c \in C$.
In the constraint in line 5 (i.e., $Q \hat{x}_{rb} \leq \sum_{c \in C} \hat{x}_{rc}$), since $\hat{x}_{rb} = 1$, and the right-hand side is zero, this constraint is violated.

The second set of constraints is added (in line 13) if $|D|<Q$ and $r \notin D$.  
Here, line 11 first selects a vertex $b$ in a component $G_0$ of $G_r-D$.
Subsequently, line 12 finds a minimal $r,b$-separator $C$ in $G$ such that $C \subseteq N(V[G_0])$. 
To compute $\sum_{c \in C} \hat{x}_{rc}$, we must identify the vertices of $C$ that are assigned to root $r$ (i.e., those with $\hat{x}_{rc}=1$). For any $u \in C$, there are two cases:
\begin{itemize}
\item $u \in D$: Since $D \subseteq V_r$, we have $u \in V_r$.
\item $u \notin D$: Recall that $C \subseteq N(V[G_0])$, and hence $u$ has a neighbor $v \in V[G_0] \subseteq V_r$. Since $G_0$ is a component of $G[V_r]$, then if both $u$ and $v$ are in $V_r$, they must be in the same component of $G[V_r]$; this cannot occur, since we cannot have both $u \in V[G_0]$ and $u \in N(V[G_0])$. Hence, we have $u \notin V_r$.
\end{itemize}
%We now show that if a vertex assigned to root $r$ is a part of $C$, then it is a part of $D$, i.e., $C\cap V_r  \subseteq D$.
%Assume the contrary that there exists a vertex $u \in (C\cap V_r) \backslash D$.
%Similar to the proof for the first set of constraints, Line 22 implies that $u \in N(V[G_0])$, and hence does not belong to $G_0$, but is a neighbor of the vertices in $G_0$.
%The fact that $u \in V_r\backslash D$ is a contradiction since $G_0$ is a component of $G_r - D$.
Hence, $C\cap V_r  \subseteq D$, and therefore $\sum_{c \in C} \hat{x}_{rc}\leq |D| < Q$.
In the constraint in line 13 (i.e., $Q \hat{x}_{rb} \leq \sum_{c \in C} \hat{x}_{rc}$), since $\hat{x}_{rb} = 1$, and the right-hand side is less than $Q$, this constraint is violated.

The third set of constraints is added (in line 18) if $|D|<Q$ and $r \in D$. 
Here, line 16 first selects vertices $a$ and $b$ from components $G_1$ and $G_2$, respectively, of $G_r-D$.
Line 17 finds a minimal $a,b$-separator $C$ in $G$.
First, we see that $C\cap V_r  \subseteq D$ using the same proof by contradiction for the second set of constraints.
Therefore, $\sum_{c \in C} \hat{x}_{rc}\leq |D| < Q$.
In the constraint in line 18 (i.e., $Q(\hat{x}_{ra}+\hat{x}_{rb}-1) \leq \sum_{c \in C} \hat{x}_{rc}$), since $\hat{x}_{ra} = \hat{x}_{rb} = 1$, and the right-hand side is less than $Q$, this constraint is violated.

We now show that each of the three sets of constraints is a subset of constraints (\ref{eq_qconnected}). 
For the first and second sets of constraints added in lines 5 and 13, respectively, the claim holds when we set $a=r$ and note that $\hat{x}_{ra} \hat{x}_{rb}= \hat{x}_{rr} \hat{x}_{rb}= \hat{x}_{rb}$, since $r$ is a root.
For the third set of constraints added in line 18, the claim holds from substituting $\hat{x}_{ra} \hat{x}_{rb}$ with $y^r_{ab}$ in constraint (\ref{eq_qconnected}), and then combining with constraint (\ref{eq_McCormick3}).
Hence, all three sets of constraints are $Q$-connectivity violated constraints.
$\square$

\subsection{Proof for Theorem \ref{thm_separationalgo_time}}

\begin{exe_thm} \label{thm_appendix_separationalgo_time}
Given a graph $G=(V,E)$ with $|V|=n$ and $|E|=m$, integers $Q \geq 1$ and $K\geq 2$, and a solution $\hat{\textbf{x}} \in \mathcal{F'}_{HESS}$, Algorithm \ref{algo_intseparation_biconnected} takes $O(Kmn^{5/3})$ time if all the parts of $\hat{\textbf{x}}$ are root-resilient to the cutsets in line 8 or $K\geq \sqrt[3]{n}$; otherwise, it takes $O(mn^2)$ time. 
\end{exe_thm}
\textit{Proof.}
First, finding a minimal separator from \cite{fischetti2017thinning} takes $O(m)$ time \citep{validi2020}. 
Since every vertex $b$ (in line 3) can be visited at most once among all roots $r$ and all components $G_0$, lines 1-5 take $O(mn)$ time.
Next, in lines 6 and 7, finding a minimum cutset in $G_r$  takes $O(mn^{5/3})$ time using Algorithm 11 in \cite{esfahanian2013connectivity}.
Lines 6 and 7 take $O(Kmn^{5/3})$ time when iterating over all the $K$ roots.
Next, in lines 9-13, every vertex $b$ can be visited at most once since each $b$ is present in exactly one component of $G_r-D$ among all roots $r$.
Hence, lines 9-13 take $O(mn)$ time using the same reasoning for lines 1-5.
If all the parts are root-resilient, only lines 1-13 are executed, and hence the overall run-time is $O(Kmn^{5/3})$.

If some part is not root-resilient, lines 15-18 are also executed.
Here, each pair of vertices $a$ and $b$ are visited exactly once among all roots $r$ and all pairs of components $G_1$ and $G_2$, resulting in $O(n^2)$ iterations of the for-loop in line 15, where each iteration takes $O(m)$ to generate a minimal separator and add the cut.
Lines 15-18 take $O(mn^2)$ time.
Hence, if some parts are not root-resilient, executing lines 1-18 takes $O(Kmn^{5/3}+mn^2)$ time, which is $O(Kmn^{5/3})$ time if $K \geq \sqrt[3]{n}$; otherwise, it takes $O(mn^2)$ time.
$\square$

\section{Pseudocodes for the Ear Construction Heuristic} \label{appendix_pseudocode}

This appendix provides pseudocodes for the ear construction heuristic from Section \ref{sec_ear_construction_heur}.

\subsection{Overview}

\begin{algorithm}[H]
\SetKwInOut{Input}{Input}
\SetKwInOut{Output}{Output}
\SetKwProg{myproc}{Procedure}{}{}   
$\mathcal{P} \leftarrow$ Construct an Initial Partition ($G$, $K$, $L$) \\
%$\mathcal{P}_{merged} \leftarrow$ Merge districts in $\mathcal{P}_{ED}$ to get $K$ districts \\
\If{$\mathcal{P}$ is not $2$-proper}{
	$\mathcal{P} \leftarrow$ Repair $2$-Connectivity ($G$, $\mathcal{P}$) \\
	\If{$\mathcal{P}$ is not $2$-proper}{
		Go to line 1\\	
	}
}
\If{$\mathcal{P}$ is not $L,U$-balanced}{
	$\mathcal{P} \leftarrow$ Local Search ($G$, $\mathcal{P}$, $\phi_{bal}$)  \\
	\If{$\mathcal{P}$ is not $L,U$-balanced}{
		Go to line 1\\
	}
}
$\mathcal{P} \leftarrow$ Local Search ($G$, $\mathcal{P}$, $\phi_{comp}$) \\
\Return{$\mathcal{P}$}
\caption{Ear-Construction Heuristic ($G$, $K$, $L$, $U$)} \label{algo_heuristic_overview}
\end{algorithm} 

Algorithm \ref{algo_heuristic_overview} provides an overview of the heuristic.
Line 1 executes the construct stage described in Section \ref{subsec_constructinitial}, lines 2-5 execute the repair stage described in Section \ref{subsec_repair}, and lines 6-10 execute the improvement stages described in Section \ref{subsec_improve}.
The pseudocodes for these stages are provided in the following subsections.

\subsection{Ear Construction}
\begin{algorithm}[H]
\SetKwInOut{Input}{Input}
\SetKwInOut{Output}{Output}
\SetKwProg{myproc}{Procedure}{}{} 
%$T \leftarrow \emptyset$ Set of traversed vertices \\ 
$P_0 \leftarrow$ Find a cycle in $H$\\
%$(u,v) \leftarrow$ Select an edge from $G$ \\
%$C \leftarrow$ Smallest cycle in $G$ containing $(u,v)$ \\
$T \leftarrow P_0$: Set of traversed vertices \\ 
\While{$size(T) < L$}{
	$T^C \leftarrow V[H]\backslash T$: Set of untraversed vertices \\
	Randomly shuffle $T$ \\
	\For{$(u,v) \in T \times T$ with $u\neq v$}
	{
		%	$\overline{G}(\overline{V},\overline{E}) \leftarrow G-T$: Graph of untraversed vertices \\
%	$u,v \leftarrow$ Select two vertices in $T$ \\
	Assign $H_{u,v} \leftarrow H[T^C\cup\{u,v\}]$ and remove edge $(u,v)$ from $H_{u,v}$ \\
	\If{there exists a $u,v$-path in $H_{u,v}$}{
		$P \leftarrow $ Shortest $u,v$-path in $H_{u,v}$ \\
		$T \leftarrow T \cup P$ \\
%		$\overline{G} \leftarrow \overline{G}-p$ \\
		Go to line 3 \\
		}
	}

	\If{no $u,v$-path exists in $H_{u,v}$ for any $(u,v) \in T \times T$ with $u\neq v$}
	{
		\Return $T$
	}
}
\Return{$T$}
\caption{Ear Construction ($H$, $L$)} \label{algo_2_connected_subgraph}
\end{algorithm} 

Algorithm \ref{algo_2_connected_subgraph} outlines a pseudocode to construct a $2$-connected subgraph of a graph $H$ as described in Section \ref{subsec_earconstruction}. 
%We also ensure that the population of the $2$-connected subgraph is at least a given parameter $L$, or is maximally populated (i.e., no vertices can be added to this subgraph while retaining its $2$-connectivity). 
This procedure follows the construction procedure in Definition \ref{defn_open_ear_decomp}.
Line 1 finds a random cycle as follows. First, the procedure  randomly selects two neighboring vertices $u$ and $v$ in $H$ and finds a shortest path in $H$ after removing the edge $(u,v)$; if no such path exists, this process is repeated with different pairs of neighboring vertices. 
The vertices traversed in the resulting path are assigned as the cycle $P_0$.
If $H$ does not contain any cycles (i.e., $P_0=\emptyset$), indicating that it is a forest, the algorithm returns an empty graph since no $2$-connected subgraph exists.
If a cycle is identified, lines 3-11 iteratively search for open ears and add them to the traversal.
In each iteration, consistent with the procedure in Definition \ref{defn_open_ear_decomp}, the procedure first selects a pair of vertices $u$ and $v$ in the traversed vertices such that there is a $u,v$-path in the remaining graph.
If such a path exists, the procedure finds the shortest vertex-weighted $u,v$-path and add those vertices to the traversal; this path must include at least one vertex from $V[H]\backslash T$, since the edge $(u,v)$ does not appear in $H_{u,v}$.
This procedure continues until either the total size of the traversed vertices is at least $L$, or if no more open ears can be added.
Hence, Algorithm \ref{algo_2_connected_subgraph} either returns the set of vertices in a $2$-connected subgraph of $H$ or an empty set.
%Algorithm \ref{algo_2_connected_subgraph} provides a pseudocode for this procedure, and Lemma \ref{lemma_2conn_subgraph} ensures the $2$-connectivity of the subgraph given a cyclical graph $G$.

\subsection{Construct an Initial Partition}

\begin{algorithm}[!ht]
\SetKwInOut{Input}{Input}
\SetKwInOut{Output}{Output}
\SetKwProg{myproc}{Procedure}{}{} 
$G'(V',E') \leftarrow G(V,E)$: Remaining graph \\
$K' \leftarrow 0$: Initiate the label for the current part\\
\While{$G'$ contains a cycle}{ 
	$K' \leftarrow K'+1$ \\ 
	$V^{(K')} \leftarrow$ Ear Construction($G'$, $L$) \\	 
	$G' \leftarrow G' - V^{(K')}$ \\
}
\For{tree $T$ of $G'$}{
	$K' \leftarrow K'+1$	\\
	$V^{(K')} \leftarrow T$ \\
}
\If{$K' < K$}{
	Go to Line 1 \\
}
\While{$K' > K$}
{
	Select two neighboring parts and merge; proceed in the non-decreasing order of pair-wise total part sizes \\
} 
\Return{$\{V^{(k)}\}_{k=1}^{K}$}  
\caption{Construct an Initial Partition ($G$, $K$, $L$)} \label{algo_find_2con_districts}
\end{algorithm}

Algorithm \ref{algo_find_2con_districts} outlines a pseudocode for the procedure described in Section \ref{subsec_constructinitial}.
Throughout this procedure, let $G'$ (initialized with $G$) denote the remaining graph.
Lines 3-6 iteratively find a set of $2$-connected subgraphs using Algorithm \ref{algo_2_connected_subgraph}.
After these iterations, if $G'$ is a forest, lines 7-9 designate each tree in the forest as an individual part.
If $K'<K$, line 11 directs the algorithm to restart.
If $K' > K$, line 13 proceeds to merge parts iteratively until $K'=K$. 
Finally, Algorithm \ref{algo_find_2con_districts} returns a partition with $K$ connected parts, some of which may be $2$-disconnected.

\subsection{Repair $2$-Connectivity}

\begin{algorithm}[H]
\SetKwInOut{Input}{Input}
\SetKwInOut{Output}{Output}
\SetKwProg{myproc}{Procedure}{}{} 
\While{$\{V^{(k)}\}_{k=1}^K$ is not $2$-proper}{
	$k \leftarrow$ a random $2$-disconnected part \\
	$c \leftarrow$ a random cut vertex in $G[V^{(k)}]$ \\
	$S \leftarrow$ smallest component of $G[V^{(k)}]-\{c\}$ \\
	\If{$S$ has a neighbor in another part}{
		$k' \leftarrow$ a random part neighboring $S$ \\
		$V^{(k)} \leftarrow V^{(k)} \backslash S$ \\
		$V^{(k')} \leftarrow V^{(k')} \cup S$ \\
	}
	\If{$S$ has no part neighbors or cycling is detected}{
		Exit the while loop  
		\tcp{the algorithm restarts from the construct stage}
	}
}
\Return{$\{V^{(k)}\}_{k=1}^K$}
\caption{Repair $2$-Connectivity ($G$, $\{V^{(k)}\}_{k=1}^K$)} \label{algo_repair_heuristic}
\end{algorithm} 

Algorithm \ref{algo_repair_heuristic} outlines a pseudocode for the repair stage  as described in Section \ref{subsec_repair}.
If the given partition is not $2$-proper, line 2 randomly selects a $2$-disconnected part $k$, line 3 randomly selects a cut vertex $c$, and line 4 selects a smallest component $S$ produced by removing the cut vertex; if there are ties, it selects a random such component.
The vertices in $S$ are reassigned to a randomly chosen neighboring part $k'$
% (if one exists) 
 in lines 6-8.
Note that an alternative choice for selecting the neighboring part could be to select a part that maintains $2$-connectivity after receiving the vertices, albeit at the expense of needing to recompute each potential recieving part's connectivity. 
%Even though
%Assuming that there is no cycling, note that the while loop will have less than $n$ iterations.
Then, the algorithm assesses a premature termination criterion in line 9, and it exits the loop.
%Here, the cycling condition in line 9 is heuristically assessed by checking whether the previous $100$ iterations are exclusively composed of repeated reassignments. 
Hence, the algorithm either terminates with a $2$-proper partition and proceeds to the next stage, or without a $2$-proper partition, in which case, the heuristic restarts from the construct stage.

\subsection{Local Search Improvement}

\begin{algorithm}[H]
\SetKwInOut{Input}{Input}
\SetKwInOut{Output}{Output}
\SetKwProg{myproc}{Procedure}{}{} 
$\phi_{best} \leftarrow \phi(\{V^{(k)}\}_{k=1}^K)$ \\
\While{termination criterion is not True}{
	$\mathcal{T} \leftarrow$ Randomly ordered set of reassignments \\
	\tcp{each element  in $\mathcal{T}$ is a vertex $u$ and a part $k'$ neighboring $u$}  
%	$u \leftarrow $ a random vertex currently in part $k$ that neighbors part $k'$ \\
	\For{$(u,k') \in \mathcal{T}$}
	{
		\If{$G[V^{(k)}]-\{u\}$ and $G[V^{(k')}\cup\{u\}]$ are both $2$-connected}{ 
			\If{$\phi$ improves after moving $u$ to $k'$}{
				$V^{(k)} \leftarrow V^{(k)}\backslash \{u\}$ \\
				$V^{(k')} \leftarrow V^{(k')}\cup\{u\}$ \\
	$\phi_{best} \leftarrow \phi(\{V^{(k)}\}_{k=1}^K)$ \\	
	%			$\phi_{best} \leftarrow \phi(\{V^{(k)}\}_{k=1}^K) $\\ 
	Go to line 2
			}
		}
	}
}
\Return{$\{V^{(k)}\}_{k=1}^K$}
\caption{Local Search ($G$, $\{V^{(k)}\}_{k=1}^K$, $\phi$)} \label{algo_localsearch}
\end{algorithm} 

Algorithm \ref{algo_localsearch} outlines the local search method  described in Section \ref{subsec_improve} to optimize an objective $\phi \in \{\phi_{bal},\phi_{comp}\}$.
In each iteration, line 3 creates a random ordering of potential reassignments.
Line 4 iterates through each potential reassignment of a vertex $u$ to a neighboring part $k'$. 
Line 5 checks whether this reassignment retains $2$-connectivity in both parts, and line 6 checks whether this reassignment improves $\phi$.
Additionally, when optimizing compactness, each iteration must ensure that the size balance objective remains zero.
If these conditions are true, lines 7-9 reassign $u$ to $k'$ and update the current best objective value. 
% and the procedure continues.
Furthermore, the termination criterion in line 2 is now described.
When optimizing size balance, the algorithm terminates when $\phi_{bal}$ is zero (i.e., the partition in $L,U$-balanced), or when no more feasible reassignments improve $\phi_{bal}$ (i.e., the local search has not balanced the parts, in which case the heuristic will restart from the construct stage).
When optimizing compactness, the algorithm terminates when no feasible reassignments improve $\phi_{comp}$, i.e., the partition is locally optimal.

%\newpage
\section{Tables}

This appendix provides tabular information on the computational analysis from Section \ref{sec_computationalresults}. 

\subsection{Overview of Instances}\label{appendix_overview_table}
Table \ref{table_dataset_full} depicts an overview of the $42$ graph instances, along with their connectivities ($\kappa$), average and minimum degrees, as well as the number of vertices removed or edges added in the pre-processing stage.
%
% \begin{table}[!ht]
%\centering
%\csvreader[ 
%before reading=\footnotesize\sisetup{round-mode=places,round-precision=2,round-integer-to-decimal},
%tabular=cccccccccc,
%table head= \toprule  
%Instance &	$|V|$ &	$|E|$  & $K$  & $OPT$ & \multirow{2}{*}{\begin{tabular}[c]{@{}c@{}} $OPT$ \\ time (s) \end{tabular}} & Heuristic 	& \multirow{2}{*}{\begin{tabular}[c]{@{}c@{}} Heuristic \\ time (s) \end{tabular}} & $OPTWS$  & \multirow{2}{*}{\begin{tabular}[c]{@{}c@{}} $OPTWS$ \\ time (s) \end{tabular}} \\ \\
%\midrule,
%  late after last line=\\  \midrule,
%%    late after line=\\\midrule,
%]{results/chicago/table_CHIfullcagosubgraphs_n500_processed.csv}{}%
%{\csvcoli & \csvcolii & \csvcoliii & \csvcolvi & \csvcolxi & \csvcolxiii & \csvcolxv & \csvcolxvi & \csvcolxvii & \csvcolxix}
%\caption{Optimal solutions for HCGP obtained from $30$ transportation network graph instances, number of districts $K=1,2,3,4$ and $Q=1,2$. \label{table_soc_transposubgraphs_results}}
%\end{table}

 \begin{table}[!ht]
 \caption{An overview of $42$ graph instances from seven domains with their sizes, densities and the vertices removed/edges added in the pre-processing stage.
\label{table_dataset_full}}
\centering
\csvreader[ 
before reading=\footnotesize\sisetup{round-mode=places,round-precision=2,round-integer-to-decimal},
tabular=ccccccccc,
table head = \toprule  
Domain & Graph $(G)$ & $|V|$ & $|E|$ & $\kappa(G)$ 
 & {\begin{tabular}[c]{@{}c@{}} Avg. \\ deg.\end{tabular}} 
 & {\begin{tabular}[c]{@{}c@{}} Min. \\ deg.\end{tabular}}
 & {\begin{tabular}[c]{@{}c@{}} Vertices \\ removed \end{tabular}}
 & {\begin{tabular}[c]{@{}c@{}} Edges \\ added \end{tabular}}
\\
\midrule,
  late after last line=\\  \midrule, 
]{tables_tex/all_instances_raw.csv}{}%
{\csvcoli & \csvcolii & \csvcolxii & \csvcolxiii & \csvcolxiv & \csvcolxv & \csvcolxvi & \csvcolxx & \csvcolxxi}
\end{table}

 \clearpage
\subsection{Results for HCGP}\label{appendix_results_HCGP}

Table \ref{table_crosstabulated_Q2} presents the distribution of instances for HCGP with $Q=2$ (among the $252$ instances solved in Section \ref{subsec_results_BCGP}), cross-tabulated for the two methods.
For example, $23$ instances were inconclusive for the exact method and yet feasible (but suboptimal) for the heuristic.

\begin{table}[!ht]
\centering
\caption{Cross-tabulated distribution of $252$ instances for HCGP with $Q=2$ based on solution outcomes for the two methods: optimal, feasible (and suboptimal), infeasible, or inconclusive.}
\begin{tabular}{@{}cccccc@{}}
\toprule
%\cmidrule{3-6}
                                  &     & \multicolumn{4}{c}{Exact method} \\ \cmidrule{3-6}
                                  &     & Optimal       & Feasible    & Infeasible    & Inconclusive    \\ \cmidrule{1-6}
\multirow{4}{*}{{\begin{tabular}[c]{@{}c@{}}Heuristic \\ method \end{tabular}}} & Optimal   & 9       & 0    & 0      & 0      \\
                                  & Feasible   & 187     & 9    & 0      & 23     \\
                                  & Infeasible & 0       & 0    & 0      & 0      \\
                                  & Inconclusive & 8       & 4    & 9      & 3      \\ 
%                                  \cmidrule(l){2-6} 
                                  \bottomrule
\end{tabular}  \label{table_crosstabulated_Q2}
\end{table} 

Tables \ref{table_results_Q=2},  \ref{table_dataset_full_higherconnected} and \ref{table_results_Q1} present results for HCGP with $Q=2$, $Q\in \{3,4\}$, and $Q=1$, respectively. 
Each table includes the optimal compactness value ($OPT$) and the corresponding computational time taken by the branch-and-cut method ($OPT$ time) in seconds.
An asterisk (*) or a double asterisk (**) in the $OPT$ column indicates that the solver reached the time limit, thus representing the best known lower bound on the optimal solution.
Furthermore, an asterisk denotes that a feasible solution was found by the solver, while a double asterisk indicates that no feasible solution was obtained within the given time limit.
If the solver produced an infeasibility certificate for a particular instance, the $OPT$ column displays ``inf".
The computational times reported in these tables are rounded to the nearest first decimal place; an entry with ``0.0" indicates a computational time less than 0.05 seconds.

In Table \ref{table_results_Q=2}, the compactness objective of the solution obtained from the ear construction heuristic is reported in the $HEUR$ column.
The $HEUR$ time column reports the time taken by the heuristic in seconds.
If the heuristic terminated without a feasible solution within the time limit, the $HEUR$ column displays ``N.S.", abbreviating ``no solution".
The approximation gap and the heuristic time ratio are defined in Section \ref{subsec_results_BCGP}.
%The Approx. Gap column displays ``N.A." when the heuristic terminated without a feasible solution or when the  
%$OPT$ column displays ``$0$**".
The Approx. Gap column displays ``N.A." when either the exact method, the heuristic method, or both, terminated without a feasible solution.
When the heuristic found an optimal solution, the Approx. Gap column displays ``-".

{\small
}

%\clearpage
%\subsection{Results for Higher Connected Partitioning}

In Table \ref{table_dataset_full_higherconnected}, the column ``Callback time" reports the total time spent in the callbacks, measured in seconds.
When the branch-and-cut method does not enter any callback, this column reports a ``-".
Further, the ``Conn." column reports the individual connectivities of the $K$ parts.

%%%  {\footnotesize\setlength{\tabcolsep}{3pt}
{\small  
}

%\section{Table}
% 
%\begin{table}[!ht]
%\centering
%\begin{tabular}{ccccc}
%\toprule
%\multirow{2}{*}{$\tau$} & \multirow{2}{*}{$K$} & \multicolumn{3}{l}{Number of instances with optimal solutions for given $Q$} \\ \cmidrule(l){3-5} 
%                         &                    & $Q \in \{1,2\}$ & $Q \in \{1,3\}$            & $Q \in \{1,4\}$     \\ \midrule
%$\infty$                 & 2                  & 42                  & 18                  & 15                 \\
%$\infty$                 & 3                  & 40                  & 16                  & 13                 \\
%$\infty$                 & 4                  & 37                  & 16                  & 14                 \\
%0.1                      & 2                  & 32                  & 19                  & 15                 \\
%0.1                      & 3                  & 28                  & 14                  & 8                  \\
%0.1                      & 4                  & 25                  & 14                  & 8                  \\ \bottomrule
%\end{tabular}
%\caption{An example table with adjusted column widths.}
%\label{table_n_instances_Q}
%\end{table}

\end{appendices}

\end{document}